\documentclass[12pt,preprint]{aastex}

\shorttitle{Structure and Dynamics of Cold Water Super-Earths}
\shortauthors{Levi, Sasselov, and Podolak}

\usepackage{graphicx}
\usepackage{textcomp}
\usepackage{amssymb}
\usepackage{natbib}
\usepackage{amsmath}
\usepackage{subfigure}
\usepackage{multirow}\usepackage[section] {placeins}

\begin{document}
\pagenumbering{arabic}

\title{Structure and Dynamics of Cold Water Super-Earths: The Case of Occluded CH$_4$ and its Outgassing}
\author{A. Levi}
\affil{Dept. of Geophysics \& Planetary Science, Tel Aviv University, Tel Aviv, Israel 69978}
\affil{Harvard-Smithsonian Center for Astrophysics, 60 Garden Street, Cambridge, MA 02138, USA}
\email{amitlevi.planetphys@gmail.com}
\author{D. Sasselov}
\affil{Harvard-Smithsonian Center for Astrophysics, 60 Garden Street, Cambridge, MA 02138, USA}
\and
\author{M. Podolak}
\affil{Dept. of Geophysics \& Planetary Science, Tel Aviv University, Tel Aviv, Israel 69978}

\maketitle

\section*{ABSTRACT}

In this work we study the transport of methane in the external water envelopes surrounding water-rich super-Earths. We investigate the influence of methane on the thermodynamics and mechanics of the water mantle. We find that including methane in the water matrix introduces a new phase (filled ice) resulting in hotter planetary interiors. This effect renders the super-ionic and reticulating phases accessible to the lower ice mantle of relatively low mass planets ($\sim 5$\,M$_E$) lacking a H/He atmosphere. We model the thermal and structural profile of the planetary crust and discuss five possible crustal regimes, which depend on the surface temperature and heat flux. We demonstrate the planetary crust can be conductive throughout or partly confined to the dissociation curve of methane clathrate hydrate. The formation of methane clathrate in the subsurface is shown to inhibit the formation of a subterranean ocean. This effect results in increased stresses on the lithosphere making modes of ice plate tectonics possible. The dynamic character of the tectonic plates is analysed and the ability of this tectonic mode to cool the planet estimated. The icy tectonic plates are found to be faster than those on a silicate super-Earth. A mid-layer of low viscosity is found to exist between the lithosphere and lower mantle. Its existence results in a large difference between ice mantle overturn time scales and resurfacing time scales. Resurfacing time scales are found to be $1$\,Ma for fast plates and $100$\,Ma for sluggish plates, depending on the viscosity profile and ice mass fraction.
Melting beneath spreading centres is required in order to account for the planetary radiogenic heating. The melt fraction is quantified for the various tectonic solutions explored, ranging from a few percent for the fast and thin plates to total melting of the upwelled material for the thick and sluggish plates.  
Ice mantle dynamics is found to be important for assessing the composition of the atmosphere. We propose a mechanism for methane release into the atmosphere, where freshly exposed reservoirs of methane clathrate hydrate at the ridge dissociate under surface conditions. We formulate the relation between the outgassing flux and the tectonic mode dynamical characteristics. We give numerical estimates for the global outgassing rate of methane into the atmosphere. We find, for example, that for a $2M_{E}$ planet outgassing can release $10^{27}$ to $10^{29}$\,molec\,s$^{-1}$ of methane to the atmosphere. We suggest a qualitative explanation for how the same outgassing mechanism may result in either a stable or a runaway volatile release, depending on the specifics of a given planet. Finally, we integrate the global outgassing rate for a few cases and quantify how the surface atmospheric pressure of methane evolves over time. We find that methane is likely an important constituent of water planets' atmospheres.


\section{INTRODUCTION}

Planets that are intermediate in size and mass between Earth and Neptune are very common \citep{Mayor2011,Howard2012,Fressin2013}. They show a wide range of densities, with some of them indicating the existence of large water-rich interiors (water mantles), as in the case of exoplanet Kepler-68b \citep{Gilliland2013}. Such a planet would have a mass of several Earths, with about half of that mass in water - a water-rich mantle surrounding a core of heavy elements (rocky-iron composition). Such planets are unknown in our Solar System and the extremely high pressures in their mantles pose challenges to theoretical models. Interior modelling has treated the water as pure \citep[see][and references therein]{Zeng2013}. while this may be sufficient to reproduce the planetary radius for a given planet mass, the neglect of treating volatiles emerging from the core and mixed in with the water is a serious impediment to understanding the atmospheres such planets might have.

Interpreting the atmospheres of super-Earths, their structure and composition as gleaned from spectroscopic observations, is now upon us with the high-quality data for the nearby exoplanet GJ1214b \citep[see][and references therein]{Kreidberg2013}. Modelling such atmospheres in the habitable zone, as is the case of exoplanet Kepler-78f \citep{Kaltenegger2013}, is crucial in determining the outer boundary of the zone and whether a particular exoplanet is potentially habitable. The success of such modelling depends heavily on the gas fluxes emerging from the planet's interior (as well as returning, in most cases). The volatile transport inside massive planets that are rich in water is poorly studied and understood, and to that end we have initiated a study of the entrapment of gases in the water-ice matrix \citep{Levi2013}. 

Methane and other abundant gases trapped inside water super-Earth planets will be occluded in the solid-state water matrix while traversing the deep water mantle between the silicate-rich planet core and the planet's surface. Chemical occlusion of methane takes the form of methane hydrate, also known as methane clathrate, or filled ice (at very high pressure). Filled ice, a novel occlusion structure \citep{loveday01}, is particularly relevant for water super-Earths because of the high pressures reached in their water mantles. We studied the relevance of filled ice to super-Earths in \cite{Levi2013}; this paper is a natural continuation of that work into understanding the phase transitions and dynamics near the planet's surface for a complete physical picture of the outgassing process.

The paper consists of $9$ sections and $3$ appendices. In section $2$ we discuss the planetary crust and its dynamics by following the common path of describing the thermodynamics (of clathrates), their rheology, and the convective stability; we end by describing five different structural regimes of the crust. In section $3$ we describe briefly the deeper planetary structure. In section $4$ we discuss the thermal profile of the (water) mantle, and in section $5$ we use all these results to discuss the tectonics of water planets. We summarize our findings by describing resulting methane outgassing in section $6$. In section $7$ we discuss a number of details - possible caveats and improvements of the model developed here, and finally we summarize our findings in section $8$.

\section{THE PLANETARY CRUST}

In this section we present a thermal model for a planetary crust which is composed of a structure I (SI) methane clathrate hydrate.  A full solution of the near surface heat transfer problem would involve modeling convection cells, which, in turn, requires a detailed simultaneous treatment of the thermodynamic and hydrodynamic behavior of the system.  To avoid this complexity, we rely on a parameterization of the problem using dimensionless parameters, such as the Nusselt and Rayleigh numbers \citep{holman}, and scaling techniques.  Such an approach is generally in good agreement with more complex numerical modeling.  In addition, the physical insight gained in this way can be extremely valuable \citep{solomatov1995}. 
      
Perhaps the most important dimensionless parameter describing this region is the Rayleigh number, $Ra$, a measure of the importance of buoyancy in driving convection, which is given by:
\begin{equation}\label{rayleighnum}
Ra=\frac{g\chi\Delta Td^3}{\nu\alpha}
\end{equation}
Here $g$ is the acceleration of gravity, $\chi$ is the volume thermal expansivity, $d$ is the convective length scale, $\Delta T$ is the temperature difference driving the convection, $\alpha$ is the thermal diffusivity, and $\nu$ is the kinematic viscosity.

\subsection{Clathrate Thermodynamics}

\begin{figure}[ht]
\centering
 \includegraphics[scale=0.5]{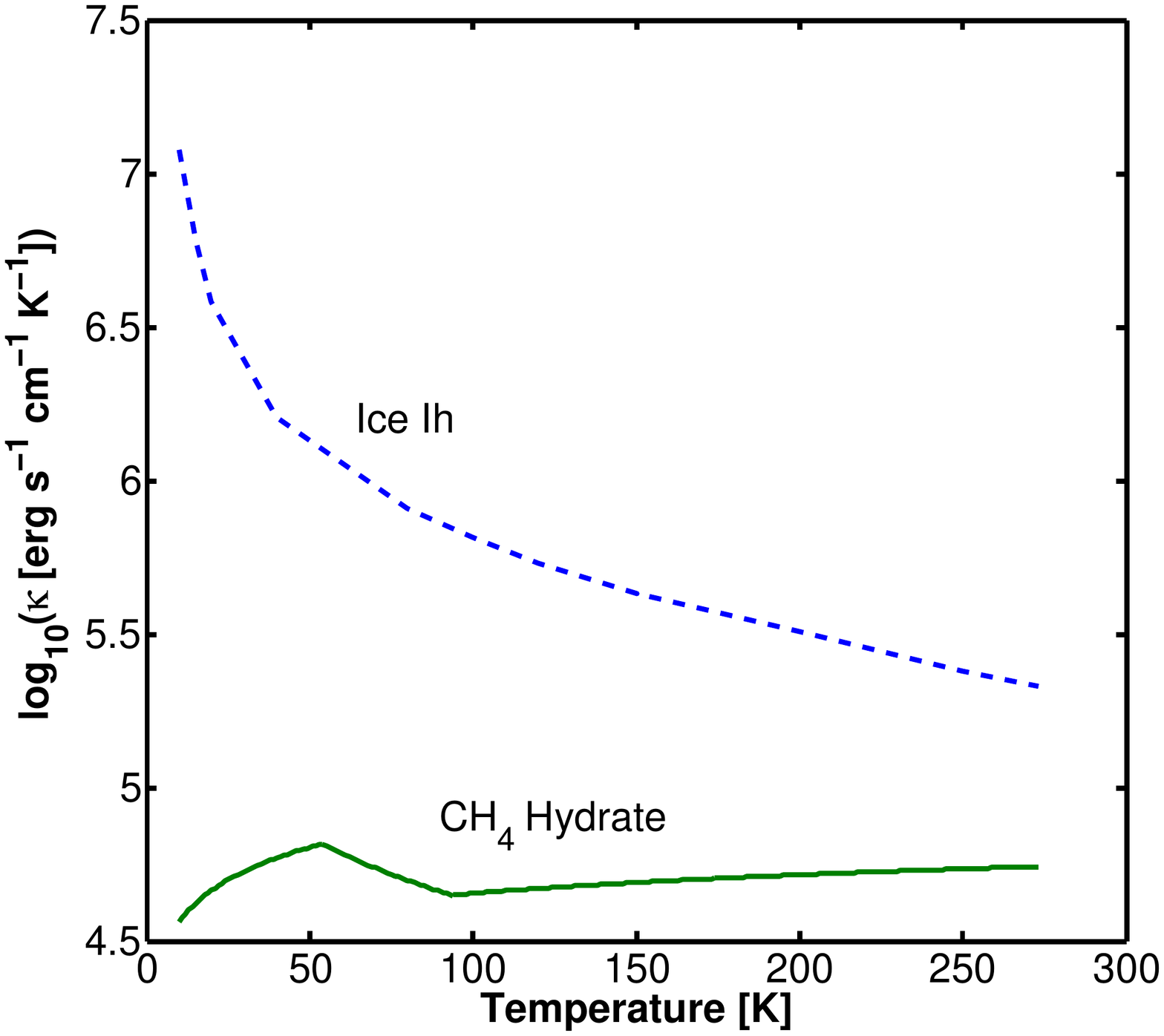}
\caption{\footnotesize{Thermal conductivity of methane hydrate as deduced from experiment \citep{krivchikov2006}. For comparison we also show the thermal conductivity of ice Ih \citep{slack1980}.}}
\label{fig:f1}
\end{figure}

We first assess the numerical values of the different thermodynamic parameters for a SI methane clathrate. Clathrates, though crystalline in structure, have thermal conductivities that are closer to those of amorphous solids \citep[see, e.g.][and the references therein]{tse1997,krivchikov2005}. This is due to the "rattling" effect of the guest molecule, whose rattling frequencies may overlap the host network acoustic frequencies, resulting in a large guest-host thermal interaction. The thermal conductivity of methane clathrate is experimentally found to obey the following rules \citep{krivchikov2006}:
\begin{equation}\label{ThermalConductivityClathrate}
\kappa_{cl}(T)=\tilde{\kappa} \left(\frac{T}{\tilde{T}}\right)^{\hat{n}} 
\end{equation}
$$
6K<T<54K \quad \hat{n}=0.35 \quad \tilde{T}=40K \quad \tilde{\kappa}=5.90\times 10^4
$$
$$
54K<T<94K \quad \hat{n}=-0.7 \quad \tilde{T}=80K \quad \tilde{\kappa}=4.98\times 10^4
$$
$$
94K<T \quad \hat{n}=0.2 \quad \tilde{T}=100K \quad \tilde{\kappa}=4.52\times 10^4
$$
Where $\tilde{\kappa}$ is in erg s$^{-1}$ cm$^{-1}$ K$^{-1}$. In Fig.\,\ref{fig:f1} we show the thermal conductivity of methane clathrate hydrate as compared with the thermal conductivity of water ice Ih, the latter given by \cite{slack1980}. 

We describe the density, $\rho_{cl}$, of the methane hydrate by a third order Birch-Murnaghan EOS. \cite{cox} gives a value of $0.912$~g\,cm$^{-3}$ for the bulk mass density of an SI methane hydrate under low pressure.  We use this as the zero pressure density. We take a bulk modulus of $8$~GPa and a bulk modulus pressure derivative of $7.61$ based on the experimental data of \cite{shimizu2002}.

\begin{figure}[ht]
\centering
\mbox{\subfigure{\includegraphics[width=7cm]{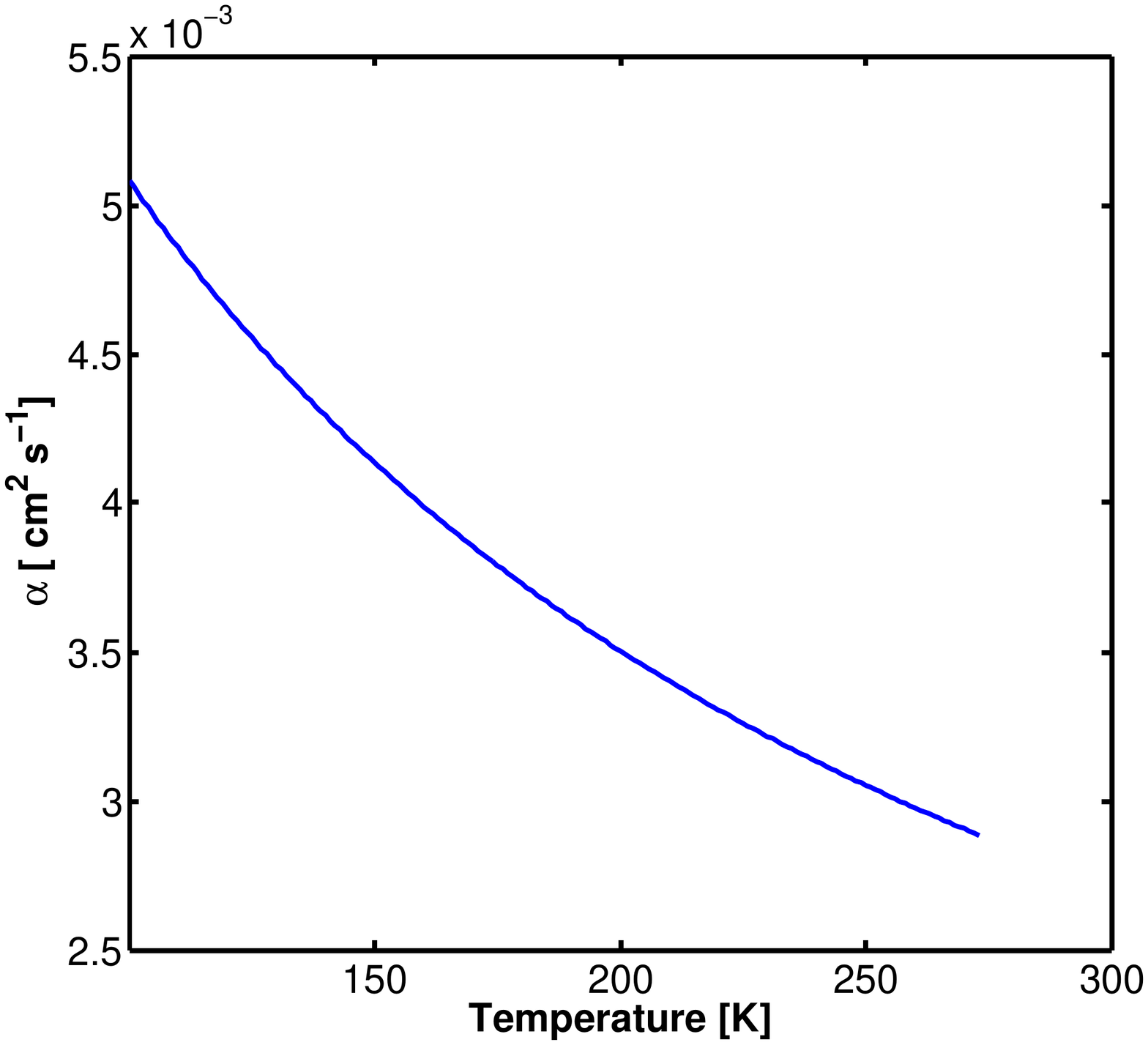}}\quad \subfigure{\includegraphics[width=7cm]{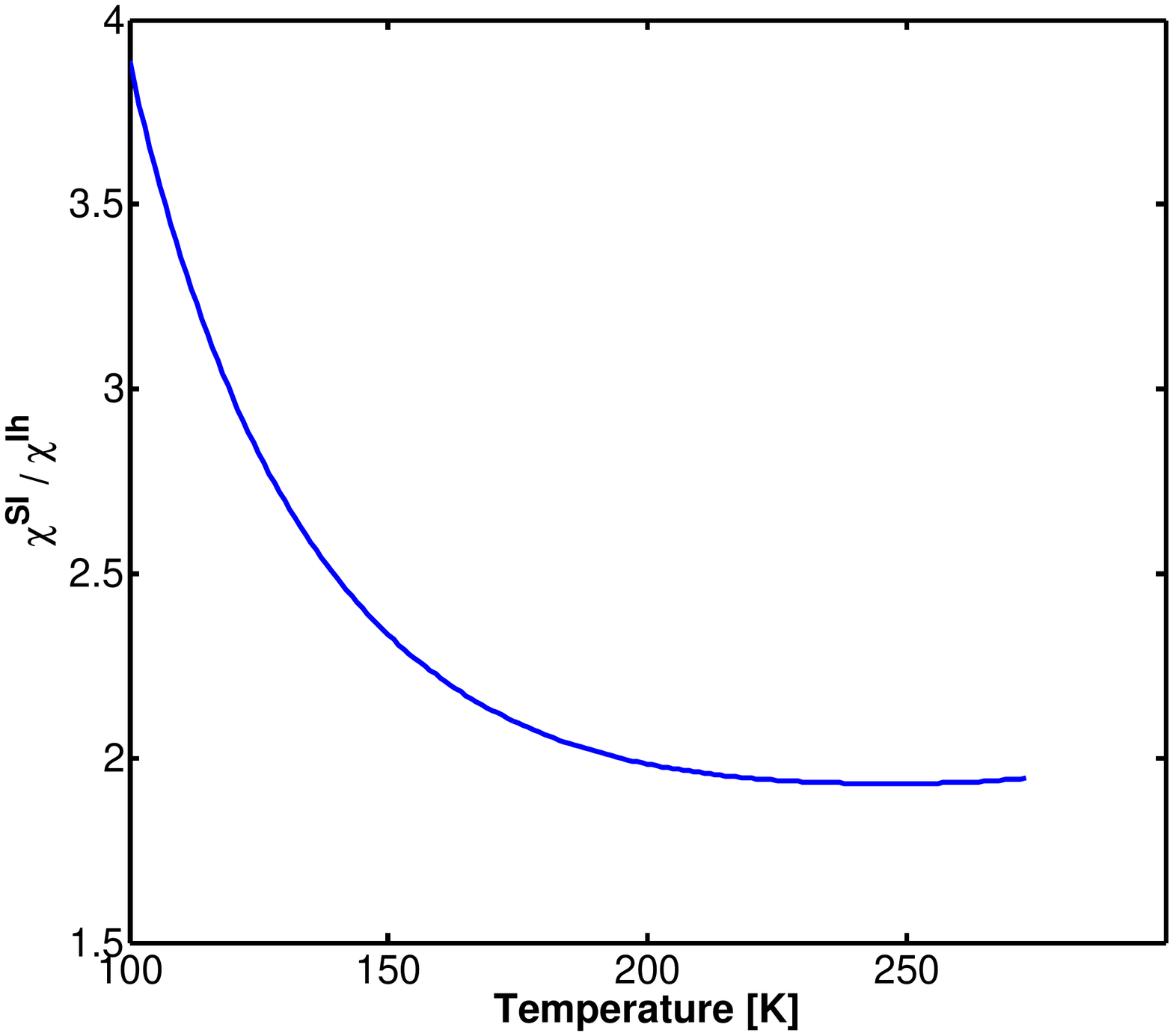}}}
\caption{\footnotesize{Thermal diffusivity of SI methane hydrate (left panel) and ratio of volume expansivity of SI hydrate to water ice (right panel). SI hydrate data from \citep{hester2007} and water ice Ih data from \cite{feistel06}.}}
\label{fig:f2}
\end{figure}

From \cite{handa1986} the heat capacity for SI methane hydrate is represented by:
\begin{equation}\label{heathanda}
C_p^{cl}=6.5\times 10^4T+3.25\times 10^6 \quad {\rm erg~ g^{-1}~ K^{-1}}
\end{equation}
The thermal diffusivity is defined as \citep{hobbs}:
\begin{equation}
\alpha\equiv\frac{\kappa}{\rho C_p}
\end{equation}
Using the experimental data mentioned above we derive the thermal diffusivity for SI methane hydrate (see Fig.\,\ref{fig:f2}). 

The linear thermal expansivity, for SI and SII (structure II) hydrates, was determined experimentally by \cite{hester2007} for different guest species. These authors have shown that it is the guest molecule contribution to the anharmonic part of the crystal intermolecular potential that enhances the thermal expansivity of hydrates above that of hexagonal water ice. The linear thermal expansivity was further found to be weakly dependent on the guest species, unless an excessively large guest is considered, and a small dependency on the hydrate structure was suggested. The linear expansivity, $L_{ex,sI}$, suggested for a SI hydrate crystal is \citep{hester2007}:
$$
L_{ex,sI}=1.1280\times 10^{-4}+3.6006\times 10^{-7}(T-298.15)-4.7694\times 10^{-11}(T-298.15)^2
$$ 
where $T$ is in Kelvins.  The volume thermal expansivity, $\chi$, is three times the linear expansivity. For its dependency on pressure we adopt the model of \cite{fei1993} to finally give:
\begin{equation}\label{ExpansivityGeneral}
\chi(T,P)=3L_{ex}\left(\frac{B+\tilde{B}P}{B+\tilde{B}P_0}\right)^{-\xi}
\end{equation}
where $L_{ex}$ is the linear thermal expansivity, $B$ is the bulk modulus and $\tilde{B}$ is its pressure derivative. We assign a value of unity to $\xi$ \citep[see][]{ashcroft}.
In Fig.\,\ref{fig:f2} we also plot the volume thermal expansivity ratio between that for a SI clathrate and that for water ice Ih. From the plot it is clear that the ratio is well approximated by $2$, deviating from this value by no more than $10\%$ for temperatures above $160$\,K. 

\subsection{Rheology}

In order to construct the near surface thermal profile for a water planet we need to evaluate the kinematic viscosity for a clathrate layer. Unfortunately, little work has been done on the viscous behavior of clathrates both experimentally and theoretically (the exceptions are \cite{stern1996} and \cite{durham2003}). Experimentally it is not known whether the rheology of the clathrate is guest-dependent or not, and whether there is a rheological difference between SI and SII clathrates. We argue that the clathrate viscosity should depend on the type of guest molecule and clathrate structure.  Because convective instability is highly dependent on the value of viscosity, we develop the expected relationships in detail below.

The dynamic viscosity $\mu$ relates the deviatoric stress tensor $\tau_{ij}$ to the strain rate tensor $\dot{e}_{ij}$ in the following way:
\begin{equation}
\tau_{ij}=2\mu\dot{e}_{ij}
\end{equation}
Experiment and theory indicate that both diffusion and dislocation creep yield the following relation between stress and strain rate \citep{schubert2001}:
$$
\tau_{ij}=C^{-1}\frac{1}{\tau^{n-1}}\exp\left({\frac{E^*+PV^*}{kT}}\right)\dot{e}_{ij}
$$
\begin{equation}\label{strainrate}
C\equiv A\frac{1}{\hat{\mu}^n}\left(\frac{|\textbf{b}|}{a}\right)^m
\end{equation}
Where $A$ is the pre-exponential factor, $\hat{\mu}$ is the shear modulus, $a$ is the grain size, $\textbf{b}$ is the Burgers vector, $\tau$ is the second invariant of the deviatoric stress tensor, $E^*$ and $V^*$ are an activation energy and volume respectively, $P$ and $T$ are the pressure and temperature and $k$ is Boltzmann's constant. 

Quite generally a strain rate equation of this type is based on the idea that solid state creep is intimately connected with the thermal creation of crystal imperfections and their thermal ability to migrate under applied stress, hence the Boltzmann factor.  
The parameters in Eq.(\ref{strainrate}) may vary for different thermodynamic regimes and for different applied stresses, as diffusional processes of different physical nature become active and dominate the solid state creep. Commonly, a single diffusion mechanism will dominate under given pressure, temperature and stress conditions \citep{durham1997}. 

The kinematic viscosity is defined as:
\begin{equation}\label{viscosity1}
\nu\equiv\frac{\mu}{\rho}=\frac{1}{2C\rho}\tau^{1-n}\exp\left(\frac{E^*+PV^*}{kT}\right)
\end{equation}
\cite{durham2003} found for methane clathrate hydrate SI, for a confining pressure of $100$~MPa and stresses of order $10$~MPa, the following values for the parameters in eq.($\ref{viscosity1}$):
$E^*_{cl}=90,000\pm 6000$~J~mol$^{-1}$, $V^*_{cl}=19\pm 10$\,cm$^3$\,mol$^{-1}$, $n_{cl}=2.2$, and $C_{cl}=10^{8.55}$\,MPa$^{-n_{cl}}$\,s$^{-1}$, where the subindex $cl$ stands for methane clathrate hydrate SI. 
These parameters were derived from an experiment on a laboratory made clathrate sample. Deciding whether these are transferable to a natural setting requires some thought of the dependency of the rheology on grain size. Due to the short time scale of an experiment one expects a laboratory sample to be composed of grains smaller than those composing a naturally formed sample that has time to ripen. \cite{durham2003} report that their sample was composed of methane clathrate grains in the size range of $20-40\mu$m, an order of magnitude smaller than the grain size composing naturally formed methane clathrate bulk (see discussion below on methane clathrate grain sizes). 
When quantifying convective instability we will estimate the viscosity at temperatures higher than $2/3$ of the melting temperature (i.e. dissociation temperature) of methane clathrate. This temperature criterion is also maintained in the experiment of \cite{durham2003}. This high temperature regime suggests the parameters above represent a viscosity whose rate-controlling step is dominated by dislocation climb which is fairly insensitive to grain size \citep{Kohlstedt2007}. Therefore the transferability of the experimental parameters to our larger grain size case may be considered permissible. Although the fact that $n_{cl}=2.2$ and not $3$ as expected from a creep solely dominated by dislocation climb \citep{Kohlstedt2007} hints that other possible creep mechanisms may have also been at work during the experiment of \cite{durham2003}. One such possibility is grain boundary sliding which is probably enhanced due to the small grain sizes of the laboratory sample. Such a creep mechanism indeed encourages $n\approx 2$ \citep{Kohlstedt2007}. If that is the case then the creep measured in \cite{durham2003} is grain size sensitive and not easily transferable to a sample composed of much larger grain sizes. One must consider though that the larger grains composing a naturally formed sample of methane clathrate will make grain boundary sliding less efficient. In this case if grain boundary sliding is indeed folded in the experimentally derived parameters of \cite{durham1997} then the viscosity they represent is a lower bound on the naturally forming methane clathrate dislocation viscosity whose larger grains make it stronger. An example of such strengthening due to increased grain size was measured in clinopyroxene \citep[see][and references therein]{Kohlstedt2007}.              

Another point that must be considered is that methane hydrate survives to pressures as high as $\sim 1$\,GPa \citep{sloan}, where the viscosity parameters of \cite{durham2003} may no longer be applicable. In order to make extrapolations of the viscosity to higher pressures, we utilize a scheme proposed by \cite{weertman1970}.

Low pressure experiments have shown that one may write the following:
\begin{equation}\label{OmegaHomologous}
\frac{E^*}{k}\equiv\Omega T_m
\end{equation}
where $T_m$ is the melting temperature and $\Omega$ is a dimensionless constant which depends on the crystal structure \citep[see][and references therein]{weertman1975}. Weertman then further proposed the following extension to higher pressures:
\begin{equation}\label{transmelting}
\frac{E^*+PV^*}{k}\equiv\Omega T_m(P)
\end{equation} 
This last transformation requires the melting curve to contain the information of how the activation energy and volume change with pressure. This has some support as, for a given stress, contours of constant viscosity which are functions of pressure and temperature, do seem to correspond fairly well to the contour of the melting curve. This is seen to be true for water ice \citep{durham1997} and for other substances as well \citep{poirier1985}. Indeed this method has been utilized by \cite{borch1987} in extrapolating the viscosity data for olivine to conditions in the Earth's upper mantle and by \cite{spohn2003} for pure water ice crusts in the Galilean satellites. 

Inserting Eq.(\ref{transmelting}) into Eq.(\ref{viscosity1}) yields for the viscosity:
\begin{equation}\label{viscosity2}
\nu=\frac{1}{2C\rho}\tau^{1-n}\exp\left(\Omega\frac{T_m(P)}{T}\right)
\end{equation} 
The melting (i.e. dissociation) curve for clathrates is both guest molecule and crystal structure dependent, and this dependence enters into the viscous behavior of a given clathrate.  We need to solve for the thermodynamic stability regime for a SI methane hydrate in order to extrapolate its viscosity. A full discussion of how to derive a clathrate hydrate thermal stability field and how to extrapolate it to high pressure ($\sim 1$\,GPa) is beyond the scope of this paper. For an in-depth explanation of thermal stability calculations we refer the reader to the works of \cite{waalplat} and \cite{sloan}.  

\begin{figure}[ht]
\centering
\includegraphics[scale=0.5]{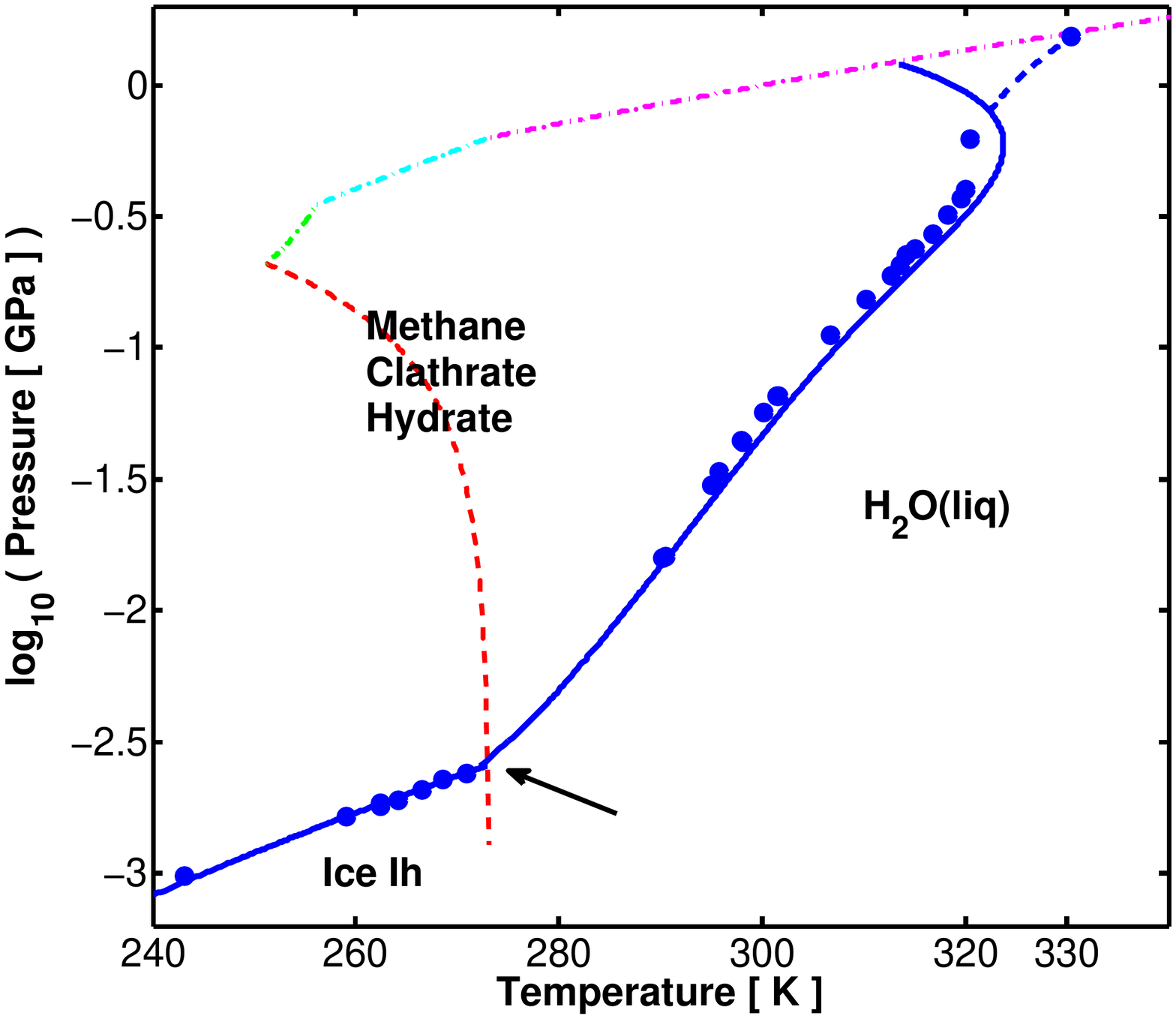}
\caption{\footnotesize{Clathrate hydrate SI dissociation curve (thick solid curve, blue in the on-line version) and clathrate hydrate structure-H (SH) dissociation curve (blue dashed curve in the on-line version). Filled circles represent available experimental data for the dissociation curve. Dashed line (red in the on-line version) is the ice-Ih melting curve. Dashed-dotted lines (green, cyan and magenta in the on-line version) are the melting curves for ice III, V and VI respectively. Arrow is pointing to the position of the quadruple point, where ice-Ih, methane clathrate hydrate, liquid water and methane vapour coexist. The stability field for methane clathrate hydrate is to the left of its dissociation curve.}}
\label{fig:f4}
\end{figure}

Solving for the case of a methane clathrate hydrate we find the pressure at the first quadruple point to be $25.71$\,bar. By first quadruple point we mean the point where a clathrate hydrate transforms from being in equilibrium with ice Ih to being in equilibrium with liquid water. Thus the four phases: clathrate, liquid water, water ice Ih and methane vapour coexist. We further find that the dissociation (i.e. melting) curve beyond this pressure is well represented by the following polynomial:
$$
T_{m,cl}(x)=-0.016145x^6+0.54446x^5-7.5525x^4+55.353x^3
$$
\begin{equation}\label{meltcurveSI}
-226.85x^2+503.39x-203.6
\end{equation}
where $x\equiv \ln [P\,({\rm bars})]$.
For clarity we have plotted this melting curve, together with the three phase hydrate-ice Ih-vapour curve, on top of a phase diagram for pure water (see Fig.\,\ref{fig:f4}). 

For pressures below the first quadruple point we set the melting temperature equal to that of pure water ice for purposes of viscosity estimation.  With the aid of Eq.\,(\ref{meltcurveSI}) and Eq.\,(\ref{transmelting}) we solve for $\Omega$, taking into account that the experiment of \cite{durham2003} was conducted at a confining pressure of $100$\,MPa.  We find $\Omega_{cl}=36\pm 1$.
The non-Newtonian viscosity, as expressed in Eq.(\ref{viscosity2}) with $n_{cl}=2.2$, is not an intrinsic quantity of a crystal but rather is dependent on external conditions. In other words it depends on the applied deviatoric stress. Estimating the importance of the non-Newtonian dislocation creep in the planetary crust, therefore, requires an estimation of the second invariant of the deviatoric stress, $\tau$.  \cite{golitsyn1979} has shown that in steady state convection, where the rate of the work done by buoyancy exactly equals the rate of energy dissipation via friction, one has the following relation:
\begin{equation}\label{energybalance}
\int\tau_{ij}\dot{e}_{ij}dv=\frac{\chi gd}{C_p}FS
\end{equation}
where the integral on the LHS is the rate of kinetic energy dissipation in a convecting cell, $F$ is the heat flux entering the cell and $S$ is the cell surface through which the heat enters the cell. In deriving Eq.\,(\ref{energybalance}) it is assumed that the density is constant (i.e. a Boussinesq fluid). In the Boussinesq approximation the background adiabatic temperature is constant \citep{schubert2001} and the temperature increase, $\Delta T$, is confined to the thermal boundary layer, $\delta$, thus:
\begin{equation}
F\sim\frac{\kappa\Delta T}{\delta}
\end{equation}
For small viscosity contrasts (SVC) this will be true for both upper and lower boundary layers. For the stagnant lid regime (SL) $\delta$ is the cold boundary layer \citep{solomatov1995}.  

Given that the creep velocity under the cold thermal boundary layer is of order $u$, the strain rate is of order $u/d$, and in terms of scales Eq.(\ref{energybalance}) may be written as:
\begin{equation}
\tau\frac{u}{d}dS\sim\frac{\chi gd}{C_p}\frac{\kappa\Delta T}{\delta}S
\end{equation}
From boundary layer theory \cite[see][and references therein]{solomatov1995}:
\begin{equation}
\delta\sim\sqrt{\alpha t}\sim\sqrt{\alpha \frac{d}{u}}
\end{equation}
This results in the following estimate for the second invariant of the deviatoric stress tensor:
\begin{equation}\label{shearestimation}
\tau\sim\chi g\rho\Delta T\delta 
\end{equation}
Assuming for the thermal boundary layer (which is the planetary crust) a length scale of $1$\,km, a temperature difference of $50$\,K, a surface gravity of $10^3$\,cm\,s$^{-2}$ and the thermal properties mentioned above, we estimate $\tau$ to be of order $10^5$\,Pa.

In order to properly describe the rheological behaviour of the crust considering dislocation creep alone is not sufficient. In particular, in case the crust is acted upon by low deviatoric stresses diffusional creep may best estimate the crustal rheology. Properly estimating the crustal rheology requires a viscosity map spanning crustal conditions, stating which creep mechanism minimizes the viscosity for varying crustal stress, pressure and temperature conditions. To that aim we shall also estimate the Newtonian (diffusion) creep for methane clathrate hydrate SI.   

The microscopic manifestation of a Newtonian creep mechanism is the diffusion of lattice vacancies and interstitial molecules. The solid state viscosity related to this physical mechanism was formulated and analysed by \cite{herring1950}. A slightly different formulation, given by \cite{weertman1975}, for the same mechanism is:
\begin{equation}\label{newtonianviscosity}
\nu\equiv\frac{\tau}{2\dot{e}\rho}=\frac{1}{2\zeta\rho}\frac{a^2}{D}\frac{kT}{\tilde{v}}
\end{equation}
Where $\zeta$ is a dimensionless constant for which we adopt the numerical value of $14$ \citep{frost1982}, $a$ is the average crystal grain diameter, $\tilde{v}$ is the atomic volume, and $D$ is the creep diffusion coefficient. For the atomic volume we use the value of $2.3\times 10^{-23}$\,cm$^3$\,molec$^{-1}$ for a water molecule, derived from hydrogen bond length.
 
\begin{figure}[ht]
\centering
\includegraphics[scale=0.5]{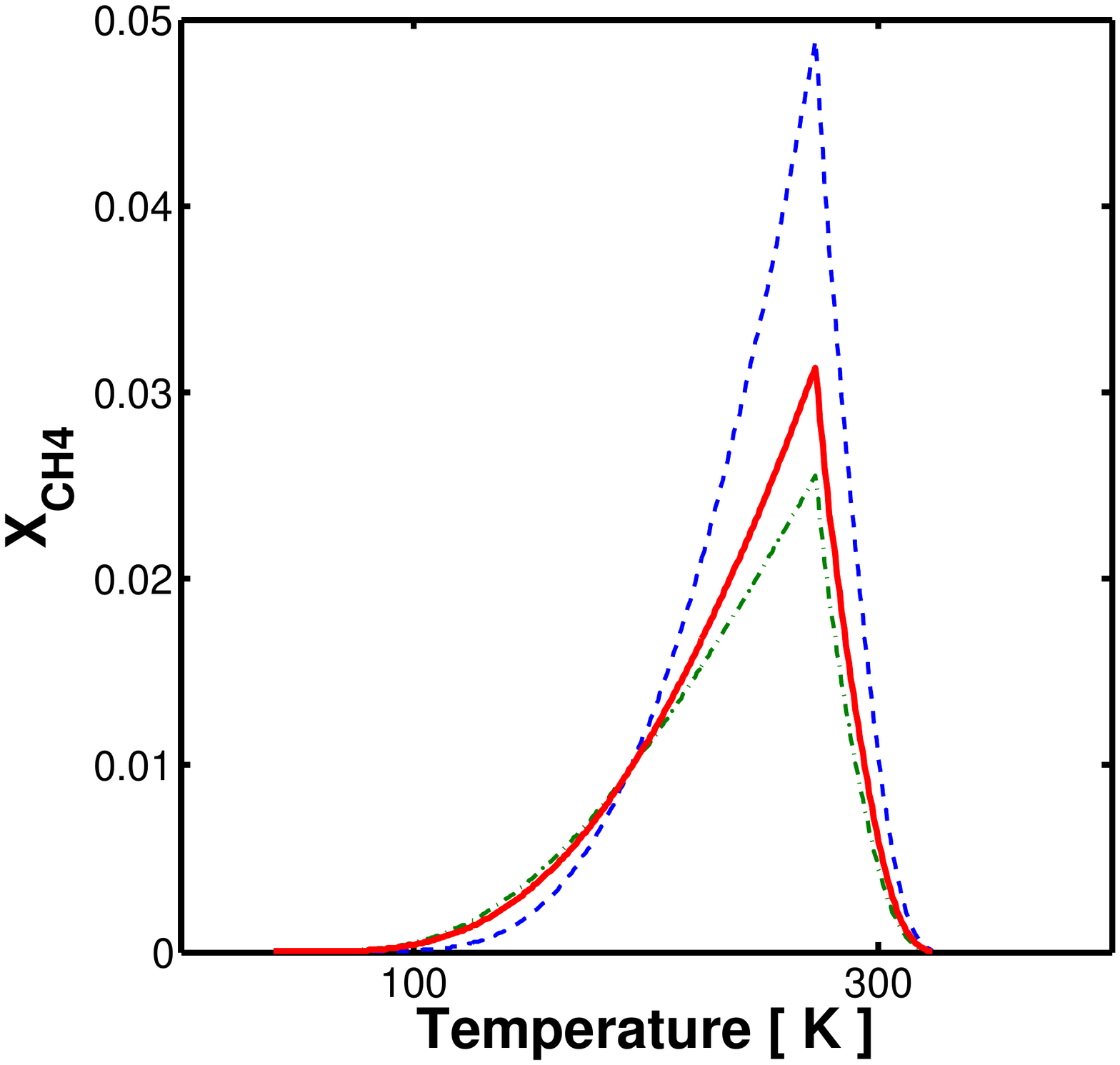}
\caption{\footnotesize{The fraction of unoccupied cages in a SI methane hydrate and its dependence on temperature. The blue (green) curve represents the degree of unoccupancy in the small (large) cage of a SI hydrate. The red curve is an averaged curve which takes into consideration that for every cubic unit cell there are eight cages, of which two are small and six are large.}}
\label{fig:f5}
\end{figure}

A question now arises about the diffusion of lattice vacancies and interstitial molecules in methane hydrate:
\cite{peters2008} studied the diffusivity of methane in a SI hydrate. The diffusion is considered to be due to thermal jumping of a methane molecule from a cage it occupies to a neighbouring vacant cage. Three jumping paths are considered, one is from a small cage to a large cage via a five membered water ring (pentagon face of a cage), and two different paths from a large cage to a neighbouring vacant large cage once via a five membered water ring and once through a six membered water ring (hexagonal face). They find the methane molecule to be too large to jump thermally through water rings without causing massive distortion to the water lattice, therefore, the authors invoke a water vacancy (defect) between the occupied and vacant cages, so as to lower the thermal barrier to jumping. The ability of the methane molecule to diffuse between cages becomes dependent not only on the degree of cage occupancy but also on the probability of water vacancy formation. From their results we may derive the following form for the diffusion coefficient of methane in a SI hydrate:
\begin{equation}\label{methanediffusion}
D_{CH_4}(T)=0.0028X_{CH_4}\exp\left(-\frac{6.042\times 10^{-13}}{kT}\right) \quad {\rm cm^2\,s^{-1}}
\end{equation}    
Here $X_{CH_4}$ is the fraction of unoccupied water cages. By solving for the thermodynamic stability regime, for a SI methane hydrate, we obtain the variation of $X_{CH_4}$ with temperature (see Fig.\,\ref{fig:f5}).
  
\cite{liang2011} proposed a mechanism for creating and migrating H$_2$O defects in the hydrate water lattice, defects that may help reduce the thermal barrier to guest molecule jumping between cages, as required by the model of \cite{peters2008}. \cite{liang2011} proposed, based on MD simulations, that some small fraction of the hydrate cages may actually become occupied by a water molecule, leaving a defect in the lattice. This defect may help a guest molecule to diffuse in the lattice. The authors also found that these interstitial water molecules represent the most mobile defect in the hydrate lattice. When a water molecule, that occupies a cage, gets too near to one of the cage boundaries, whether it is a five (pentagonal face) or six (hexagonal face) membered water ring, it creates hydrogen bonds with molecules forming the water ring, which results in a metastable structure. This metastable structure then collapses by emitting a water molecule back to the center of the cage it came from, or to a first or second neighboring empty cage. In this way they derived a coefficient for self diffusion of water molecules within the hydrate lattice. Using their results we calculate the diffusion coefficient to be:
\begin{equation}\label{waterdiffusion}
D_{H_2O}(T)=0.1898\exp\left(-\frac{6.9708\times 10^{-13}}{kT}\right) \quad {\rm cm^2\,s^{-1}}
\end{equation}    

The diffusion coefficient, $D$, in Eq.\,\ref{newtonianviscosity} is actually a weighted average of the contributions of water and methane to the diffusion. This weighted diffusion for a SI hydrate may be written as: 
\begin{equation}
D_{ave}^{(SI)}=\frac{D_{H_2O}D_{CH_4}}{\frac{46}{54}D_{CH_4}+\frac{8}{54}D_{H_2O}}
\end{equation}
In the last equation we adopt the weighing procedure for lattice diffusion in multicomponent solids \citep[see][]{weertman1975}, for the case of clathrate hydrate solid solutions. The weights in the denominator consider a cubic crystal unit cell, composed of $46$ water molecules and $8$ cages that with high probability are fully occupied, with a single methane molecule per cage.

Estimating the Newtonian viscosity also requires a value for the average diameter of the clathrate crystal grains, $a_{cl}$. Laboratory experiments show that single crystal grains of synthetically produced hydrates have diameters in order of several tens of micrometers. Smaller grains have higher growth rates. The smaller grains could be a consequence of the short time scales to which the laboratory experiment is confined. An examination carried out by \cite{klapp2007} of actual geological samples of hydrates retrieved from the Gulf of Mexico and from Hydrate Ridge, revealed that in natural samples the crystal grain size diameter was in the range of $300-600$\,$\mu$m. The fact that naturally occurring hydrates have grains an order of magnitude larger than their synthetic counterparts was explained by these authors to be due to an Ostwald ripening process. In the latter process minimization of the free energy causes big grains to grow on the expense of smaller ones over a geological time scale.  As a final note, the proper viscosity, whether it be Newtonian or non-Newtonian, is chosen to be such that the viscosity is a minimum for the given conditions.  

\subsection{Convective Stability Analysis}

We are now ready to evaluate the thermal profile, in the near surface layer of a water planet, using the values derived above. As in \cite{fu10} we define the crust of a planet as the domain where conduction is the dominant mechanism for the heat transport.
If the crust is of a radial dimension $\delta_{crust}$, we may write for it:
\begin{equation}\label{crustwidth}
\delta_{crust}=\frac{\kappa_{cl}\left(T_{b,crust}-T_s\right)}{F_s}
\end{equation}
where $T_{b,crust}$ and $T_s$ are the temperatures at the crustal base and planetary surface respectively and $F_s$ is the surface heat flux.
As in \cite{fu10} we shall leave $T_s$ and $F_s$ as independent variables. Assuming the flux is only due to radioactive decay in the planetary metallic and silicate interior, and that the power released per gram of silicates and metals equals that for Earth, one may write:
\begin{equation}\label{surfheatflux}
\frac{F_{s,planet}}{F_{s,Earth}}=\frac{M_{p}X^{Si+Fe}_p}{M_{E}}\left(\frac{R_{E}}{R_{p}}\right)^2=\frac{g_{s}}{g_{s,E}}X^{Si+Fe}_p
\end{equation}
where $F_{s,Earth}=0.087$\,W\,m$^{-2}$ \citep{Turcotte2002}, $M_{E}$ and $M_{p}$ are the mass of Earth and the studied planet respectively, $X^{Si+Fe}_p$ is the mass fraction of silicates and metals in the studied planet, $R_{E}$ and $R_{p}$ are the planetary radii respectively and $g_{s}$ and $g_{s,E}$ are the appropriate surface accelerations of gravity.

Both scaling analysis and assuming the thermal boundary layer is on the verge of convective instability are independent and equivalent techniques \citep{solomatov1995}.  The termination of the planetary crust occurs at some deep sublayer whose Rayleigh number is maximal and equals a critical value ($Ra_{crit}$). By deriving the width of the sublayer, that maximizes its Rayleigh number, the transition between the small viscosity regime and the stagnant lid regime was found by \cite{solomatov1995} to occur when:
\begin{equation}\label{ViscoContrast}
\frac{\nu(T_s)}{\nu(T_{b,crust})}\equiv e^\theta =e^{4(n+1)}
\end{equation}
where the LHS is the ratio of viscosities across the cold boundary layer and $n$ is the deviatoric stress power [see Eq.\,(\ref{viscosity1})].

First we assume a Newtonian viscosity.  The condition that the crustal layer be on the verge of convective instability may be written, with the aid of Eq.\,(\ref{rayleighnum}), as: 
\begin{equation}\label{convinstab}
\frac{g_{s}\chi_{cl}(T_{b,crust}-T_s)\delta^3_{crust}}{\nu_{cl}\alpha_{cl}}=Ra_{crit}
\end{equation}
where the subindex $cl$ means each parameter is assigned the appropriate value for methane clathrate hydrate SI.
Further, assuming the crust is in hydrostatic equilibrium, one may write:
\begin{equation}\label{hydrostatic}
P_{b,crust}=P_s+\rho_{cl} g_{s}\delta_{crust}
\end{equation}
where $P_{b,crust}$ and $P_s$ are the crust bottom pressure and planetary surface pressure respectively.
For the small viscosity contrast (SVC) we estimate the viscosity at the mid-layer temperature  \citep{solomatov1995}, using Eqs.\,(\ref{hydrostatic}) and (\ref{crustwidth}).
\begin{equation}
\bar{T}(P_{b,crust})=\frac{T_{b,crust}+T_s}{2}=T_s+\frac{F_s}{2\kappa_{cl}}\frac{P_{b,crust}-P_s}{\rho_{cl} g_{s}}
\end{equation}
The condition for convective instability (eq.\ref{convinstab}) can then be written as:
\begin{equation}
\frac{\chi_{cl}}{\nu_{cl}\alpha_{cl}\kappa_{cl}\rho^4_{cl}}\frac{F_s}{g^3_{s}}\left(P_{b,crust}-P_s\right)^4=Ra_{crit}
\end{equation}
This last equation is a univariant equation for the pressure at the crustal base. 
We further assume $Ra_{crit}=2000$ for the case of the small viscosity contrast. This value is appropriate for $\theta<8$ (SVC regime for Newtonian fluids) as was shown for various wave numbers \citep[see][and references therein]{schubert2001}. It is not likely that $\theta$ will be so low as to turn this value excessively large because the thermal conductivity of clathrates is low.

For the stagnant lid regime ($\theta>8$ for Newtonian fluids) the viscosity is estimated at the temperature of the bottom of the crust:
\begin{equation}
T_{b,crust}(P_{b,crust})=\frac{F_s}{\kappa_{cl}}\left(\frac{P_{b,crust}-P_s}{\rho_{cl} g_{s}}\right)+T_s
\end{equation} 
The critical Rayleigh number for the stagnant lid regime is \citep{schubert2001}:
\begin{equation}\label{Racritstalid}
Ra_{crit}=20.9\theta^4
\end{equation} 
The relation between the viscosity contrast, $\theta$, and the creep mechanism enthalpy of activation, $H^*$, was shown to be \citep{mckinnon1999}:
\begin{equation}\label{theta1}
\theta=\frac{H^*}{k}\frac{T_{b,crust}-T_s}{T_{ad}^2}
\end{equation}
Where $T_{ad}$ is a characteristic adiabatic temperature in the convecting sub-layer.
It is important to note here that in the stagnant lid regime the actual temperature difference across the stagnant lid ($T_{b,crust}-T_s$) does not constitute the temperature difference which drives the convection. Rather, between the convection cell and the stagnant lid there exists a boundary layer in which the viscosity increases exponentially with decreasing depth while at the same time the strain rate decreases exponentially from its value at the convective cell to its negligible value at the stagnant lid. The temperature difference across this boundary layer (referred to as the rheological temperature difference, $\Delta T_{reo}$ ) is what drives convection in the stagnant lid regime \citep{solomatov1995}. It was shown by \cite{solomatov1995} that the rheological temperature difference obeys:
\begin{equation}\label{Treo1}
\Delta T_{reo}=\frac{T_{b,crust}-T_s}{\theta}
\end{equation}
It is also roughly given by:
\begin{equation}\label{Treo2}
\Delta T_{reo}\approx T_{b,crust}-T_{ad}
\end{equation}
From Eqs.\,(\ref{theta1}-\ref{Treo2}) one may obtain a second order polynomial for $\theta$, whose roots are:
\begin{equation}
\theta(P_{b,crust})=\frac{T_{b,crust}-T_s}{2T^2_{b,crust}}\left[2T_{b,crust}+\frac{H^*}{k}\pm\sqrt{\left(2T_{b,crust}+\frac{H^*}{k}\right)^2-4T^2_{b,crust}}\quad\right]
\end{equation}  
To choose the physical root we note that $H^*_{cl}/k\sim 10^4$. Therefore for the scenario where $T_s\ll T_{b,crust} \ll H^*/k$ we expect $\theta\gg 1$ [see Eq.\,(\ref{ViscoContrast})], and this is satisfied by the root with the plus sign in front of the square root.  The other root will go to unity. The last equation is used in conjunction with Eq.\,(\ref{Racritstalid}) for the cases where $\theta>8$.

\subsection{Results and Discussion on the Planetary Crust}

For the Newtonian creep mechanism, discussed above, we find there are five possible structures for the near surface thermodynamic behavior of water planets.  

\textbf{Regime I} - In this case the surface temperature is low and so is the surface heat flux (low metallic and silicate content for a given surface gravity). For a low enough surface temperature even a very small surface atmospheric pressure will suffice to stabilize methane clathrates on the planetary surface (the dissociation pressure of methane clathrate at $150$~K is $\approx 6$~kPa). In this regime the low surface heat flux will result in relatively small increases in temperature with depth even though the thermal conductivity of clathrates is very small. Therefore, in this regime we expect a conductive planetary crust composed of methane clathrate beginning from the surface and ending in the depth where convective instability is reached.

\textbf{Regime II} - In this regime the surface temperature is still low enough so that the corresponding clathrate dissociation pressure is so low that probable surface atmospheric pressures ($\sim 1$ bar) will suffice to stabilize clathrates on the planetary surface. Contrary to regime I the surface heat flux is now high (a large mass fraction of metals and silicates for a given surface gravity). In this case the very low thermal conductivity of clathrate hydrates will result in a steep temperature increase with depth. The resulting conductive thermal profile, in this case, will reach the hydrate dissociation (i.e. melt) curve before becoming convectively unstable, and will try to intrude into the liquid water phase region. Liquid water, having a very low viscosity, will introduce convection resulting in an adiabatic profile whose gradient is much steeper than that of the hydrate dissociation curve. This will drive the system immediately back to the hydrate stability regime which will again try to penetrate the liquid water regime. The result is a planetary crust whose upper part has a conductive profile and its lower part is restricted to the hydrate dissociation curve until convective instability is reached. The part of the crust with the on-melt behavior is expected to have a small viscosity contrast due to the fact that the viscous topology follows the melt curve, as explained above. We shall refer to this on-melt layer as the {\it dissociation boundary layer} (DBL).  

\textbf{Regime III} - This regime is the counterpart of regime I, except that the surface temperature is now high enough so that the appropriate hydrate dissociation pressure could be higher than the atmospheric surface pressure (in case $P_s<25$~bar, the quadruple point pressure). In this case ice Ih rather then methane hydrate will be stable on the planetary surface. Since the first quadruple point for methane hydrate is $\approx 25$~bar, then for a surface gravitational acceleration of $10$~m~s$^{-2}$ the depth at which clathrate hydrates will become stable is, at most, of the order of $100$~m. Since in this regime (as in regime I) the surface heat flux is low, and since the thermal conductivity of ice Ih is much higher than that of hydrates, the temperature will increase slowly with depth reaching the stability field for methane hydrates at a depth of, at most, $\sim 100$~m. As hydrates become stable their lower thermal conductivity will lead to a faster temperature increase with depth and the system will converge to the situation described in regime I. 

\textbf{Regime IV}  - In this regime, as in regime III, the upper part of the crust is a thin sheet of water ice Ih and only at a depth of, at most, an order of $100$~m, is methane hydrate stabilized. In this regime the high surface heat flux means the near surface behavior will converge to that depicted in regime II.

\textbf{Regime V} - In this regime the crust is made of hexagonal ice, and the surface temperature and/or the surface heat flux are high enough so that even in a relatively thin sheet of ice Ih the temperature may rise fast enough with depth to reach melting before entering the stability field for methane clathrate hydrate, where the solution would have been stuck in a regime III or IV type behaviors. In this regime we expect to find a subterranean ocean. Since a liquid ocean has a higher mass density than methane clathrate hydrate, any sublayer of clathrate hydrates trying to form will travel upward due to buoyancy consequently experiencing depressurization and decomposition.

\begin{figure}[ht]
\centering
\includegraphics[scale=0.5]{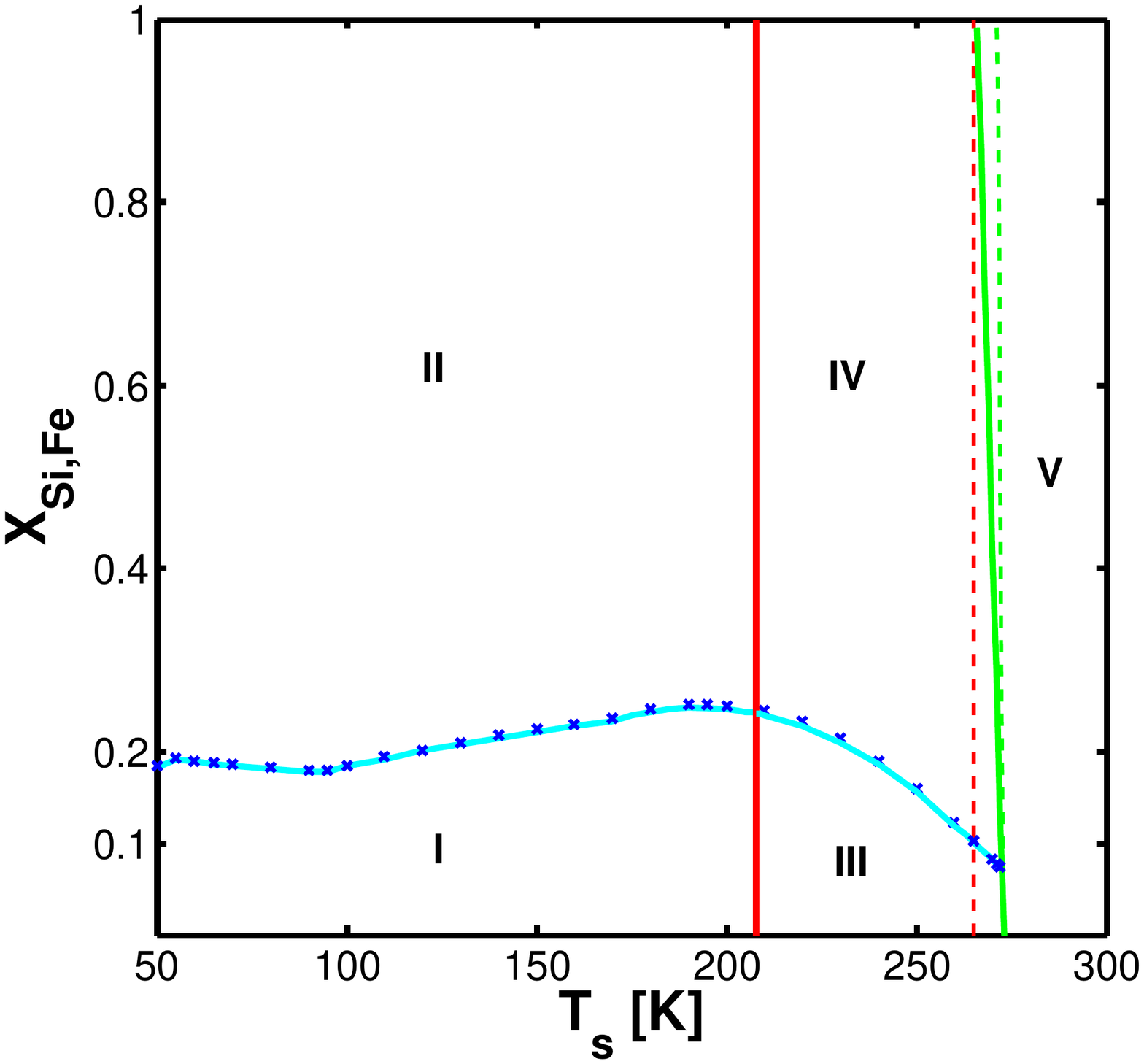}
\caption{\footnotesize{The parameter space for each of the five near surface thermodynamic regimes described in subsection $2.4$. Regimes I and III will have a conductive upper thermal boundary layer ending with initiation of convection. In regimes II and IV a layer confined to the SI hydrate melt curve will separate the conductive boundary from the convection cell. In regime V an ocean will be formed, a configuration where the less dense pure methane hydrate will be unstable. The vertical solid line (red in the on-line version) marks the stability for methane hydrate on the planetary surface assuming a $2$\,bar surface atmospheric pressure. Markers and dashed curves represent the change in the parameter space of each regime when assuming a $20$\,bar surface atmospheric pressure.}}
\label{fig:f6}
\end{figure}

In Fig.\,(\ref{fig:f6}) we plot the domains of the five crustal structures as a function of the planetary surface temperature and surface heat flux, expressed in silicate and metal mass fraction [see Eq.\,\ref{surfheatflux}], for a surface gravitational acceleration of $10$\,m\,s$^{-2}$. The solid lines represent the boundaries between the crustal regimes for a $2$\,bar surface atmospheric pressure while the dashed lines and the markers are for the $20$\,bar surface atmospheric pressure scenario.          

The transition from regime I to regime II and the transition from regime III to IV are hardly affected by the increase in the assumed surface atmospheric pressure, from $2$ to $20$~bar. This is because these regime transitions occur deeper in the crust where the effects of surface pressure are negligible. The most pronounced change, due to the increase in surface atmospheric pressure, is the shift to higher temperature (from $208$ to $265$\,K) in the boundary between regime I and III and from II to IV.
As we have already mentioned, in order to be in regime V, the thermal profile in the ice Ih crust (whose depth is at most $\sim 100$~m and whose thermal conductivity is high, relative to that of hydrates) must reach the melting curve for liquid water before stabilizing methane hydrates. This is hard to accomplish in such a narrow ice Ih layer and requires high surface heat fluxes or very high initial surface temperatures. From Fig.\,(\ref{fig:f6}) we see that for the $2$~bar atmosphere such an effect becomes possible for a minimum surface temperature of $266$\,K. For the $20$~bar atmosphere the minimum stands at $271$\,K, as the higher surface atmospheric pressure results in an even thinner ice Ih layer which must reach the liquid water melting curve before stabilizing hydrates.  We point out that for the ice Ih viscosity we adopt the formalism and parameters given in \cite{spohn2003}.  We now wish to venture deeper into the planet, beginning with its internal structure.

\section{THE PLANETARY STRUCTURE}

We consider in our model a differentiated body composed of four distinct regions: An Fe core, a surrounding silicate mantle, and an outer water mantle which is itself divided into two regions. As we are interested in the details of the transport and cycling of CH$_4$ in water planets we mainly focus on the fine structure of the water mantle where such transport exists. We divide the water mantle into a high pressure region composed of {\it filled ice}, and a lower pressure region where methane clathrate hydrate is stable.  The filled ice is a water ice polymorph created under high pressure in the presence of methane (see \cite{Levi2013} - hereafter called paper I). With regards to the iron core and silicate mantle we will restrict ourselves to making simpler though adequate assumptions. 

For the iron core we use the Vinet EOS with the parameters given in \cite{seager2007}, which, according to these authors, describes the $\epsilon$ phase of Fe up to a pressure of $2.09\times 10^4$~GPa. This pressure is never exceeded in any of the planets we consider. The external water layer we assume ensures that the contribution from low pressure silicate phases is negligible. We therefore model the silicate mantle to be solely composed of the perovskite phase of MgSiO$_3$, where for the EOS we use the fit suggested in \cite{seager2007}, which smoothly connects a fourth-order Birch-Murnaghan EOS with the Thomas-Fermi-Dirac EOS. 

For methane filled ice Ih we adopt a third order Birch-Murnaghan EOS with bulk modulus, $B_{FI}=10$\,GPa, derived from \cite{hirai2003}. We also use this EOS for the methane clathrate hydrate, with $B_{cl}=8$\,GPa and $\tilde{B}_{cl}=7.61$ determined experimentally by \cite{shimizu2002}. In tables \ref{tab:2MEtable}-\ref{tab:10MEtable} we present various internal structure results for our $2M_E$, $5M_E$ and $10M_E$ planets.  For each mass we assume both a $25\%$ and a $50\%$ water mass fraction. For the case of the $2M_E$ planet we examined a wider range of ice mass fractions ranging from $60\%$ to as low as $3\%$ in order to better understand the dependence of mantle convection on this parameter.

\begin{deluxetable}{cccccccc}
\tablecolumns{8}
\tablewidth{0pc}
\tablecaption{Internal Structure for 2M$_E$ Planet}
\tablehead{
\colhead{Ice Fraction} & \colhead{g$_s$}  & \colhead{P$_{Si-H_2O}$}    & \colhead{D$_M$} &
\colhead{P$_{center}$}  & \colhead{R$_{core}$}  & \colhead{P$_{Fe-Si}$}  & \colhead{R$_{Si-H_2O}$} \\
\colhead{\%} & \colhead{(m~s$^{-2}$)}  & \colhead{(GPa)}  & \colhead{(km)} &
\colhead{(GPa)}  & \colhead{(km)}  & \colhead{(GPa)}  & \colhead{(km)}
}
\startdata
3  & 13.4  & 7   & 377  & 822 & 3899 & 300 & 7363 \\
5  & 12.9  & 11  & 576  & 814 & 3874 & 300 & 7299 \\
10 & 12.1  & 22  & 990  & 794 & 3808 & 301 & 7143 \\
15 & 11.5  & 32  & 1370 & 771 & 3741 & 301 & 6988 \\
20 & 10.9  & 41  & 1732 & 747 & 3672 & 300 & 6833 \\
25 & 10.4  & 50  & 2082 & 722 & 3600 & 298 & 6675 \\
30 & 10.1  & 59  & 2382 & 696 & 3525 & 295 & 6512 \\
35 & 9.7   & 67  & 2753 & 669 & 3446 & 291 & 6346 \\
40 & 9.4   & 75  & 3050 & 641 & 3363 & 287 & 6172 \\
45 & 9.2   & 83  & 3351 & 612 & 3275 & 282 & 5991 \\
50 & 8.9   & 90  & 3680 & 582 & 3182 & 276 & 5800 \\
55 & 8.7   & 97  & 4011 & 550 & 3082 & 270 & 5598 \\
60 & 8.4   & 104 & 4363 & 517 & 2973 & 262 & 5383
\enddata
\tablecomments{g$_s$-surface gravity, P$_{Si-H_2O}$-silicate and water mantle boundary pressure, D$_M$-water mantle depth, P$_{center}$-pressure at the center, R$_{core}$-iron core radius, P$_{Fe-Si}$-iron core and silicate mantle boundary pressure, R$_{Si-H_2O}$-distance from center to water mantle.}
\label{tab:2MEtable}
\end{deluxetable}

\begin{deluxetable}{cccccccc}
\tablecolumns{8}
\tablewidth{0pc}
\tablecaption{Internal Structure for 5M$_E$ Planet}
\tablehead{
\colhead{Ice Fraction} & \colhead{g$_s$}  & \colhead{P$_{Si-H_2O}$}    & \colhead{D$_M$} &
\colhead{P$_{center}$}  & \colhead{R$_{core}$}  & \colhead{P$_{Fe-Si}$}  & \colhead{R$_{Si-H_2O}$} \\
\colhead{\%} & \colhead{(m~s$^{-2}$)}  & \colhead{(GPa)}  & \colhead{(km)} &
\colhead{(GPa)}  & \colhead{(km)}  & \colhead{(GPa)}  & \colhead{(km)}
}
\startdata
25 & 16.3 & 119 & 2520 & 1786 & 4537 & 716 & 8566 \\
50 & 14.2 & 228 & 4472 & 1442 & 4012 & 679 & 7402
\enddata
\tablecomments{g$_s$-surface gravity, P$_{Si-H_2O}$-silicate and water mantle boundary pressure, D$_M$-water mantle depth, P$_{center}$-pressure at the center, R$_{core}$-iron core radius, P$_{Fe-Si}$-iron core and silicate mantle boundary pressure, R$_{Si-H_2O}$-distance from center to water mantle.}
\label{tab:5MEtable}
\end{deluxetable}

\begin{deluxetable}{cccccccc}
\tablecolumns{8}
\tablewidth{0pc}
\tablecaption{Internal Structure for 10M$_E$ Planet}
\tablehead{
\colhead{Ice Fraction} & \colhead{g$_s$}  & \colhead{P$_{Si-H_2O}$}    & \colhead{D$_M$} &
\colhead{P$_{center}$}  & \colhead{R$_{core}$}  & \colhead{P$_{Fe-Si}$}  & \colhead{R$_{Si-H_2O}$} \\
\colhead{\%} & \colhead{(m~s$^{-2}$)}  & \colhead{(GPa)}  & \colhead{(km)} &
\colhead{(GPa)}  & \colhead{(km)}  & \colhead{(GPa)}  & \colhead{(km)}
}
\startdata
25 & 23.4 & 246 & 2900 & 3763 & 5330 & 1483 & 10180 \\
50 & 20.9 & 493 & 5088 & 3055 & 4713 & 1436 & 8739
\enddata
\tablecomments{g$_s$-surface gravity, P$_{Si-H_2O}$-silicate and water mantle boundary pressure, D$_M$-water mantle depth, P$_{center}$-pressure at the center, R$_{core}$-iron core radius, P$_{Fe-Si}$-iron core and silicate mantle boundary pressure, R$_{Si-H_2O}$-distance from center to water mantle.}
\label{tab:10MEtable}
\end{deluxetable}

\section{THE MANTLE THERMAL PROFILE}

For the purposes of this section we define the planetary water mantle as the water ice layer bounded from above by the crust and from below by the silicate mantle.  Starting from the planetary crust and making our way deeper into the planet we need to make a distinction between crustal regimes I/III and II/IV.  In regimes I and III the surface heat flux is low enough so that the crust becomes unstable with respect to convection within the methane hydrate stability field. In these scenarios a convective cell, still in the methane hydrate layer, lies immediately underneath the conductive crust. As explained above, in regimes II and IV, underneath the conductive crust lies a layer confined to the hydrate dissociation (i.e melt) curve, which separates the crust from the underlying convective cell. We shall refer to this layer as the dissociation boundary layer (DBL) whose radial dimension, $\delta_{DBL}$, we constrain by assuming it is on the verge of convective instability \cite[see][and references therein]{fu10}. Since the DBL follows the dissociation curve we do not expect a large viscosity contrast across its length. 

\begin{figure}[ht]
\centering
\includegraphics[scale=0.5]{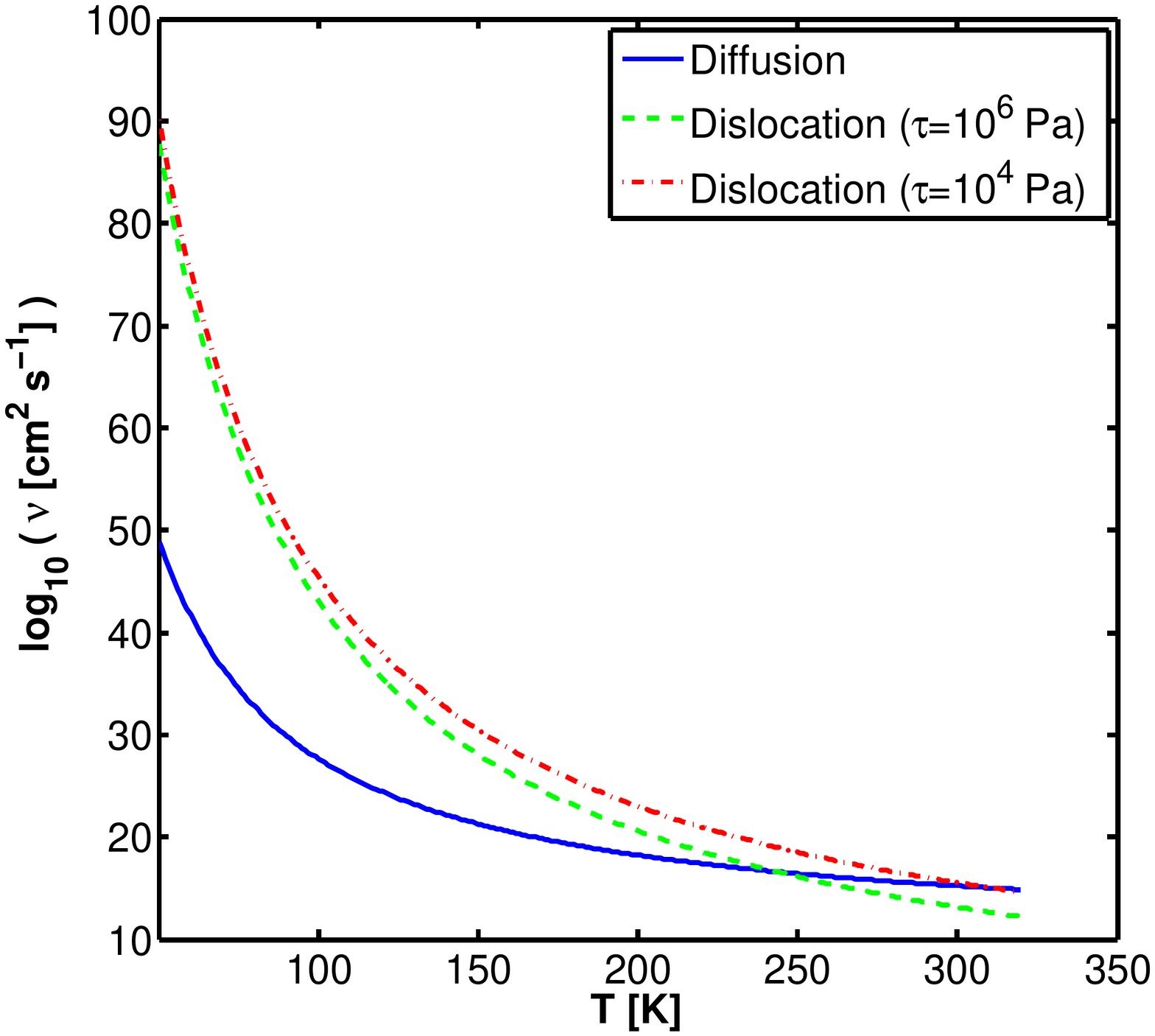}
\caption{\footnotesize{Diffusion creep (solid blue curve) and dislocation creep (dash and dash-dot curves) dependence on temperature for a solid of CH$_4$ clathrate hydrate and a reference pressure of $100$~bar. Up to a temperature of $250$K diffusion yields the lower viscosity while at higher temperatures dislocation creep yields the lower viscosity. Non-Newtonian viscosity is stress dependent and the viscosity curves shown are confined between stresses of $10^4$-$10^6$~Pa.}}
\label{fig:f7}
\end{figure}

Assuming the DBL is on the verge of convective instability brings up the question of the proper formalism for the convective threshold calculation. In Fig.\,\ref{fig:f7} we present a map of the Newtonian and non-Newtonian viscosities of methane hydrate as a function of temperature for a confining pressure of $100$~bar and the parameters given in section $2$. The lower viscosity of the two general mechanisms (i.e diffusion versus dislocation creep) is the proper viscosity \citep{schubert2001}. It is clear from Fig.\,\ref{fig:f7} that for the viscosity parameters adopted and for the temperatures expected at the DBL, dislocation creep will be slightly more efficient than diffusion creep, resulting in a non-linear stability problem. Tough, as discussed above, the parameters adopted for describing dislocation creep in methane hydrate could actually represent a lower bound on the dislocation viscosity due to the grain size difference between experimental and naturally forming clathrate samples. Also, for the viscosity parameters adopted and at the probable temperatures of the DBL both diffusion creep and dislocation creep do not show many orders of magnitude difference.  In addition, the non-linear dislocation creep requires finite disturbances whose existence is not certain.   For these reasons we shall treat the DBL stability problem with the formalism of the linear diffusion creep viscosity.

Assuming the DBL is on the verge of convective instability yields the following relation [see Eq.\,\ref{convinstab}]:
\begin{equation}
\frac{g_s\chi_{cl}\left[T_{m,cl}(P_{b,DBL})-T_{b,crust}\right]\delta_{DBL}^3C^{cl}_p\rho_{cl}}{\nu_{cl}(\bar{T})\kappa_{cl}}=Ra_{crit}
\end{equation}
where we take advantage of the fact that the thermal profile in the DBL follows the hydrate dissociation curve. Therefore, the temperature at its bottom is the hydrate dissociation temperature at the pressure prevailing at the base of the DBL layer, $P_{b,DBL}$. The temperature at the base of the crust is $T_{b,crust}$.
If the dissociation boundary layer is, relatively, narrow (which will prove to be the case) then Eq.\,\ref{hydrostatic} may be used to yield a univariant equation for the DBL base pressure:
\begin{equation}
\frac{\chi_{cl}\left(T_{m,cl}(P_{b,DBL})-T_{b,crust}\right)C^{cl}_p\left(P_{b,DBL}-P_{b,crust}\right)^3}{\nu_{cl}(\bar{T})\kappa_{cl}\rho^2_{cl}g^2_s}=Ra_{crit}
\end{equation}
where $P_{b,crust}$ is the pressure at the base of the crust.  Due to the mild viscosity contrast that is expected across the DBL (because it follows the methane hydrate melt curve) we assume a value of $1000$ for its critical Rayleigh number \cite[see][]{Turcotte2002}. The viscosity is approximated by its value at the average temperature of the DBL layer.        

For our choices for the ice mass fractions we find that all the planets we investigated have a dissociation boundary layer. The radial dimension of the DBL is listed, for six of our studied planets, in table\,\ref{tab:CrustandDBL}. $\delta_{DBL}$ is approximately $1$\,km for all six bodies. The radial dimension of the crust varies from a few hundred meters up to a kilometer for the six bodies listed. Changing the composition from $25\%$ to $50\%$ water, for a given body, widens the crust by a factor of about two. The lower mass planet has a slightly thicker crust than the more massive planet.

\begin{deluxetable}{lcc}
\tablecolumns{3}
\tablewidth{0pc}
\tablecaption{Crust and DBL data}
\tablehead{
\colhead{Planet} & \colhead{$\delta_{crust}$}   & \colhead{ $\delta_{DBL}$} \\    
\colhead{Parameters}    & \colhead{(km)}   & \colhead{(km)} } 
\startdata
M$_p$=2M$_E$ \quad 25$\%$ H$_2$O   & 0.61   & 1.36  \\
M$_p$=2M$_E$ \quad 50$\%$ H$_2$O   & 1.13   & 1.52   \\
M$_p$=5M$_E$ \quad 25$\%$ H$_2$O   & 0.39   & 1.11  \\
M$_p$=5M$_E$ \quad 50$\%$ H$_2$O   & 0.71   & 1.22   \\
M$_p$=10M$_E$ \quad 25$\%$ H$_2$O  & 0.27   & 0.94  \\
M$_p$=10M$_E$ \quad 50$\%$ H$_2$O  & 0.48   & 1.02   \\
\enddata
\tablecomments{\footnotesize{For six characteristic planets, out of our studied water planets, we give the radial dimension of the planetary crust ($\delta_{crust}$) and the radial dimension of the dissociation boundary layer ($\delta_{DBL}$).}}
\label{tab:CrustandDBL}
\end{deluxetable}

Regardless of whether the convecting cell underlies the conductive crust or the DBL it will follow a thermodynamic adiabatic profile into the abyss. The adiabatic temperature gradient is \citep{schubert2001}:
\begin{equation}\label{adiabatgrad}
dT=\frac{\chi(T,P) T}{\rho(T,P)C_p(T)}dP
\end{equation}
The heat capacity measurements of \cite{handa1986} [see Eq.\,\ref{heathanda}] are indeed appropriate for the planetary crust as these measurements were conducted at a low confining pressure (approximately $3$~MPa). The low pressure increases the probability that the sample studied contained pores filled with methane gas. The higher confining pressure (about $100$~MPa) in the experiments of \cite{waite2007} indicate a higher compaction and a lower probability for the existence of gas filled pores, which is more appropriate for the higher pressures in the convecting cell.  We therefore assume for the \textit{clathrate hydrate section of the convecting cell} the following heat capacity:
\begin{equation}
C^{cl}_p=6.71\times 10^4\left(T-273.16\right)+2.38\times 10^7 \quad {\rm erg\,g^{-1}\,K^{-1}}
\end{equation} 

From the base of the DBL we follow the adiabat in the convecting clathrate hydrate cell to find that in all the six planets we studied a point of intersection is reached where the adiabat tries to cross the SH methane hydrate dissociation curve into the liquid water regime. For all our planets this happens at a pressure of between $1-1.5$\,GPa.
To understand the thermal behavior beyond this point one has to calculate the adibatic profile in liquid water.  The temperature dependence of the volume thermal expansivity can be calculated from the tables of \cite{kell1975}. We calculate its pressure dependence using a scheme similar to what we used for the clathrate expansivity:
$$ 
\chi^w_{liq}(T,P)=\left[a_1(T-273.16)^3+a_2(T-273.16)^2+a_3(T-273.16)+a_4\right] 
$$
\begin{equation}
\times\left(\frac{B^w_{liq}+\tilde{B}^w_{liq}P}{B^w_{liq}+\tilde{B}^w_{liq}P_0}\right)^{-\xi}
\end{equation}
where $a_1=4.8777\times 10^{-10}$, $a_2=-1.1722\times 10^{-7}$, $a_3=1.4888\times 10^{-5}$ and $a_4=-5.1928\times 10^{-5}$. For the bulk modulus of liquid water and its derivative we assume numerical values of $2.17$~GPa and $7.0$ respectively \citep{manghnani}. For $\xi$ we again suggest a value of unity as for hydrates. The reference pressure here is $1$~bar.

For the heat capacity of liquid water we adopt without change the formulation in \cite{waite2007}. To the liquid water bulk mass density dependence on temperature \citep[see][and references therein]{waite2007} we add a pressure dependency, yielding the following form:
$$
\rho^w_{liq}(T,P)=\left[0.9999+5.330\times 10^{-5}(T-273.16)-6.834\times 10^{-6}(T-273.16)^2\right]
$$
\begin{equation}
\times\left(\frac{B^w_{liq}+\tilde{B}^w_{liq}P}{B^w_{liq}+\tilde{B}^w_{liq}P_0}\right)^{\frac{1}{\tilde{B}^w_{liq}}}
\end{equation}  
Calculating the adiabat in the liquid water, beyond the point of intersection just mentioned, we find its gradient ($dP/dT$) to be even larger than the gradient of the SH methane hydrate dissociation curve. In that case, as in the DBL, the penetration into the liquid water regime drives the thermal profile back to the SH methane hydrate stability field which in turn will try to re-penetrate the liquid regime. We are once more in a situation where the thermal profile is confined to the methane hydrate dissociation curve, only now it is the dissociation curve for the SH methane hydrate. \textit{This implies that, under our assumptions, our studied planets do not have a liquid subterranean ocean}. 

The SH methane hydrate dissociation curve is based on the experimental data points of \cite{dyadin1997} which are given up to a pressure of $1.5$\,GPa. Extrapolating these experimental data points we find the SH hydrate dissociation curve crosses the pure water ice VI melting curve at $1.6$\,GPa and $331$\,K. Going even deeper into the planet the SH methane hydrate dissociation curve is now set by the chemical potential equality of the SH methane hydrate and water ice VI. Unfortunately, no experimental data points are available for this regime of the methane hydrate dissociation curve.  Beyond a pressure of $2$\,GPa the SH methane hydrate will transform to a filled ice-Ih structure \citep{lovedaynat01} on which we elaborate below. Therefore, the range of uncertainty in the location of the SH hydrate dissociation curve spans a pressure difference of only about $0.4$\,GPa. Since the pressure range from $1.6$\,GPa to the introduction of the filled ice-Ih structure is relatively narrow,  it will not make a major impact on our results if we assume for it an adiabatic profile of hydrates or of water ice VI.

We expect that an adiabatic profile with SH methane hydrate characteristics prevails in the pressure range from $1.6$\,GPa to $2$\,GPa because the filled ice-Ih structure is able to maintain within it more methane per water molecules than the SH methane hydrate. It is unlikely that along the adiabat somewhere between $1.6$ and $2$\,GPa the SH hydrate would dissociate to pure water ice VI and solid methane, only to incorporate methane with even a greater efficiency within the water structure due to a small increase in pressure of the order of $0.1$\,GPa. On that ground we tentatively assume that a direct transition from clathrate hydrate to filled-ice occurs not only at room temperature, where it is seen experimentally, but also up to $340$~K (the adiabatic temperature at $2$~GPa for our planets).             
 
At a pressure of about $2$~GPa the methane clathrate hydrate will transform into a filled ice-Ih structure. An informative depiction of the filled ice-Ih crystal structure may be found in \cite{loveday01}. In the filled ice-Ih structure the methane molecules occupy the widened channels of the filled ice lattice instead of the quasi-spherical cages they occupy in classic clathrate hydrate crystals. For more information on filled ice-Ih we refer the reader to paper I where we have estimated the filled ice-Ih thermodynamic stability field, a point to which we shall return after obtaining the water mantle thermal profile.

The introduction of the classical clathrate hydrate to filled ice phase transition raises the important issue of phase change induced partitioning of the convective cell. Such a partitioning may result in higher temperatures inside the planet, as the partitioning introduces conductive boundary layers between the various convective cells. We have already mentioned above the pronounced low thermal conductivity of methane hydrate.  In appendix A we derive the clathrate hydrate to filled ice-Ih phase transition curve and show that the mantle convective cell is not likely to partition due to this phase transformation.

We therefore follow the adiabat in the SH methane hydrate layer until the transition to the filled ice phase, where we continue along an adiabat for the latter phase. Estimating the adiabat in the filled ice layer requires knowledge of its equation of state, for which we adopt the formalism derived in paper I. We also require the volume thermal expansivity for filled ice, which is experimentally unknown. As we explain in paper I, it is expected to be intermediate to the values for water ice VII and pure solid methane. By analogy with clathrate hydrates (see Fig.\,\ref{fig:f2}), which also represent a methane-water solid solution, we assign filled ice a volume thermal expansivity twice the value determined for water ice VII \citep{fei1993}.
 
The heat capacity of the filled-ice mantle is taken to be a linear combination of the values for water ice VII and pure solid methane, weighted according to their abundances in the crystal, $2/3$ and $1/3$ respectively. The heat capacity for water ice VII is taken from \cite{fei1993} and the heat capacity for a homogeneous system of methane is taken from \cite{Chase1998}.  We point out that the data used here for the heat capacity of methane is from experiments on the gaseous phase.  In other words we assume the entrapment of methane in the water ice lattice does not restrain the degrees of freedom of the methane molecules. Though this assumption is probably correct for low pressure \citep[see][and references therein]{sloan} it will gradually lose validity as the pressure increases, when venturing deeper into the planet. At high pressure the methane molecule may partly lose its ability to rotate freely and its heat capacity will decrease. Therefore, the adiabatic temperature across the ice mantle will rise [see eq.\ref{adiabatgrad}].  However, we do not expect the resulting uncertainty in the heat capacity to have a large influence on our results.  Rather we find the uncertainty associated with the probable range of values for the thermal expansivity of filled ice to dominate the overall uncertainty in the filled ice layer adiabat.

Following the adiabatic thermal profile, in the methane filled-water ice Ih layer deeper into the planet, we assume it terminates at a boundary layer which connects the water mantle with the silicate-metal interior. This boundary layer at the bottom of the filled ice mantle is henceforth referred to as the BBL. As for the case of the DBL, we also assume the BBL is on the verge of convective instability. Before formulating the BBL's appropriate scaling Rayleigh number we first wish to estimate its kinematic viscosity ($\nu_{bbl}$).

As we have already discussed above, a methane molecule will find it hard to diffuse through a clathrate structure water ring without causing local deformation of the surrounding water lattice. The filled-ice water lattice, which is far more compressed than the water lattice of cage clathrates, will impose even greater limitations on the ability of methane to diffuse.  We therefore assume that diffusion creep is subdued in the methane filled water ice mantle. In conjunction with the high stress acting in the deep mantle, the viscosity associated with non-Newtonian mechanisms should be the dominant creep mechanism. 

The non-Newtonian viscosities of high pressure water ice poly-morphs, such as ice VII and X, are not known experimentally.  This is also the case for highly pressurized solid solutions such as filled ice. We can remedy this lack of data, to some modest extent, by making use of the algorithm given in subsection $2.2$, where the melting curve is assumed to correctly describe the dependency of the viscosity activation enthalpy on pressure and temperature. The caveat here is that the factor $\Omega$ [see eq.\ref{OmegaHomologous}] is assumed to be a constant which depends on the crystal structure, whereas in reality it is also a function of pressure. Therefore, while over limited pressure ranges $\Omega$ may be assumed constant, over large pressure ranges extending over the entire ice mantle, $\Omega$ should be allowed to vary with pressure. Since the physical basis for $\Omega$ is not yet well formulated, the method of the homologous temperature is somewhat lacking in its ability to predict viscosity for cases where no experimental data exists. This said, we shall adopt the non-Newtonian viscosity for the highest pressure water ice poly-morph whose viscosity was determined, i.e. water ice VI, and try to adjust its activation energy and volume to better suit the characteristics of methane filled ice.   
 
We assume that the viscosity of filled ice, which comprises the BBL, has characteristics analogous to those of methane clathrate hydrate. This assumption stems from the general point of view that both clathrates and filled ice are solutions to the basic problem of methane-water solid solubility and therefore probably share similar characteristics. In addition, the inclusion of methane in both crystal structures introduces voids and represents a similar impurity inserted into the water lattices, which tends to increase the viscosity \citep{Durham1992}.

It is interesting to note that while the activation {\it energy} for cage clathrates ($90$~kJ\,mol$^{-1}$) is lower than the activation energy of water ice VI ($110$~kJ\,mol$^{-1}$), the activation {\it volume} for cage clathrates ($19$~cm$^3$\,mol$^{-1}$) is higher than the activation volume of water ice VI ($11$~cm$^3$\,mol$^{-1}$) \citep{durham1997,durham2003}. A tentative explanation for these values would be that the introduction of methane into the water lattice introduces weaker methane-water bonds and some distortion of the water lattice, resulting in a reduced activation energy. In addition, the gliding of crystal planes one along the other requires breaking followed by re-connection of molecular bonds in the new location. As molecular bonds break the local molecules tend to expand resulting in an activation volume. \cite{raghavendra2008} have shown that the non-bonded radius of the oxygen atom in pure water is $1.9\times 10^{-8}$~cm, while the bonded radius is $1.3\times 10^{-8}$~cm. This represents a volume change of $11.76$~cm$^3$\,mol$^{-1}$, remarkably close to the experimentally determined activation volume for pure water ice VI. 

\cite{raghavendra2008} further give the penetration distance of methane into water due to the formation of a weak hydrogen bond, and that can be compared with methane's van der Waals radius (see paper I) to yield a volume difference of $23.51$~cm$^3$\,mol$^{-1}$. For the case of filled ice we adopt an activation volume which is weighted according to the abundance of the two different constituents:
\begin{equation}
V^\ast_{0,FI}=23.51\times\frac{1}{3}+11.76\times\frac{2}{3}=15.67\quad {\rm cm^3\,mol^{-1}}
\end{equation}
Following \cite{Oconnell1977} the activation volume is assumed to decrease with pressure as a lattice vacancy, with an effective bulk modulus ($B_{eff}$) given by:
$$
V^\ast_{FI}(P)=V^\ast_{0,FI}\left(\frac{B^{FI}_{eff}+\tilde{B}_{FI}P}{B^{FI}_{eff}+\tilde{B}_{FI}P_{ref}}\right)^{-1/\tilde{B}_{FI}}
$$
\begin{equation}\label{VolActFI}
B^{FI}_{eff}=\frac{2(1-2\Lambda)}{3(1-\Lambda)}B_{FI}
\end{equation}
Where $\Lambda$ is the Poisson ratio estimated to be $0.35$ \citep{fu10}. $B_{FI}$ and $\tilde{B}_{FI}$ are methane filled ice bulk modulus and its pressure derivative, which experimentally are found to be $10$\,GPa and $4$ (see paper I), respectively.

For the activation energy of filled ice, also expected to be lower than that for water ice VI, we simply adopt the value from cage  clathrates of $90$~kJ mol$^{-1}$. All the other parameters are adopted from water ice VI \citep{durham1997}, giving for the filled ice kinematic viscosity the following form:

\begin{equation}\label{VisFInum1}
\nu_{FI}=\frac{1}{2C_{FI}\rho_{FI}}\tau^{1-n_{FI}}e^{\frac{E^\ast+PV^\ast}{kT}}
\end{equation}
where the index $FI$ refers the parameter to the filled ice structure. For $n_{FI}$ we adopt the value of $4.5$ and for $C_{FI}$ we assume $10^{6.7}$~MPa$^{-4.5}$~s$^{-1}$, both from ice VI viscosity measurements \citep{durham1997}. 

Now that we have an estimate for the viscosity at the BBL, we can analyze its stability with respect to convection.  Here we need consider the complication of the viscosity being non-Newtonian. This effect was analyzed by \cite{solomatov1995}, whose scaling for the non-Newtonian Rayleigh number, in combination with the viscosity formalism as expressed in Eq.\,($\ref{VisFInum1}$), gives the following Rayleigh instability criterion for the BBL:
\begin{equation}
Ra_{bbl}\equiv\frac{\chi_{bbl}\rho_{bbl} g_{bbl}\left(T_{Si-H_2O}-T_{ad}^{FI}\left(P_{bbl}^{up}\right) \right)d_{bbl}^{\frac{n+2}{n}}}{\alpha_{bbl}^{\frac{1}{n}}\left(2C\right)^{-\frac{1}{n}}e^{\left(\frac{E^\ast+PV^\ast}{nkT}\right)}}=\left(1568\right)^{\frac{1}{n}}\left(20\right)^{\frac{n-1}{n}}
\end{equation} 
where the subscript $bbl$ refers the parameter to its value at the bottom boundary layer.  

The temperature difference across the BBL is here represented by the temperature at the water/silicate boundary ($T_{Si-H_2O}$) and the temperature along the filled ice adiabat at the pressure prevailing in the outer boundary of the BBL, $T_{ad}^{FI}\left(P_{bbl}^{up}\right)$.
The length scale of the BBL, $d_{bbl}$, and the heat flux at the BBL, $F_{bbl}$, are related through the relation:
\begin{equation}\label{BBLflux}
d_{bbl}=\frac{\kappa_{bbl}\left(T_{Si-H_2O}-T_{ad}^{FI}\left(P_{bbl}^{up}\right)\right)}{F_{bbl}}
\end{equation}
Using the last relation to eliminate the temperature difference across the BBL, yields:
\begin{equation}
\frac{\chi_{bbl}\rho_{bbl}^{\frac{n+1}{n}} g_{bbl}F_{bbl}C_{p,bbl}^\frac{1}{n}d_{bbl}^{\frac{2(n+1)}{n}}}{\kappa_{bbl}^{\frac{n+1}{n}}\left(2C\right)^{-\frac{1}{n}}e^{\left(\frac{E^\ast+PV^\ast}{nkT}\right)}}=\left(1568\right)^{\frac{1}{n}}\left(20\right)^{\frac{n-1}{n}}
\end{equation}
Further, assuming the BBL is narrow enough so that it may be represented using a constant density and acceleration of gravity [see Eq.\,\ref{hydrostatic}], the last equation transforms to:
\begin{equation}\label{stabilitynonNewton}
\frac{\chi_{bbl}F_{bbl}C_{p,bbl}^\frac{1}{n}\left(P_{Si-H_2O}-P_{bbl}^{up}\right)^{\frac{2(n+1)}{n}}}{\kappa_{bbl}^{\frac{n+1}{n}}\left(2C\right)^{-\frac{1}{n}}e^{\left(\frac{E^\ast+PV^\ast}{nkT}\right)}\rho_{bbl}^{\frac{n+1}{n}}g_{bbl}^{\frac{n+2}{n}}}=\left(1568\right)^{\frac{1}{n}}\left(20\right)^{\frac{n-1}{n}}
\end{equation}  
where $P_{Si-H_2O}$ is the pressure at the water/silicate boundary.  

Assuming the water mantle has neither heat sources nor sinks we may scale the heat flux at the surface to the water/silicate boundary:
\begin{equation}
F_{bbl}=F_{s,planet}\left(\frac{R_{p}}{R_{Si-H_2O}}\right)^2
\end{equation}
The acceleration of gravity at the BBL obeys:
\begin{equation}
g_{bbl}=\frac{GM_{p}X_p^{Si+Fe}}{R_{Si-H_2O}^2}
\end{equation}
Which, in combination with Eq.\,(\ref{stabilitynonNewton}), yields a univariant equation for the pressure at the outer boundary of the BBL:
\begin{equation}\label{stabilitynonNewton2}
\frac{\chi_{bbl}C_{p,bbl}^\frac{1}{n}\left(P_{Si-H_2O}-P_{bbl}^{up}\right)^{\frac{2(n+1)}{n}}F_{s,planet}R_{p}^2R_{Si-H_2O}^{\frac{4}{n}}}{\kappa_{bbl}^{\frac{n+1}{n}}\left(2C\right)^{-\frac{1}{n}}e^{\left(\frac{E^\ast+PV^\ast}{nkT}\right)}\rho_{bbl}^{\frac{n+1}{n}}\left[GM_{p}X_p^{Si+Fe}\right]^{\frac{n+2}{n}}}=\left(1568\right)^{\frac{1}{n}}\left(20\right)^{\frac{n-1}{n}}
\end{equation}
The thermal conductivity at the BBL, $\kappa_{bbl}$, is estimated in appendix C. When solving Eq.\,($\ref{stabilitynonNewton2}$) we set the temperature to the average temperature at the BBL:
\begin{equation}
\bar{T}_{bbl}\left(P_{bbl}^{up}\right)=T_{ad}^{FI}\left(P_{bbl}^{up}\right)+\frac{F_{s,planet}R_{p}^2\left(P_{Si-H_2O}-P_{bbl}^{up}\right)}{2\kappa_{bbl}\rho_{bbl}GM_{p}X_p^{Si+Fe}}
\end{equation}

\begin{figure}[ht]
\centering
\includegraphics[scale=0.5]{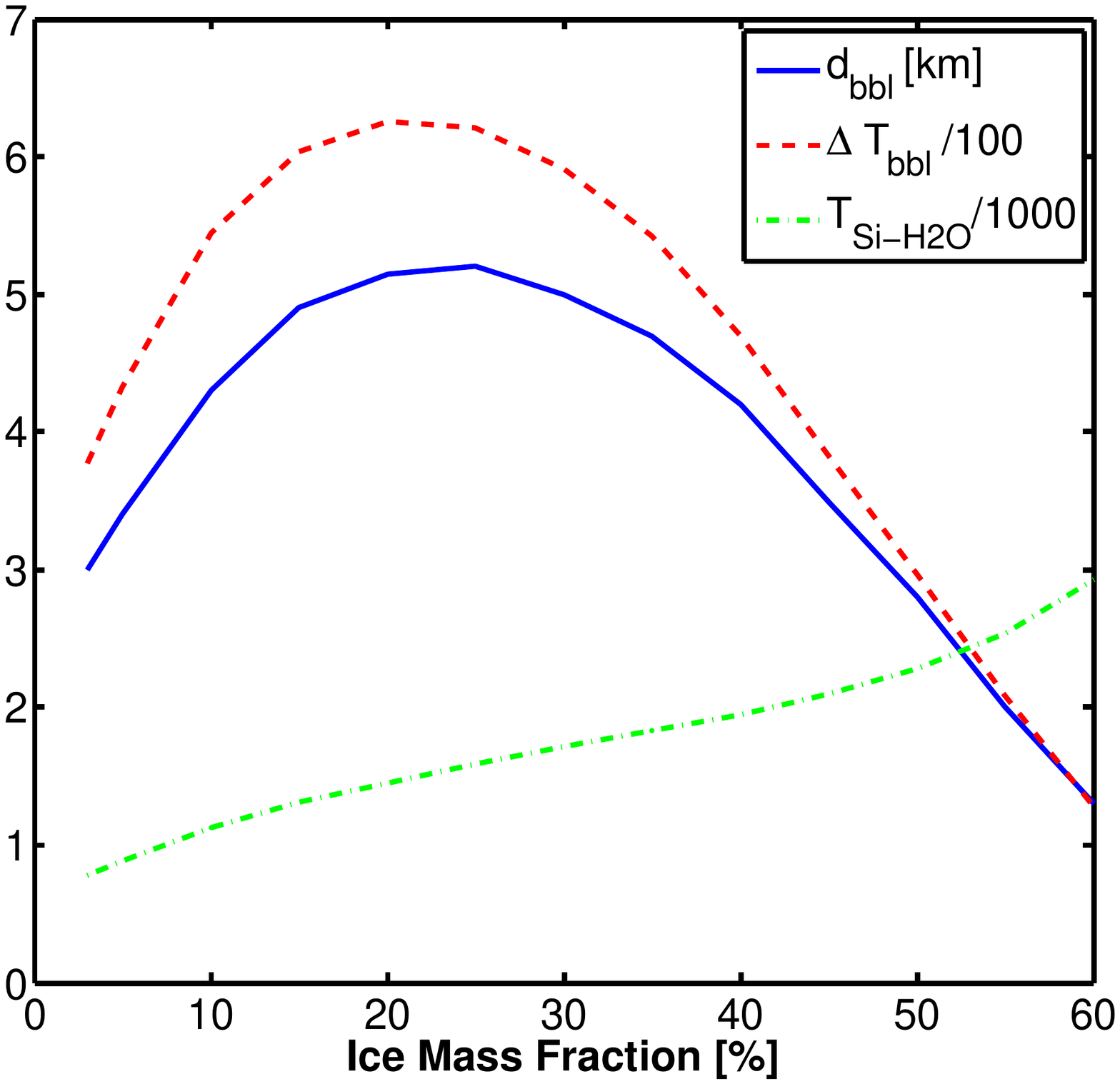} 
\caption{\footnotesize{Bottom boundary layer, BBL, parameters for the $2M_E$ planet for a varying ice mass fraction. The solid (blue) curve is the length scale of the BBL, in km. The dashed (red) curve is the temperature difference across the BBL in kelvins (normalized by $100$) and the dashed-dotted curve (green) is the estimated temperature on the transition to the silicate mantle in kelvins (normalized by $1000$).}}
\label{fig:f8}
\end{figure}

In Fig.\,\ref{fig:f8} we show the variation of $d_{bbl}$ as a function of the ice mass fraction for the $2M_E$ planet.
It is interesting that the bottom boundary length scale has a maximum for an ice mass fraction between $20\%$ to $30\%$.  The situation is more complicated for the more massive planets. \textit{Quite generally, we find that the higher thermal expansivity of the filled ice mantle compared to a pure water ice mantle results in hotter planetary interiors, i.e. less steep adiabatic profiles}. 

\begin{figure}[ht]
\centering
\mbox{\subfigure{\includegraphics[width=7cm]{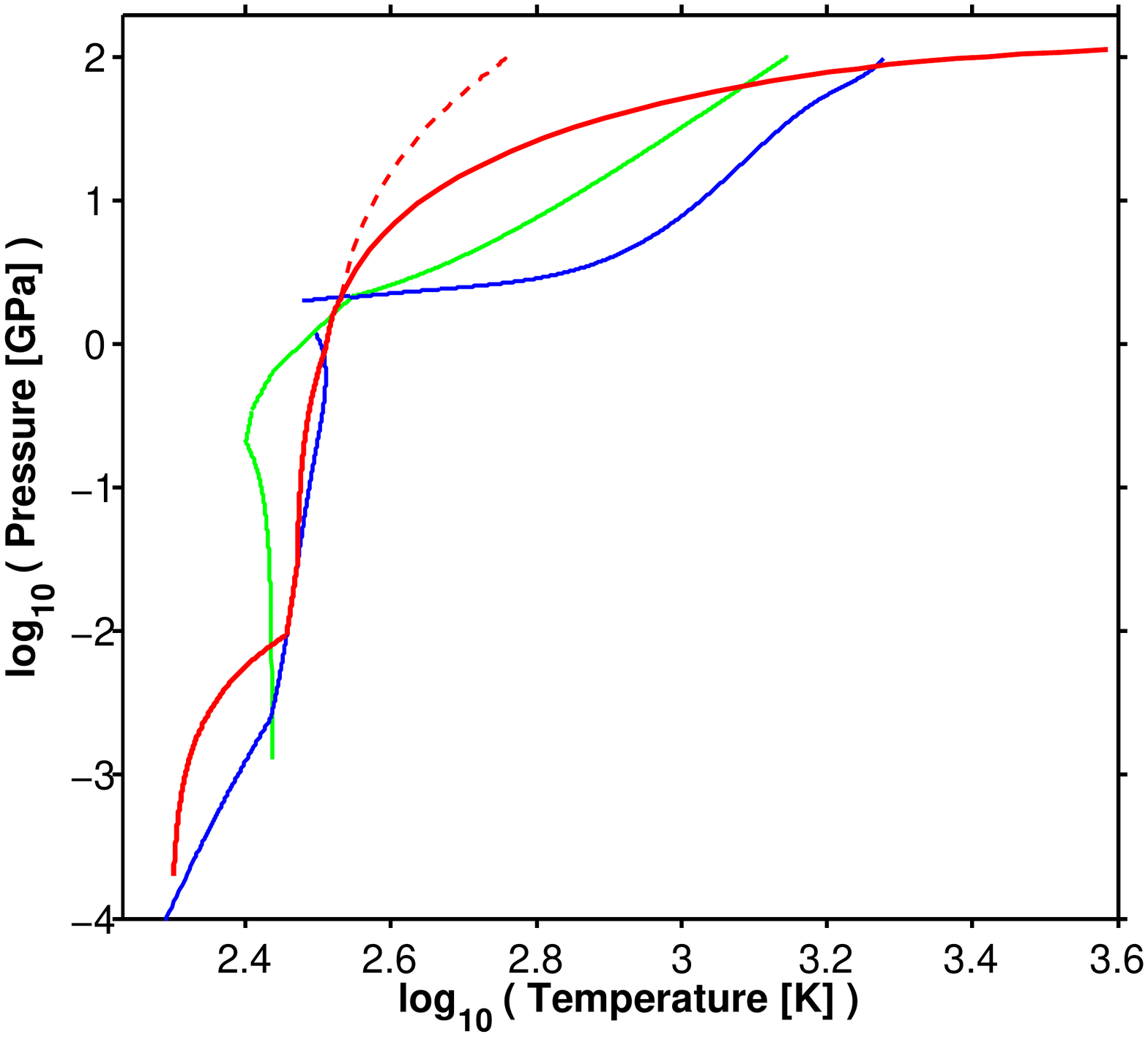}}\quad \subfigure{\includegraphics[width=7cm]{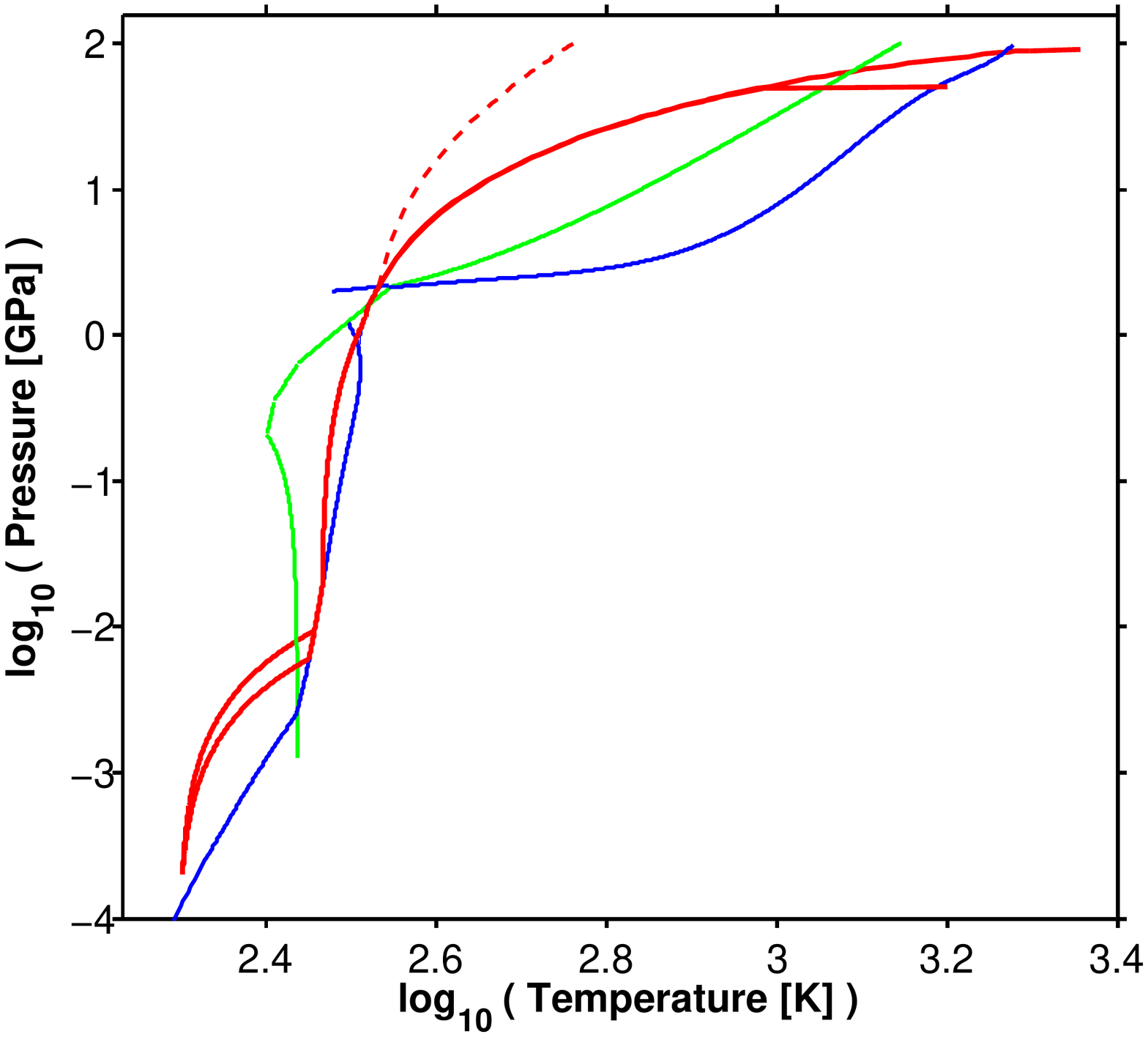}}}
\caption{\footnotesize{Thick (red) curves represent the thermal profile in the clathrate hydrate and filled-ice mantle of the $10M_E$ planet with a $50\%$ ice mass fraction (left panel) and a $2M_E$ planet with $50\%$ and $25\%$ ice mass fractions (right panel). Thin (green) curves represent the melting curve for pure molecular water ice \citep[see][]{Lin2004,Goncharov2009}. Thin (blue) curve represents the dissociation boundary for SI methane clathrate hydrate , SH methane clathrate hydrate and methane filled ice Ih, in order of increasing pressure, respectively. The dashed (red) curve is the adiabat for the case where filled ice is replaced with water ice VII.}}
\label{fig:f9}
\end{figure}  

In Fig.\,$\ref{fig:f9}$ we show the thermal profile (thick red curve) in the ice mantle for the $10M_E$ planet, with $50\%$ ice mass fraction, against our estimated water-methane phase diagram (blue curve) and the melting curve of pure molecular water (green curve). The dashed (red) curve is the adiabat for the case where filled-ice is replaced with water ice VII. For the water ice VII mantle the interior temperatures are shown to be lower.  In addition, as shown in the figure, the thermal profile in the mantle of the $10M_E$ planet crosses the estimated stability field for methane filled ice (filled ice is not stable to the right of the blue curve). The extension of the mantle adiabat to the right of the filled ice stability boundary is therefore not appropriate. A proper extension of the adiabat beyond this point of intersection would require knowledge of the thermodynamic properties of the mixtures that exist beyond the stability of the methane filled ice.  For the case of the $5M_E$ planet the general behavior of the icy mantle thermal profile is similar to that of the $10M_E$ planet.  We find a point of intersection with the methane filled ice stability boundary to occur for both the $5M_E$ and $10M_E$ planets, regardless of whether the ice mass fraction is $25\%$ or $50\%$.  

For these more massive water planets, a question now arises, of what lies beyond the filled-ice stability regime. According to \cite{Benedetti1999}, at high temperatures the C-H bond may break, resulting in the dissociation of the methane molecules. At high pressure condensation of the freed carbon atoms may ensue. It is of particular interest to compare our model, which we confine to the molecular solid regime, with the phase diagram for synthetic Uranus, derived experimentally by \cite{Chau2011}. The introduction of carbon atoms to a water surrounding, at pressure above $100$~GPa and temperature beyond $1000$~K, introduces a super-ionic phase whose extent of stability is narrower than the corresponding phase for a pure water system. This is because the introduction of carbon atoms increases diffusivity among the oxygen atoms, resulting in destruction of the super-ionic phase. At even higher temperatures ($2000$-$4000$~K), depending on pressure, a reticulating phase is introduced, where, methane dissociates, releasing excess hydrogen. The relatively long lifetime of the C-C bond results in the formation of dense carbon clusters that should tend to segregate and sink \citep{Chau2011}. 

It is possible that these high temperature and high pressure phases are present in the \textit{lower} part of the icy mantle of our $5M_E$ and $10M_E$ planets, underlying an \textit{upper} mantle composed of methane filled-ice acting as a thermal insulator to keep the interior warm. If that is indeed the case, then the creation of carbon clusters and their segregation will limit the ability of carbon expelled from the silicate interior from reaching the filled ice upper mantle, thus hindering its convection upward to the surface and the atmosphere.

It is tempting to generalize these arguments and say that for the more massive planets we describe, where the interior temperatures and pressures expected are higher, the lower part of the ice mantle may indeed be in the reticulating phase. For a less massive body the decrease in the expected interior temperature and pressure may result in a lower ice mantle in the super-ionic regime, leading to different consequences for carbon transport. For low mass planets (as we will show for our $2M_E$ planets) the much lower interior temperatures and pressures could lead to an icy mantle which is entirely in the molecular crystal regime of filled-ice. For these low mass planets, the transport of methane expelled from the interior may be entirely due to its incorporation in the filled ice phase.

Due to the difficulties just mentioned from this point onward we will continue with emphasis on the $2M_E$ planet alone.
We have already derived the thickness of the BBL for the $2M_E$ planet above. The temperature difference across the BBL, which is a conductive layer, obeys:
\begin{equation}
\Delta T_{bbl}=\frac{F_{s,planet}}{\kappa_{bbl}}\left(\frac{R_{p}}{R_{Si-H_2O}}\right)^2d_{bbl}
\end{equation}      
In Fig.\,\ref{fig:f8} we give the temperature difference across the BBL and the expected temperature at the ice/silicate boundary for the $2M_E$ planet for various ice mass fractions. As expected, as $d_{bbl}$ increases so does the temperature difference across the BBL. The temperature at the transition to the silicate mantle increases monotonically with increasing ice mass fraction, even though the thickness of the BBL decreases. This is because the total scale of the ice mantle increases with the ice mass fraction. 

In the right hand panel of Fig.\,$\ref{fig:f9}$ we give the thermal profiles in the icy mantles of the $25\%$ and $50\%$ water mass fraction, $2M_E$ planet. The $25\%$ water mass fraction scenario is to the left of the filled ice dissociation curve (blue), and therefore its entire ice mantle is probably composed of filled ice molecular solid. In the case of the $2M_E$ and $50\%$ water mass fraction planet, a point of intersection with the filled-ice stability curve exists, as for the more massive planets. Though, contrary to the case of the more massive planets, the uncertainty in determining the exact location of the filled-ice dissociation curve is great enough so that we cannot rule out the possibility that its entire mantle is composed of filled-ice as well.  After deriving probable thermal profiles in the icy envelopes of our water planets we wish to estimate the surface outgassing flux of methane into the atmosphere. Determining the outgassing mechanism is intimately linked to the geophysical behavior of the lithosphere, and therefore depends on the active tectonic mode. This issue is addressed in the following section.  

\section{TECTONICS IN WATER PLANETS} 

It has been suggested that due to their larger masses super-Earths are even more likely than Earth to establish plate tectonics \citep{Valencia2007}. This stems from the hypotheses that the lithospheres of more massive super-Earths will be thinner and experience higher applied stresses, therefore having a greater ability to deform.  This is a vital condition for plate tectonics.
On the other hand, \cite{ONeill2007} argue that the increased fault strength due to scaling up of the planetary mass will make stagnant lid more probable. The reason for this apparently contradictory behavior was recently shown to stem from the fact that tectonic modes may have multiple solutions for the same parameter space. This was shown both analytically \citep{Crowley2012} and numerically \citep{Lenardic2012}.

The multiple solution nature of tectonics tells us that listing a planet's parameters (e.g. mass, composition, viscosity, etc.) does not guarantee a unique tectonic mode.  Rather, the geologic and climatic history must also be taken into account \citep{Lenardic2012}. Certainly such a detailed history is not known for any water planet. Therefore, in this section we try to map the characteristics of the different multiple tectonic mode solutions, bearing in mind that all the derived modes are possible since we presently lack the knowledge required to rule out particular modes. If the different modes result in sufficiently different atmospheric regimes, it may be possible to distinguish among the different possibilities observationally.  A first step towards this goal will be addressed in the next section.
    
Although the theory for multiple tectonic modes was originally developed for rocky planets, we assume that similar forces are responsible for maintaining plate motion and subduction in planets with ice layers. We take into consideration the fact that our icy mantle will have a rheological profile with depth which is very different from that assumed for the Earth's mantle.    

Above, we mentioned the existence of two layers whose thermal profiles are confined to the melting curves of SI methane clathrate  and SH methane clathrate. Being confined to the melting curve implies these layers have relatively low viscosities. Indeed, thermal profiles depicted in Fig.\,\ref{fig:f9} reveal a mid-layer whose thermal profile is fairly close to the local melting (dissociation) curve, suggesting that a low viscosity layer exists between the planetary lithosphere and lower mantle.  This corresponds to the asthenosphere in the Earth.  Considering the effect of clathrates on the planetary thermal profile, such a low viscosity mid-layer may be the rule rather then the exception in these icy worlds. 

The ratio of the asthenospheric to lower mantle dynamic viscosities ($\mu_A/\mu_M$) in our case is somewhat difficult to constrain, since, as discussed above, the viscosities are probably non-Newtonian. Therefore, the viscosity ratio depends on the applied stress profile with depth, which, in turn, depends on the vertical and horizontal velocity profiles along with the temperature and pressure profile.  Using boundary layer theory [see Eq\,.$\ref{shearestimation}$], we find that the shear stress second invariant, $\tau\sim10$~MPa for the $2M_E$ planet. In conjunction with our estimated non-Newtonian viscosity model for filled ice [see Eq.\,\ref{VisFInum1}] this gives an average lower mantle viscosity ranging from $2\times 10^{21}$~Pa\,s to $4\times 10^{21}$~Pa\,s for the $2M_E$ planet, when varying the ice mass fraction between $50\%$ to $25\%$ respectively. 

\begin{figure}[ht]
\centering
\includegraphics[scale=0.5]{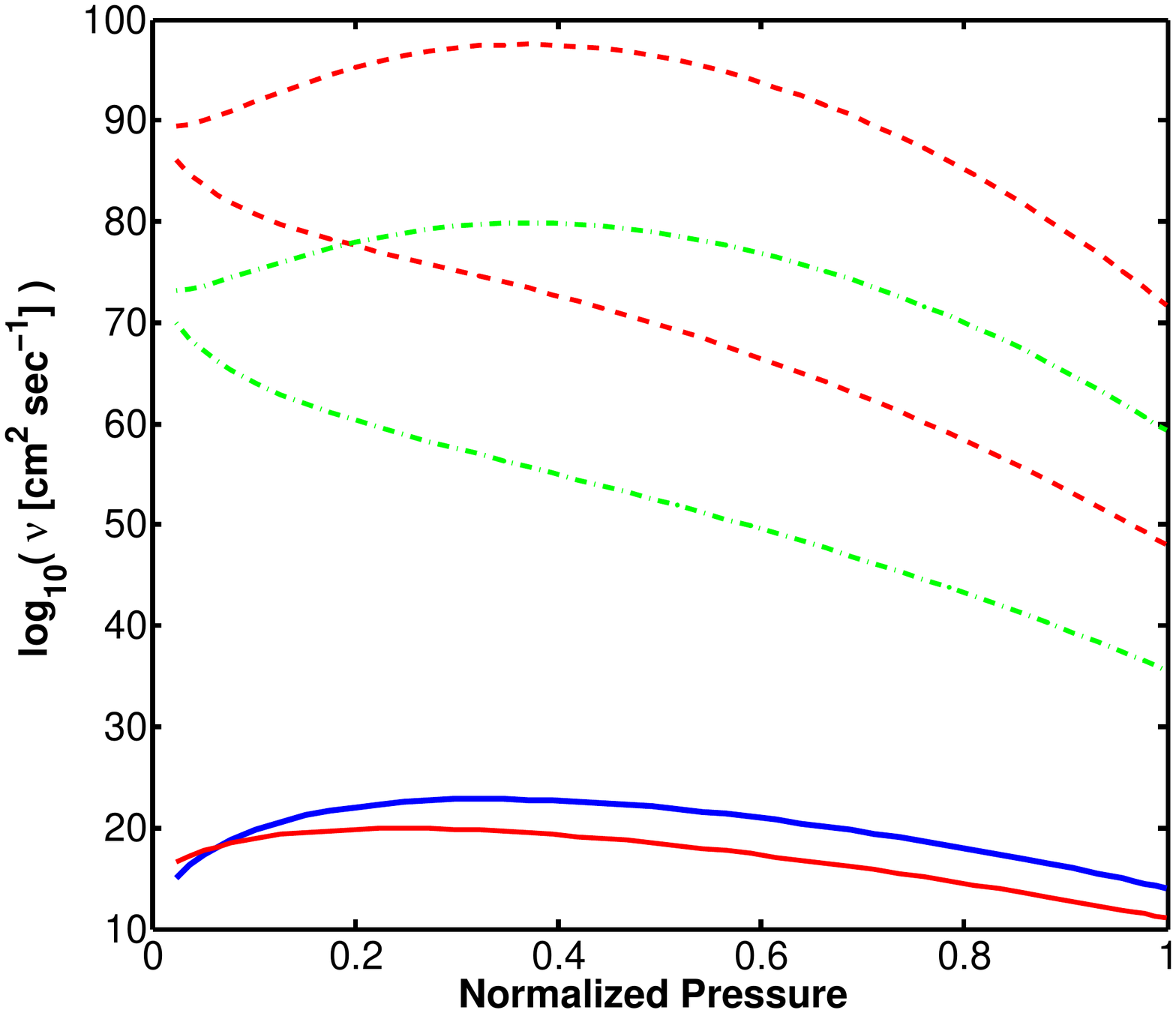}
\caption{\footnotesize{We compare the kinematic viscosities along the adiabat, for the $2M_E$ planet, assuming $50\%$ ice mass fraction, for four different material viscosities, using the same planetary conditions. Thick solid (blue) curve is assuming a filled ice mantle. Dashed (red) curves confine the kinematic viscosity assuming dry olivine \citep{Karato1993}. Dashed-dotted (green) curves confine the kinematic viscosity assuming wet olivine \citep{Karato1993}. Thin (red) curve is the kinematic viscosity assuming water ice VI \citep{durham1997}. The x-axis is normalized pressure, therefore spans the entire mantle, where $0$ is the SH clathrate hydrate to filled-ice transition and $1$ is the transition from ice to the silicate-metal core.}}
\label{fig:f11}
\end{figure}

In Fig.\,\ref{fig:f11} we show the actual viscosity profile with depth in the filled ice mantle for the $2M_E$ planet assuming a $50\%$ ice mass fraction. Fig.\,\ref{fig:f11} also serves as a test for the filled ice viscosity model by comparing it to viscosities of high pressure water ice polymorphs and silicates under the same thermal conditions. We expect the viscosity of silicates to be much higher than filled ice, and that of filled ice to be close to that of pure water ice.  In Fig.\,\ref{fig:f11} the viscosity profiles are obtained by keeping the mantle thermal profile the same while varying the viscosities between that for olivine, both wet and dry \citep{Karato1993}, our filled ice viscosity model and the viscosity for water ice VI \citep{durham1997}. Indeed our model for the viscosity of filled ice yields a viscosity intermediate between that for ice VI and that for olivine, though it is much closer to ice VI than to olivine.

For the asthenospheric viscosity we use the form suggested for clathrate hydrates [see Eq.\,\ref{viscosity2}]. Although the viscosity of SH clathrate hydrate is unknown, one can estimate it by replacing the melting curve for clathrate SI with that for clathrate SH in Eq.\,\ref{viscosity2}, while keeping all the other parameters unchanged. Clearly these approximations for the viscosities of the lower mantle and asthenosphere are somewhat crude and more experimental data is required. Due to the possible large errors in the viscosities assigned to the different layers an exact solution is currently beyond reach. However, the viscosity formulations we use indicate, at least qualitatively, that an asthenosphere exists and that its viscosity may easily be at least three orders of magnitude smaller than the viscosity of the lower mantle. The ratio of asthenospheric to lower mantle viscosity in our water planets may therefore be significantly lower than what is assumed for Earth. As a result this asthenosphere will be of great significance for the ability to develop plate tectonics.  

A plate tectonic theory that has the ability to account for the effect of an asthenosphere was recently developed by \cite{Crowley2012}. It has the ability to quantify both an active and a sluggish lid and to qualitatively describe the transition to a stagnant lid scenario. Adopting their model we solve for the dynamics of the lithospheric plate, except that we use the composition for icy worlds rather then rocky planets. Due to the large uncertainties in the viscosities of our icy crystal structures we solve for ratios of asthenospheric to lower mantle viscosity ranging $10^{-3}$ to $10^{-6}$. This is done by keeping the lower mantle viscosity constant at the values mentioned above while varying the asthenospheric viscosity accordingly. The purpose of this exercise is to obtain a general qualitative understanding of the way plate tectonics may behave in water worlds for different viscosity ratios.  Below we briefly summarize the theory.  Further details can be found in \cite{Crowley2012}.

The theory is based on scaling considerations with the exception that it involves the derivation of the horizontal velocity profile in the convection cell. The latter may be used to derive the maximal vertical flow velocity in the cell, $V_M$. Two mechanical energy balance equations are formulated, one for the lithosphere alone and the other for the entire convection cell.  The energy balance equation for the lithosphere is:
\begin{equation}\label{EnergyBalLith}
\underbrace{\frac{\chi_lg_sd_l}{C_{p,l}}\left\langle Q_{adv}\right\rangle_l}_\textit{T1} = \underbrace{L \left(\frac{\partial P}{\partial x}\right)_{base}d_lU_p}_\textit{T2}+\underbrace{L\tau_pU_p}_\textit{T3}+\underbrace{\tau_Rd_lU_p}_\textit{T4}
\end{equation}  
Here $\chi_l$ and $C_{p,l}$ are the lithospheric plate thermal expansivity and isobaric heat capacity respectively. The gravitational acceleration is $g_s$. The thermal thickness of the lithosphere is $d_l$, estimated by \cite{Crowley2012} to obey the half space cooling model.  This gives $d_l=2.32\sqrt{\alpha_l t}$, where $\alpha_l$ is the lithospheric thermal diffusivity and $t$ is a characteristic time scale.  If $L$ is the length of the plate and $U_p$ is its horizontal speed, then $t\approx L/U_p$. $\left\langle Q_{adv}\right\rangle_l$ is the advective rate of heat transfer through a horizontal cross section of the lithosphere averaged along the lithospheric depth. $\left(\partial P/\partial x\right)_{base}$ is the flow horizontal pressure gradient estimated at the base of the plate. $\tau_p$ is the shear stress operating on the base of the plate due to coupling with the underlying flow (upper part of the asthenosphere). The net resistive stress is $\tau_R$, which obeys:
\begin{equation}
\tau_R=\tau_{bend}+\tau_F-\tau_{sp}
\end{equation} 
where $\tau_{bend}$ is an effective bending stress representing a weighting factor for the action of plate bending at subduction and its ability to dissipate plate kinetic energy. $\tau_F$ is the fault stress from friction with the overlying plate during subduction.  It too is a weighting factor for the dissipative efficiency of this mechanism. The final weighting factor is $\tau_{sp}$, the normal stress associated with slab pull, this term weighs the ability of the pulling slab to generate kinetic energy in the plate.
 
In Eq.\,\ref{EnergyBalLith} the different terms represent the work of buoyancy ($T1$) which is a source of kinetic energy for the lithosphere, the work due to the horizontal pressure gradient ($T2$) which is basically the flow pressure difference between the ridge and subduction zone, the work of traction from the underlying asthenosphere ($T3$), and the work of the net resistive forces ($T4$). Finally the energy balance equation for the entire convection cell is:
\begin{equation}\label{EnergyBalTot}
\frac{\chi_{total}g_sd_{total}}{C_{p,total}}\left\langle Q_{adv}\right\rangle_{total} = \left(\tau_F+\tau_{bend}\right)d_lU_p + \Phi_M
\end{equation}
Here the subindex $total$ replaces the subindex $l$ meaning that the parameter is now a representative  average for the entire cell rather then for the lithosphere alone. Furthermore, the lithospheric scale ($d_l$) is replaced with the depth scale of the entire cell, $d_{total}$. $\Phi_M$ is a term representing the dissipation in the lower mantle from both the horizontal flow at mid-cell and vertical flow at the convection cell corners. \cite{Crowley2012} solve for both Eqs.\,(\ref{EnergyBalLith}) and (\ref{EnergyBalTot}) simultaneously to obtain $\tau_p$ and $U_p$. 

We also solve for $\tau_p$ and $U_p$ for the case of the $2M_E$ super-Earth for ice mass fractions of $25\%$ and $50\%$. 
After producing several rheological profiles for different asthenospheric stresses we find it to be a good approximation to partition the lower mantle and the asthenosphere at the clathrate hydrate to filled ice phase transition at $2$\,GPa. This results in lower mantle depths of $1900$\,km and $3472$\,km, for the $25\%$ and $50\%$ ice mass fraction respectively.  The sinking slab, which is still attached to the lithosphere, will apply a stress due to its negative buoyancy. We estimate the normal stress due to this slab pull to be:
\begin{equation}
\tau_{sp}\approx \chi_l\rho_{cl}\Delta Tg_sh_{slab}
\end{equation}   
where $\rho_{cl}$ is the bulk mass density of clathrate hydrates, $\Delta T$ is assumed to be $500$~K and the slab length, $h_{slab}$, is assumed to be $100$~km. This gives $\tau_{sp}=110$~MPa and $94$~MPa for the $25\%$ and $50\%$ ice mass fraction respectively.

Estimating the effective bending stress ($\tau_{bend}$) is somewhat more complicated. According to \cite{Crowley2012} and \cite{Conrad1999} treating the lithosphere as a beam that experiences bending at subduction, and thus dissipation, results in the following parametrization for its effective bending stress:
\begin{equation}
\tau_{bend}=\mu_L\frac{d_l^2}{R_{curv}^3}U_p\sim\mu_L\frac{\alpha_lL}{R_{curv}^3}
\end{equation}
Here $\mu_L$ is the dynamic viscosity appropriate for the lithosphere and $R_{curv}$ is the radius of curvature of the bent lithosphere. To obtain the term on the far right hand side of the last equation one simply has to replace the lithospheric length scale, $d_l$ with its half space cooling model estimate. For a given planetary mass and ice mass fraction the terms in the expression on the far right hand side may be considered constant. This results in effective bending stresses of $75$~Pa and $24$~Pa for the $25\%$ and $50\%$ ice mass fractions respectively. These seemingly low stress values should not be surprising.  By themselves they are not physically significant, rather the physical significance is in the energy dissipation term due to plate bending at subduction whose scale is $\tau_{bend}d_lU_p$. The low values for $\tau_{bend}$ simply mean that relatively little energy is dissipated due to plate bending at subduction. This is mainly due to the dynamic viscosity difference between silicates and water ice. For Earth $\mu_L\sim 10^{23}$~Pa\,s while for a water world $\mu_L\sim 10^{17}$~Pa\,s. This means that if in a water world the bending stress is about $10-100$~Pa then for a rocky world it is $10-100$~MPa. Comparing these results with $\tau_{sp}$ shows that while in a rocky planet the dissipation term due to plate bending at subduction can almost counteract the effect of slab pull, which is a kinetic energy source for the plate, it can hardly do so in a frozen water world. If this were the whole story then \textit{plate tectonics could be said to be more likely in water planets in comparison to rocky planets}. 

One may question the proper choice for the radius of curvature, $R_{curv}$, for which we assign a value an order of magnitude less than the depth scale of the whole icy mantle, $d_{total}$. This is a reasonable estimate for Earth \citep{Crowley2012}, and there is no reason to assume $R_{curv}$ is much smaller in super-Earths. Actually numerical models suggest that the flow system will not allow $R_{curv}$ to decrease too much as the system favors minimizing the dissipation due to plate bending \citep{Capitanio2009}.   

Another complication that may increase $\tau_{bend}$ is the fact that the above theory for $\tau_{bend}$ assumes the lithosphere is a flat sheet, whereas in reality it is a 2D surface on a 3D sphere. Much like a flat slice of pizza is more easily bent at the tip than a folded slice, so too is the subducting lithosphere that has to fold into itself during down-welling \citep{Mahadevan2010}. This effect may be important in increasing the dissipation at subduction but due to its purely geometrical nature it will have the same effect on either a water or a rocky planet. Therefore, if this folding were to increase the dissipation due to plate bending substantially it would more readily stop plate tectonics on Earth, for which the flat sheet assumption gives $\tau_{bend}\approx\tau_{sp}$. Thus this 2D on 3D folding effect is probably not large enough to change the fact that in water planets $\tau_{sp}>>\tau_{bend}$. 

The dissipation in the lithosphere due to friction with the overriding plate is proportional to the fault zone stress, $\tau_F$. The numerical value of the latter is not known. Due to the exploratory nature of this section we vary its value between zero and $150$~MPa. These end values characterize two scenarios, one with a positive $\tau_R$ and the other with a negative $\tau_R$, representing an accumulative tendency of the plate bending the fault zone friction and the slab pull to either restrain or encourage plate motion, respectively.

\begin{deluxetable}{cccccccc}
\tablecolumns{8}
\tablewidth{0pc}
\tablecaption{Plate Tectonic Parameters for the $2$M$_E$ Planet and $25\%$ Ice Mass Fraction}
\tablehead{
\colhead{$\mu_A/\mu_M$} & \colhead{$\tau_R$} & \colhead{$U_p$} & \colhead{$\tau_p$} & \colhead{$d_l$} & \colhead{$V_M$} & \colhead{$F_{plate}$} & \colhead{Type} \\
\colhead{} & \colhead{sign} & \colhead{(cm yr$^{-1}$)} & \colhead{(MPa)} & \colhead{(km)} & \colhead{(cm yr$^{-1}$)} & \colhead{(erg\,cm$^{-2}$ s$^{-1}$)} & \colhead{}}
\startdata
\multirow{2}{*}{$10^{-3}$} & $\tau_R<0$ & 157.8 & 0.47  & 12.5  & 18.9 & 8.9 (13\%)  & I\\
                           & $\tau_R>0$ & 126.2 & 0.051 & 14.0  & 15.8 & 7.9 (11\%)  & I\\
\hline
\multirow{2}{*}{$10^{-4}$} & $\tau_R<0$ & 220.9 & 0.27  & 10.5  & 18.9 & 10.5 (15\%) & I\\
                           & $\tau_R>0$ & 94.7  & 0.11  & 15.6  & 15.8 & 7.1 (10\%)  & I\\
\hline
\multirow{4}{*}{$10^{-5}$} & $\tau_R<0$ & 1.3   & 0.51  & 139   & 12.6 & 0.8 (1\%)   & II\\
                           & $\tau_R<0$ & 410.2 & 0.19  & 7.9   & 18.9 & 14.0 (20\%) & II\\
                           & $\tau_R<0$ & 536.5 & 0.17  & 6.8   & 18.9 & 16.3 (24\%) & I\\
                           & $\tau_R>0$ & 2.2   & 0.24  & 103   & 18.9 & 1.1 (2\%)   & I\\
\hline
\multirow{4}{*}{$10^{-6}$} & $\tau_R<0$ & 0.9   & 0.25  & 163   & 3.2  & 0.7 (1\%)   & II\\
                           & $\tau_R<0$ & 2525& 0.081   & 3.2   & 2.5  & 34.8 (50\%) & II\\
                           & $\tau_R<0$ & 2367& 0.082   & 3.2   & 18.9 & 34.2 (49\%) & II\\
                           & $\tau_R>0$ & 0.9   & 0.085 & 159   & 18.9 & 0.7 (1\%)   & I\\
\enddata
\tablecomments{\footnotesize{Estimated plate tectonic parameters for the $2M_E$ planet assuming $25\%$ ice mass fraction. $\tau_R<0$ stands for $-110$~MPa and  $\tau_R>0$ stands for $40$~MPa. $\mu_A/\mu_M$ is the assumed asthenospheric to lower mantle viscosity ratio. $U_p$ and $\tau_p$ are the plate speed and basal deviatoric shear stress, respectively. $d_l$ is the maximal plate depth and $V_M$ is the maximal vertical velocity in the convection cell. $F_{plate}$ is the surface heat flux as allowed by plate tectonic conduction, the percent value in parenthesis is with respect to silicate core radiogenic heat flux scaled to the surface. $Type$ refers to whether the convection cell is partitioned (II) or not (I).}}
\label{tab:PlateTectonic25}
\end{deluxetable}

\begin{deluxetable}{cccccccc}
\tablecolumns{8}
\tablewidth{0pc}
\tablecaption{Plate Tectonic Parameters for the $2$M$_E$ Planet and $50\%$ Ice Mass Fraction}
\tablehead{
\colhead{$\mu_A/\mu_M$} & \colhead{$\tau_R$} & \colhead{$U_p$} & \colhead{$\tau_p$} & \colhead{$d_l$} & \colhead{$V_M$} & \colhead{$F_{plate}$} & \colhead{Type} \\
\colhead{} & \colhead{sign} & \colhead{(cm yr$^{-1}$)} & \colhead{(MPa)} & \colhead{(km)} & \colhead{(cm yr$^{-1}$)} & \colhead{(erg\,cm$^{-2}$ s$^{-1}$)} & \colhead{}}
\startdata
\multirow{2}{*}{$10^{-3}$} & $\tau_R<0$ & 946.7 & 0.29  & 7.2   & 1.9    & 16.2 (41\%)  & I\\
                           & $\tau_R>0$ & 631.2 & 0.12  & 8.0   & 1.3    & 14.6 (37\%)  & I\\
\hline
\multirow{2}{*}{$10^{-4}$} & $\tau_R<0$ & 473.4 & 0.13  & 9.6   & 31.2   & 12.1 (31\%) & I\\
                           & $\tau_R>0$ & 252.5 & 0.075 & 13.0  & 24.6   & 9.0 (23\%)  & I\\
\hline
\multirow{4}{*}{$10^{-5}$} & $\tau_R<0$ & 2.5   & 0.39  & 132   & 31.2   & 0.9 (2\%)   & II\\
                           & $\tau_R<0$ & 410.2 & 0.12  & 10.5  & 31.2   & 11.1 (28\%) & II\\
                           & $\tau_R<0$ & 946.7 & 0.084 & 6.8   & 31.2   & 17.0 (43\%) & I\\
                           & $\tau_R>0$ & 2.5   & 0.20  & 131   & 16.7   & 0.9 (2\%)   & I\\
\hline
\multirow{4}{*}{$10^{-6}$} & $\tau_R<0$ & 1.4   & 0.20  & 175   & 1041.4 & 0.7 (2\%)   & II\\
                           & $\tau_R<0$ & 3155.8& 0.046 & 3.7   & 3.0    & 31.4 (79\%) & II\\
                           & $\tau_R<0$ & 2934.9& 0.048 & 3.9   & 31.2   & 30.2 (76\%) & II\\
                           & $\tau_R>0$ & 1.5   & 0.12  & 168   & 25.9   & 0.7 (2\%)   & I\\
\enddata
\tablecomments{\footnotesize{Estimated plate tectonic parameters for the $2M_E$ planet assuming $50\%$ ice mass fraction. $\tau_R<0$ stands for $-94$~MPa and  $\tau_R>0$ stands for $56$~MPa. See table \ref{tab:PlateTectonic25} for explanation of column headers.}}
\label{tab:PlateTectonic50}
\end{deluxetable}    

In tables \ref{tab:PlateTectonic25} and \ref{tab:PlateTectonic50} we summarize the different plate tectonic solutions that conserve mass, momentum and energy, for our two planetary composition cases.  We have omitted solutions for which the lithosphere is so thick that it penetrates into the lower mantle. Such solutions have a poorly defined asthenosphere and lower mantle, and require a more elaborate technique in order to correctly account for them. These omitted solutions may actually be physical if one restricts the lithospheric plate thickening with age. For now we avoid this complication and will return to discuss it below. 

The result of this omission is that for the viscosity ratios of $10^{-3}$ and $10^{-4}$, for both ice mass fractions, no sluggish plate ($U_p\sim 1$~cm yr$^{-1}$) solutions exist.  Therefore, for these smaller viscosity contrasts, prescribing a value for $\tau_R$ results in a unique solution for the lithospheric plate that preserves the system mass, momentum and energy. This unique solution represents a fast moving plate ($U_p\sim (1-10)$~m yr$^{-1}$), where the higher end plate speeds are for the larger ice mass fraction. The derived plate velocities are fairly high, between one and two orders of magnitude faster than the fastest moving plates on Earth, the Pacific Superswell group, estimated at $U_p=10$~cm\,yr$^{-1}$ \citep{schubert2001}.

For the case of the larger viscosity contrasts, $10^{-5}$ and $10^{-6}$, when assuming $\tau_R>0$, again a unique solution for the plate motion emerges with $U_p\sim 1$~cm\,yr$^{-1}$. The reason for the unique solution stems from the balance of forces on the lithosphere. A positive $\tau_R$ acts as a sink for the plate kinetic energy and it is the plate gravitational potential, the flow pressure gradient, and the basal shear ($T1$, $T2$ and $T3$ in Eq.\,\ref{EnergyBalLith}, respectively) which keep the plate moving. The ability of the pressure gradient and the basal traction terms to be efficient kinetic energy sources for the lithosphere is reduced due to the low viscosity asthenosphere, resulting in a single solution, that of a somewhat sluggish plate. For this particular scenario we find the pressure gradient term, $T2$, is a more efficient source for kinetic energy for the plate than the plate gravitational potential and basal traction. For the $10^{-5}$ contrast the plate gravitational potential is even less important for contributing kinetic energy than the basal traction term, whereas for the $10^{-6}$ contrast the relative importance switches between the two.  Still, for the case of the larger viscosity contrasts, assuming $\tau_R<0$, yields both sluggish and fast plate dynamics. 

We now wish to test whether the applied stresses on the lithosphere can actually break it apart into plates. If $\sigma$ is the horizontal stress responsible for plate failure and $\sigma_{tens}$ is the tensile strength, then plate tectonics requires:
\begin{equation}\label{PlateFailure}
\sigma\sim\tau_p\frac{L}{d_l}>\sigma_{tens}
\end{equation}  
The tensile strength of ordinary ice at $233$~K is $1.8$~MPa and decreases with increasing temperature \citep{hobbs}. Room temperature measurements give a tensile strength of $0.2$~MPa for SI methane clathrate hydrate \citep{Jung2011}. Using the results tabulated in tables \ref{tab:PlateTectonic25} and \ref{tab:PlateTectonic50} we find the lowest possible values for $\sigma$ are in the range of a few MPa. All of them belong to the sluggish lids. Therefore, according to Eq.\,(\ref{PlateFailure}) the underlying convection is probably capable of breaking the lithosphere into plates.

It is interesting to note that the inclusion of volatiles into water ice and the formation of clathrate hydrates that result in a weak asthenosphere, in turn also yield low values for $\tau_p$. This results in $\sigma$ values for the sluggish plates that exceed the tensile strength by only a factor of a few. This point will be of particular importance if, at some point, a frozen water planet experiences a large influx of silicate dust that may get incorporated into its surface. The tensile strength of an ice-soil mixture will be larger than that for pure water ice \citep{Petrovic2003}. In such a case the underlying convection may face difficulties in breaking apart the plates and maintaining the thick and slow plate solution. Such a reaction of the planetary surface to a mass load of dust is probably temporary until the surface cleans itself. The solutions involving a fast moving thin plate will not be affected by the inclusion of dust in this way since for these plate solutions $\sigma$ is in order of $10$~MPa. Even the fortified dusty plates will probably fail under the action of this higher stress value \citep[see][for ice-soil tensile strengths]{Petrovic2003}.

It is important to note that our system has two different timescales:
\begin{equation}
t_{overturn}\sim\frac{d_{total}}{V_M} \quad ,\quad t_{resurface}\sim\frac{L}{U_p}
\end{equation}
where $t_{overturn}$ represents the mantle overturn time.  This is the time it takes material to convect from the silicate-ice boundary to the near surface region. On the other hand, $t_{resurface}$ is the time scale for plate renewal. 
   
In an isoviscous system, which is symmetrical with respect to its mid-depth level, these two time scales are similar. Therefore the rate for exposure of new material to the planetary surface is often taken to be the mantle overturn rate.  The introduction of an asthenosphere breaks the symmetry with respect to the mid-depth level and creates a difference between these two time scales.  One effect contributing to the difference between the two time scales is the need to conserve mass along the vertical section of the convection cell. This means the flux going to the left and to the right through the vertical cell section must exactly cancel. If the point where the horizontal flow vanishes is dislocated from the mid-depth level, the result of mass conservation is the acceleration of the flow through the shorter segment (lithosphere $+$ asthenosphere).  In addition, the flow from the lower mantle may prefer to return via the low viscosity asthenosphere so as to minimize dissipation, resulting in relatively weak forcing of material through the lithosphere ending with a sluggish plate.    
 
For the common assumption of an isoviscous mantle, the convective velocity in a cell of unit aspect ratio is given by two dimensional boundary layer theory as \citep{schubert2001}:
\begin{equation}
v_{isovis} = 0.233\frac{\alpha_{total}}{d_{total}}Ra^{\frac{2}{3}}
\end{equation}    
where $\alpha_{total}$ is the average ice mantle thermal diffusivity, $d_{total}$ is the length scale of the entire mantle and $Ra$ is the Rayleigh number representative of the entire ice mantle. In order to evaluate the latter, we estimate each parameter in the Rayleigh number definition [see Eq.\,\ref{rayleighnum}] at the average temperature and pressure in the filled ice mantle.
The time scale associated with the isoviscous convective velocity is:
\begin{equation}
t_{isovis}\sim\frac{d_{total}}{v_{isovis}}
\end{equation}
For our $2M_E$ planet we find $t_{isovis}=24$\,Ma and $13$\,Ma assuming $25\%$ and $50\%$ ice mass fractions, respectively.
These two time scales ought be compared with the time scales given in table \ref{tab:TimeScale}, where we give the values for $t_{overturn}$ and  $t_{resurface}$ for all the viscosity ratio cases we have considered.
 
From table \ref{tab:TimeScale} we see that relaxing the assumption of isoviscosity may result in mantle overturn time scale solutions that can be an order of magnitude larger ($\sim 100$~Ma) than that predicted by $t_{isovis}$.  For the case of the lower viscosity contrast ($10^{-3}$) the mantle overturn time for the $50\%$ ice mass fraction is an order of magnitude larger than the equivalent value for the $25\%$ ice mass fraction. This cannot be explained simply by the increase of distance traversed due to the increase of the ice mantle.  This increase contributes about a factor of two.  In addition, there is a dynamic change in the mantle vertical velocity, where it decreases with increasing ice mass fraction. For the viscosity ratios $10^{-6}-10^{-4}$ the mantle overturn time scale seems to have the same order of magnitude for the two ice mass fractions, with one exception for a sluggish lid and a viscosity ratio of $10^{-6}$.

\begin{deluxetable}{cccccc}
\tablecolumns{6}
\tablewidth{0pc}
\tablecaption{Overturn and Resurface Time Scales}
\tablehead{
\colhead{} & \colhead{} & \multicolumn{2}{c}{$25\%$ Ice Mass Fraction} & \multicolumn{2}{c}{$50\%$ Ice Mass Fraction} \\
\cline{3-4} \cline{5-6} \\
\colhead{$\mu_A/\mu_M$} & \colhead{$\tau_R$}   & \colhead{$t_{overturn}$}    & \colhead{$t_{resurface}$} &
\colhead{$t_{overturn}$ }    & \colhead{$t_{resurface}$} \\   
\colhead{} & \colhead{sign}& \colhead{(Ma)}& \colhead{(Ma)}& \colhead{(Ma)}& \colhead{(Ma)}}
\startdata
\multirow{2}{*}{$10^{-3}$} & $\tau_R<0$ & 10 & 2.4  & 182 & 0.7  \\
                           & $\tau_R>0$ & 12 & 3.0  & 267 & 1.1  \\
\hline
\multirow{2}{*}{$10^{-4}$} & $\tau_R<0$ & 10 & 1.7  & 11  & 1.5  \\
                           & $\tau_R>0$ & 12 & 4.0  & 14  & 2.6  \\
\hline 
\multirow{4}{*}{$10^{-5}$} & $\tau_R<0$ & 15 & 292  & 11  & 277  \\
                           & $\tau_R<0$ & 10 & 0.9  & 11  & 1.7  \\
                           & $\tau_R<0$ & 10 & 0.7  & 11  & 0.7  \\
                           & $\tau_R>0$ & 10 & 172  & 21  & 277  \\
\hline
\multirow{4}{*}{$10^{-6}$} & $\tau_R<0$ & 59 & 422  & 0.3 & 496  \\
                           & $\tau_R<0$ & 76 & 0.2  & 116 & 0.2  \\
                           & $\tau_R<0$ & 10 & 0.2  & 11  & 0.2  \\
                           & $\tau_R>0$ & 10 & 422  & 13  & 463  \\
\enddata
\tablecomments{\footnotesize{Mantle overturn, $t_{overturn}$, and plate resurfacing, $t_{resurface}$, time scales for the $2M_E$ planet assuming two different ice mass fractions. The different asthenospheric to lower mantle viscosity ratios correspond to all the cases tested for and whose results are given in tables \ref{tab:PlateTectonic25} and \ref{tab:PlateTectonic50}.}}
\label{tab:TimeScale}
\end{deluxetable}

As is evident from the tabulated data, $t_{overturn}$ and $t_{resurface}$  may be quite different. For some of the cases we investigated the difference between these two time scales may span two to three orders of magnitude.  \textbf{For the estimation of methane outgassing the importance lies in the rate of plate resurfacing, determined by $t_{resurface}$. We find that, with little dependency on the ice mass fraction, the resurfacing time associated with the fast plate solutions is in order of $1$~Ma for viscosity contrasts up to $10^{-5}$, and on order of $0.1$~Ma for a viscosity contrast of $10^{-6}$. For the sluggish plates the resurfacing time is in order of $100$~Ma, independent of the ice mass fraction to a good approximation as well}.   

\begin{figure}[ht]
\centering
\includegraphics[scale=0.5]{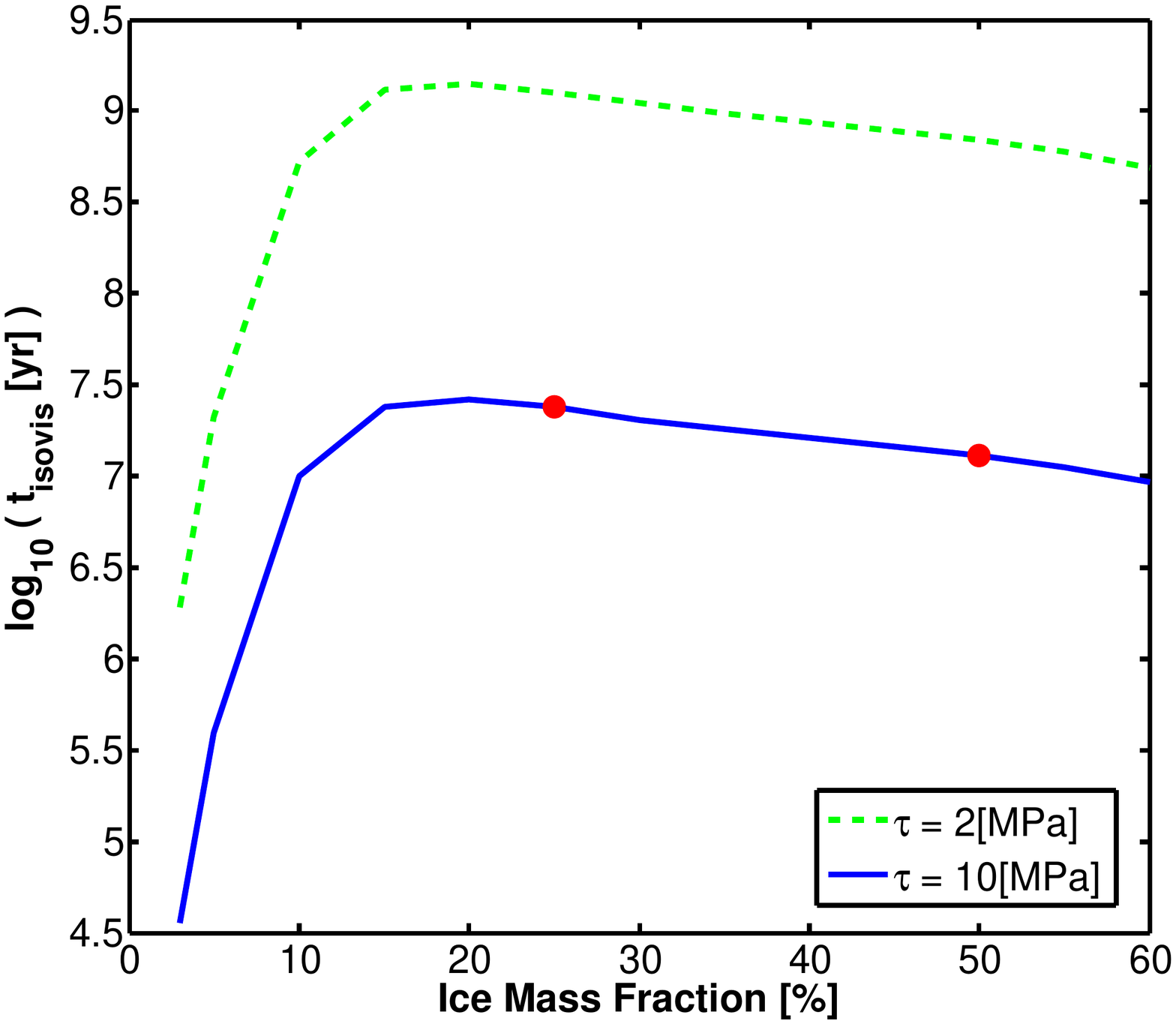}
\caption{\footnotesize{Isoviscous overturn time scale for the filled-ice mantle of the $2M_E$ planet for various ice mass fractions. Dashed (green) curve is for the assumption of an average mantle stress of $2$~MPa. Blue (solid) curve is for the assumption of an average mantle stress of $10$~MPa. The red circles highlight the cases of $25\%$ and $50\%$ ice mass fractions.}}
\label{fig:f12}
\end{figure}

Even though $t_{isovis}$ is strictly applicable only for isoviscous systems it still is interesting to study its behavior. In Fig.\,\ref{fig:f12} we plot $t_{isovis}$ for the $2M_E$ planet and for various ice mass fractions. This time scale is very sensitive to the parameters chosen for the viscosity and thus to the estimated shear stress second invariant ($\tau$). 
To obtain some insight into how sensitive $t_{isovis}$ is to the choice for $\tau$ we solve once for our estimated value of $10$~MPa and once for $2$~MPa. The iso-viscosity is approximated as the average over a more realistic viscosity which depends on the mantle thermal profile.  For ever smaller ice mass fractions (a shallower icy mantle) the time scale ought diminish to zero, as is shown in the figure.  

For small ice mass fractions the ice mantle is relatively cold and the viscosity activation volume is relatively high.  Therefore, the viscosity increases rapidly with pressure increasing the isoviscous overturn time scale. A maximum is reached at approximately $20\%$ ice mass fraction, even though the ice mantle continues to thicken with increasing ice fraction. Further increasing the ice mass fraction results at a higher average mantle temperature and a lower average activation volume (it decreases with pressure, see Eq.\,\ref{VolActFI}), helping to decrease the average viscosity and make convection more vigorous.  This reduces the isoviscous overturn time scale. Between the two scenarios for $\tau$ the isoviscous time scale can change by almost two orders of magnitude. This large change manifests the sensitivity of the dependence of $t_{isovis}$ on $\tau$. In other words the area between the two curves would have represented the area of permissible solutions if $\tau$ could obtain any value between $2$\,MPa and $10$\,MPa. Nevertheless, our results for the stress in the convection cell are more consistent with $\tau\approx 10$\,MPa. Another point of interest is that \textit{for ice mass fractions larger than approximately $20\%$ the isoviscous time scale dependency on the ice mass fraction becomes relatively weak}.

We now turn to the question of the surface heat flux.  How does the thermal flux conducted through the lithospheric plates compare with the assumed radiogenic budgets? The heat flux due to radioactive decay, $F_{s,planet}$, was scaled to the surface of our planets using radiogenic data from Earth (see explanation to Eq.\,\ref{surfheatflux}), yielding values of $69.2$~erg\,cm$^{-2}$\,s$^{-1}$ and  $39.5$~erg\,cm$^{-2}$\,s$^{-1}$ for the $25\%$ and $50\%$ ice mass fractions, respectively. In tables \ref{tab:PlateTectonic25} and \ref{tab:PlateTectonic50} we give the surface conductive heat flux, $F_{plate}$, associated with each plate dynamic scenario, followed by a percentage in parenthesis representing the ratio of $F_{plate}$ to the estimated radiogenic heat flux.  From the tables it is clear that the conductive heat flux through the different plate scenarios does not account for all the radioactive heat released from the silicate core, although, the thin plate solutions give values of $F_{plate}$ that may account for a large fraction of the radiogenic budget.  This is especially true for increasing viscosity contrasts. We wish to elaborate to some extent on this energetic discrepancy. 

In the plate tectonic theory of \cite{Crowley2012} the radiogenic budget is not taken into consideration. The conductive heat flux through the plates is simply assumed to obey:
\begin{equation}\label{halfspacecooling}
F_{plate}=\frac{\kappa_l \Delta T_l}{d_l}\approx\kappa_l\Delta T_l\sqrt{\frac{U_p}{\alpha_l L}}
\end{equation}  
where $\kappa_l$ is the lithospheric thermal conductivity and $\Delta T_l$ is the temperature difference across the lithosphere. The approximation on the far right hand side assumes the plate thickens as the square root of its age (the half space cooling model, $d_l\sim\sqrt{\alpha_lt}$) and also that the age may be estimated by its largest value, i.e. $t\sim L/U_p$. This last assumption exaggerates the thickness of the plate, effectively reducing the surface heat flux through conduction. In order to understand this last point one has to remember that under the thin crust exists a layer confined to the melting curve (the DBL), beyond which convection is instated resulting in an adiabatic profile. Therefore, the ability of $\Delta T_l$ to grow with the thickening plate is somewhat restrained and the dependency of $F_{plate}$ on the inverse of the lithospheric depth ($d_l$) has nothing to counteract it. 

To correct for the assumption that the plate age may be approximated by its oldest age, the surface heat fluxes in tables \ref{tab:PlateTectonic25} and \ref{tab:PlateTectonic50} were calculated using a model that considers the fact the plate was thinner when it was formed, allowing for higher surface heat fluxes \citep{schubert2001}. The average conductive heat flux will in this case be:
\begin{equation}\label{heatfluxave}
F_{plate} = \frac{1}{L-L_i}\int_{L_i}^{L}\frac{\kappa_l\Delta T_l}{d_{asy}}\left[1+2\sum_{n=1}^\infty e^{-\frac{\alpha_ln^2\pi^2x}{d_{asy}^2U_p}} \right]dx
\end{equation} 
where $d_{asy}$ is the asymptotic plate thickness at old age which we take to be $\sqrt{\alpha_lL/U_p}$. The model assumes a vanishing plate thickness at the ridge, so in order to avoid a divergence of the sum appearing in the integrand we start the integration at a small distance from the ridge ($L_i$), which is much smaller than the actual plate length, $L$. We find the actual divergence begins to play a substantial role only at very small distances from the ridge ($x<<1$~m) and so our choice for $L_i$ ($100$~m) both accounts for a thinner plate at younger ages and avoids an artificial divergence of the flux. 

There remains an important caveat to the corrected surface heat flux given by Eq.\,($\ref{heatfluxave}$).  We have basically allowed the lithospheric plates to freely thicken as the square root of their age without restricting their total thickness in any way. Assuming this continuous plate thickening with age for Earth, would also give us an energetic discrepancy. Surface heat flux data along Earth's oceanic lithosphere clearly shows that upon reaching an age of $60-100$~Ma the plate ceases to thicken, causing the surface heat flux curve to flatten with age \citep{JaupartMareschal}. One may argue that the Earth has a way of eliminating any energetic discrepancy by keeping its lithosphere thinner than predicted by the half space cooling model. The two widely accepted physical mechanisms that keep the lithosphere from continuously thickening are: small scale convection and hot spot formation \citep[see][chapter $6$, for an in depth discussion]{JaupartMareschal}. 

Hot spots are generally believed to originate from instabilities in Earth's lower thermal boundary layer.  When these instabilities reach a critical volume they may detach from the boundary layer and upwell. Thermal boundary layer instabilities are repeatedly generated when the convection is time dependent. Time dependency, in turn, is intrinsic to vigorous convection, i.e. supercritical Rayleigh numbers. For Earth's mantle the Rayleigh number may reach a value as high as $5\times 10^7$, which is highly supercritical \citep[see discussion in][on mantle plume formation]{schubert2001}. Our $2M_E$ planet may have an icy mantle Rayleigh numbers as high as $7\times 10^7$ and $10^9$, for the $25\%$ and $50\%$ ice mass fractions, respectively. Dynamically, therefore, both cases studied in this section may also have vigorous time-dependent convection capable of supporting mantle plumes. Thus the mechanisms responsible for keeping Earth's lithosphere thin may also be at work in our planets. Quantifying these physical mechanisms requires a rigorous derivation for the behavior of convection. Even for Earth quantifying these two mechanisms on theoretical grounds is a formidable task and the more common approach is to compare the surface heat flux predicted by the half space cooling model with the data collected from the field. In our case no field data exists and therefore approximating $d_{asy}$ with the thickness at its oldest age is reasonable. It is important, however, to note that the consequence of this assumption is that \textit{the percentage values quoted in tables $\ref{tab:PlateTectonic25}$ and $\ref{tab:PlateTectonic50}$ should be considered minimum values}. 

Another mechanism that may help cool the body is partial melting. Beneath spreading centers, clathrate hydrates carrying volatiles following an adiabatic path may cross their thermodynamic stability field, to produce liquid water and methane gas. It would be interesting to estimate how much melt is needed to account for the difference between the radiogenic energy budget and the conductive-cooling ability of the plates.  The global energy rate difference is:
\begin{equation}
Q_{deficiency}=\left(F_{s,planet}-F_{plate}\right)4\pi R^2_p
\end{equation} 
where $R_p$ is the planetary radius. The total clathrate hydrate mass that crosses the clathrate hydrate stability field under a ridge each second is:
\begin{equation}
\dot{M}=\rho_{cl}WL_{GR}V_{ascent}
\end{equation}
where $\rho_{cl}$ is the clathrate hydrate mass density, $W$ is the spreading center width at the depth of the solidus, $L_{GR}$ is the global ridge length and $V_{ascent}$ is the speed of mass ascent. If $\Delta H$ is the energy required to melt a mass of clathrate hydrates and the actual fraction that melts is $X_{melt}$, then we have:
\begin{equation}\label{meltfraction}
X_{melt}=\frac{\left(F_{s,planet}-F_{plate}\right)4\pi R^2_p}{\Delta H\rho_{cl}WL_{GR}V_{ascent} }
\end{equation}  
We will assume the ascent velocity is approximately the plate velocity and therefore from mass conservation $W$ is approximately the lithospheric depth.  The global ridge length equals the number of plates times the contribution from each plate to the ridge. Assuming each plate is a square of size $L^2$ that contributes a $2L$ ridge length, yields:
\begin{equation}
L_{GR}=\tilde{s}\frac{4\pi R^2_p}{L^2}2L
\end{equation}

For Earth the last calculation results in $3.4\times 10^5\tilde{s}$~km.  Since the mid-ocean ridge length is $6\times 10^4$~km \citep{Langmuir2012} then $\tilde{s}$ is about $0.18$.  Incorporating the last approximations into Eq.\,($\ref{meltfraction}$) gives:
\begin{equation}
X_{melt}=\frac{\left(F_{s,planet}-F_{plate}\right)}{4.64\Delta H\rho_{cl}\tilde{s}}\sqrt{\frac{L}{\alpha_lU_p}}
\end{equation}
where we have estimated the lithospheric depth scale using the half space cooling model.  In table \ref{tab:MeltFraction} we give $X_{melt}$ for all the tectonic scenarios solved for in tables \ref{tab:PlateTectonic25} and \ref{tab:PlateTectonic50}. Two interesting conclusions may be derived from the table. 

First, all the cases of fast moving plates require only partial melting under the ridge in order to account for the planetary radiogenic budget. The sluggish plates on the other hand require more than complete melting ($>100\%$). This is true for both ice mass fractions. The fast moving plate tectonic modes are therefore very efficient at losing heat as opposed to the sluggish plate modes. This suggests that if partial melting is higher than estimated here, a planet in a fast plate tectonic mode may over-cool and evolve into a sluggish plate mode. The latter will over-heat and thus melt part of its thicker lithosphere until it thins to the value required by the fast lithospheric mode which will again result in over-cooling. This tectonic cycling may continue till the planet loses a substantial fraction of its radiogenic and accretional energy, perhaps ending as a stagnant lid.

The second conclusion that may be derived from table \ref{tab:MeltFraction} is that for a given planetary mass, the lower the ice mass fraction the more melt is required to cool the body. This has consequences for the way methane is released into the atmosphere and is addressed in the next section.   

\begin{deluxetable}{cccc}
\tablecolumns{4}
\tablewidth{0pc}
\tablecaption{Melt Fractions}
\tablehead{
\colhead{} & \colhead{} & \colhead{$25\%$ Ice Mass Fraction} & \colhead{$50\%$ Ice Mass Fraction} \\
%\cline{3} \cline{4} \\
\colhead{$\mu_A/\mu_M$} & \colhead{$\tau_R$} & \colhead{$X_{melt}$} & \colhead{$X_{melt}$}     \\   
\colhead{} & \colhead{sign}& \colhead{($\%$)}& \colhead{($\%$)} }
\startdata
\multirow{2}{*}{$10^{-3}$} & $\tau_R<0$ & 25  & 5       \\
                           & $\tau_R>0$ & 29  & 7       \\
\hline
\multirow{2}{*}{$10^{-4}$} & $\tau_R<0$ & 21  & 9       \\
                           & $\tau_R>0$ & 34  & 14      \\
\hline 
\multirow{4}{*}{$10^{-5}$} & $\tau_R<0$ & 319 & 172     \\
                           & $\tau_R<0$ & 14  & 10      \\
                           & $\tau_R<0$ & 12  & 5       \\
                           & $\tau_R>0$ & 243 & 172     \\
\hline
\multirow{4}{*}{$10^{-6}$} & $\tau_R<0$ & 383 & 231     \\
                           & $\tau_R<0$ & 4   & 1       \\
                           & $\tau_R<0$ & 4   & 1       \\
                           & $\tau_R>0$ & 383 & 223     \\
\enddata
\tablecomments{\footnotesize{Percent of melt required beneath spreading centres to compensate for the difference between the radiogenic budget and the ability of the lithospheric plates to cool conductively.}}
\label{tab:MeltFraction}
\end{deluxetable}

\begin{figure}[ht]
\centering
\mbox{\subfigure{\includegraphics[width=7cm]{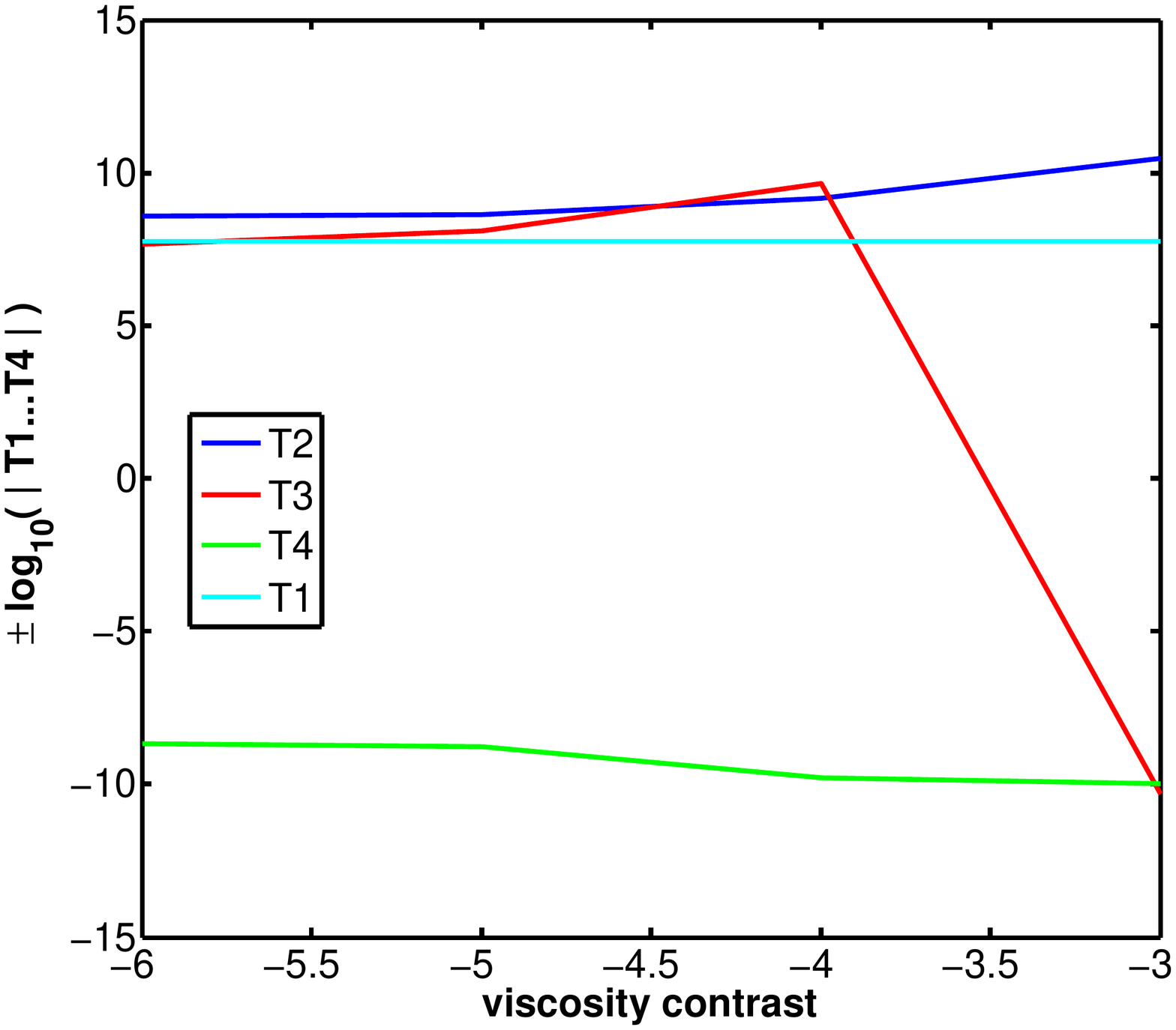}}\quad \subfigure{\includegraphics[width=7cm]{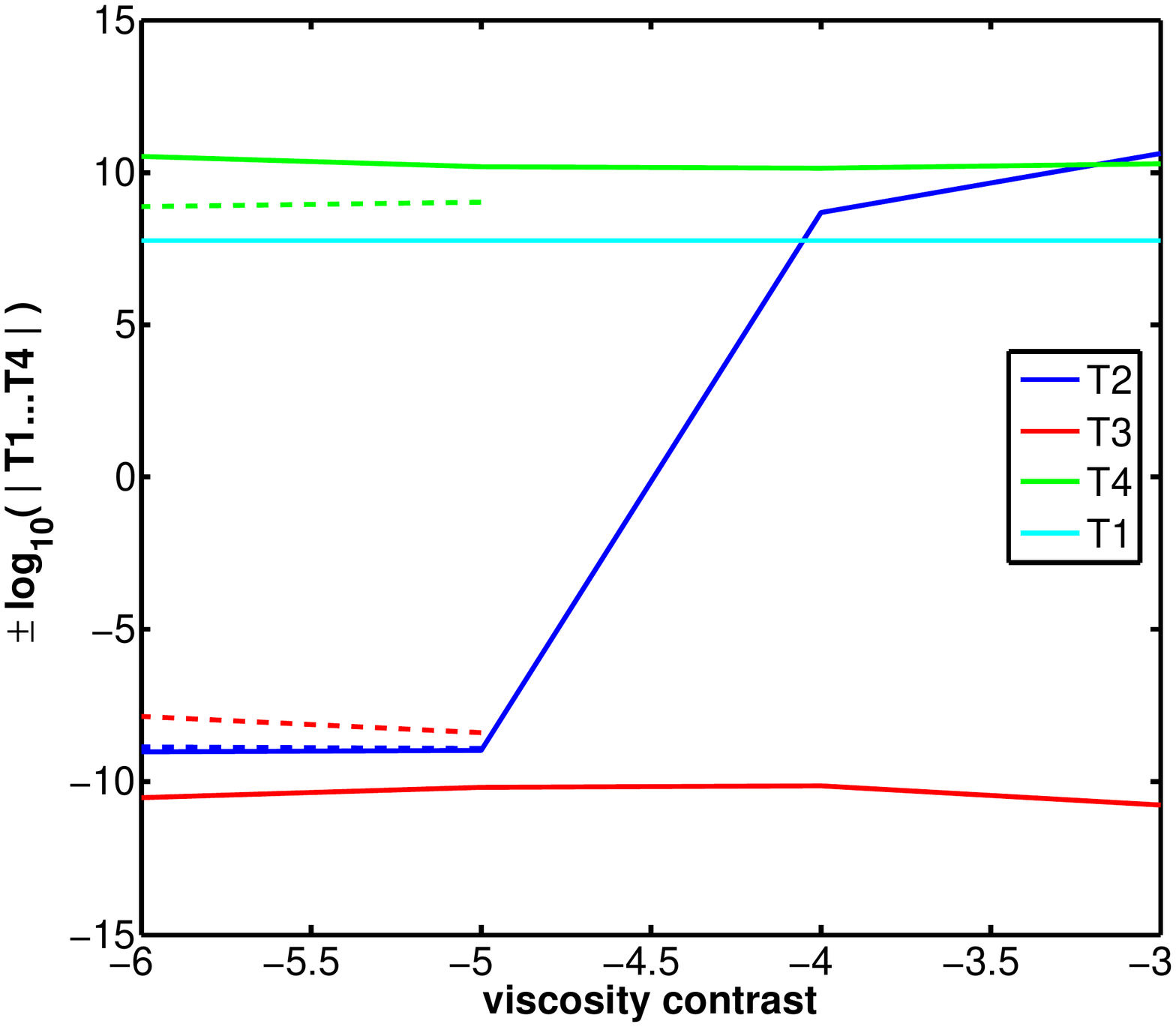}}}
\caption{\footnotesize{Tectonic forces as a function of viscosity contrast for a $2M_E$ planet with $50\%$ ice mass fraction assuming a positive net resistive stress on the plate ($\tau_R>0$) (left panel) and for a negative net resistive stress (right panel).  Shown are $T1$, the plate gravitational potential (cyan curve), $T2$ the effect of the horizontal pressure gradient (blue curve), $T3$ the traction acting on the base of the plate (red curve) and $T4$ the net resistive stress (green curve). }}
\label{fig:f13}
\end{figure}

In Fig.\,\ref{fig:f13} we plot the numerical values of $T1-T4$ for the $50\%$ ice mass fraction case, assuming $\tau_R>0$. The absolute value of the curves represent the relative importance of the different mechanisms driving the plate. A positive value means the mechanism contributes kinetic energy to the plate therefore supporting it, while a negative value means the mechanism acts as a sink for the plate kinetic energy and thus suppresses its motion. Clearly the choice of a positive net resistive stress means plate bending and fault friction are more important than slab pull and the general effect of the net resistive stress, $T4$, is to suppress the motion of the plate, so that it is negative in the figure. The lithospheric gravitational potential, $T1$, always supports the motion of the plate, but the horizontal pressure gradient is a stronger contributor of kinetic energy to the lithospheric plate. The basal traction, interestingly, shifts from a plate inhibitor to plate motion contributor upon increasing the viscosity contrast. At a viscosity contrast of $10^{-4}$ it is an even stronger contributor to plate motion than the horizontal pressure gradient.  However, its role as a source of kinetic energy for the plate diminishes with increasing viscosity contrast until eventually the asthenospheric viscosity is so low that traction contributes to plate motion even less than the plate gravitational potential.     

Also shown is the dependence of $T1-T4$ on the viscosity contrast between the asthenosphere and the lower mantle for negative net resistive stress ($\tau_R<0$). A negative resistive stress means the slab pull effect is more dominant than the dissipation due to plate bending at subduction and fault zone friction. Therefore the net resistive stress is a kinetic energy source for the plate. At the higher viscosity contrasts there are two fast plate solutions and we represent them both by the average between them (solid curves). This averaging is permissible due to the relative closeness of these two solutions. The sluggish plate solution is represented using the dashed curves. The plate gravitational potential always contributes to plate motion, but plays only a minor role in comparison to the other driving mechanisms. 

The basal traction now plays the role of a plate motion inhibitor for all the viscosity contrasts. For the small viscosity contrasts, which in our case translate to high asthenospheric viscosity, basal traction is a very efficient mechanism for dissipating plate energy. Therefore even the effect of slab pull cannot, by itself, counteract the force of traction and keep the plate at uniform motion. The pressure gradient has to adjust and become a source of kinetic energy to help maintain plate motion. For the viscosity contrast of $10^{-3}$ the contribution of the pressure gradient to plate motion is even larger than the contribution of slab pull. Increasing the viscosity contrast to $10^{-4}$ represents a decrease in asthenospheric viscosity rendering basal traction less efficient.  In this case the pressure gradient need not be so large and may decrease in importance relative to slab pull. Further increasing the viscosity contrast (lower asthenospheric viscosity) diminishes the ability of basal traction to restrain the plate motion and the pressure gradient becomes a plate motion suppressor acting together with it to restrain the pulling slab from accelerating the plate. For the sluggish plate solution the pressure gradient has a larger importance in plate motion suppression than the basal traction.  

\begin{figure}[ht]
\centering
\mbox{\subfigure{\includegraphics[width=7cm]{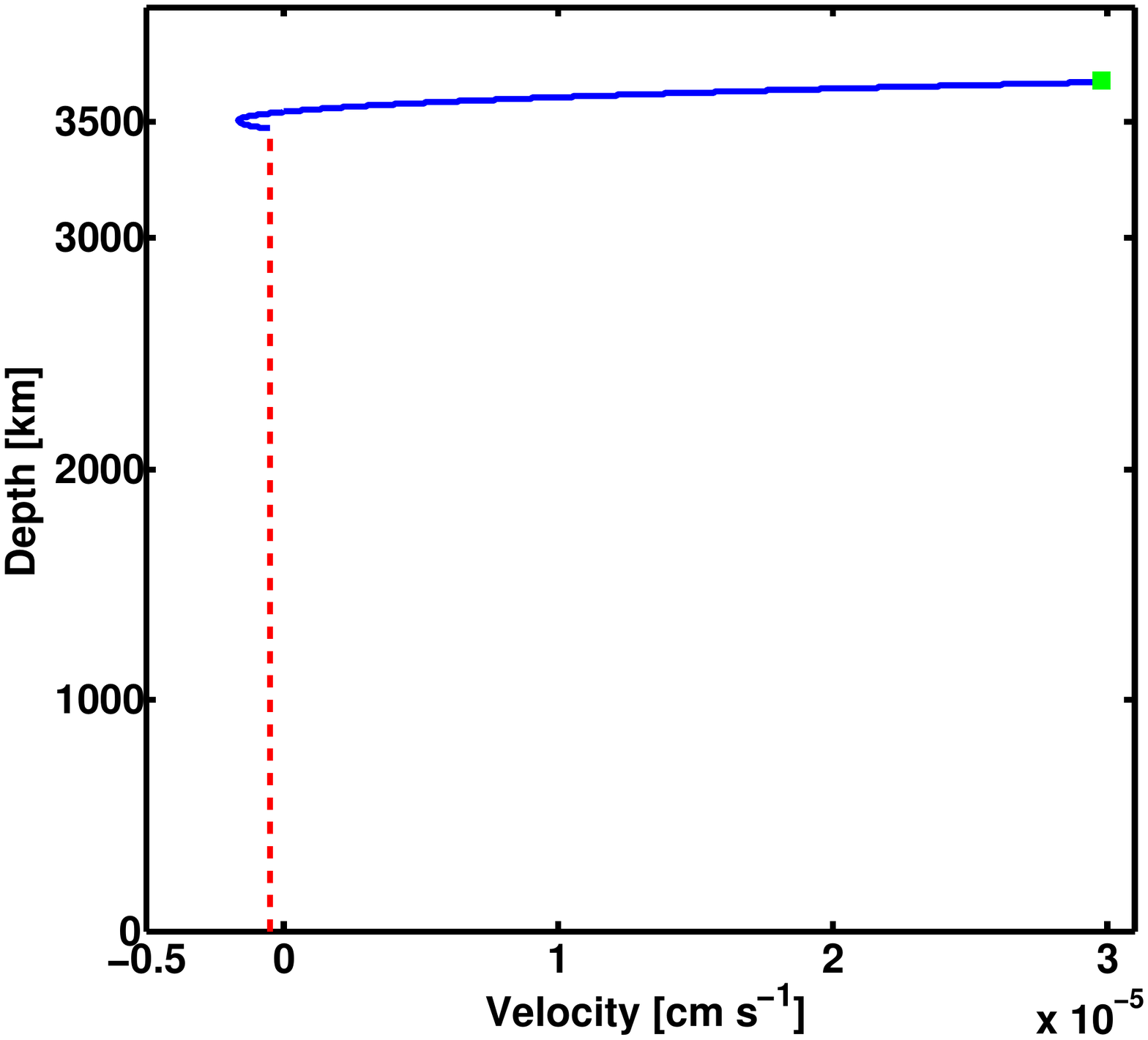}}\quad \subfigure{\includegraphics[width=7cm]{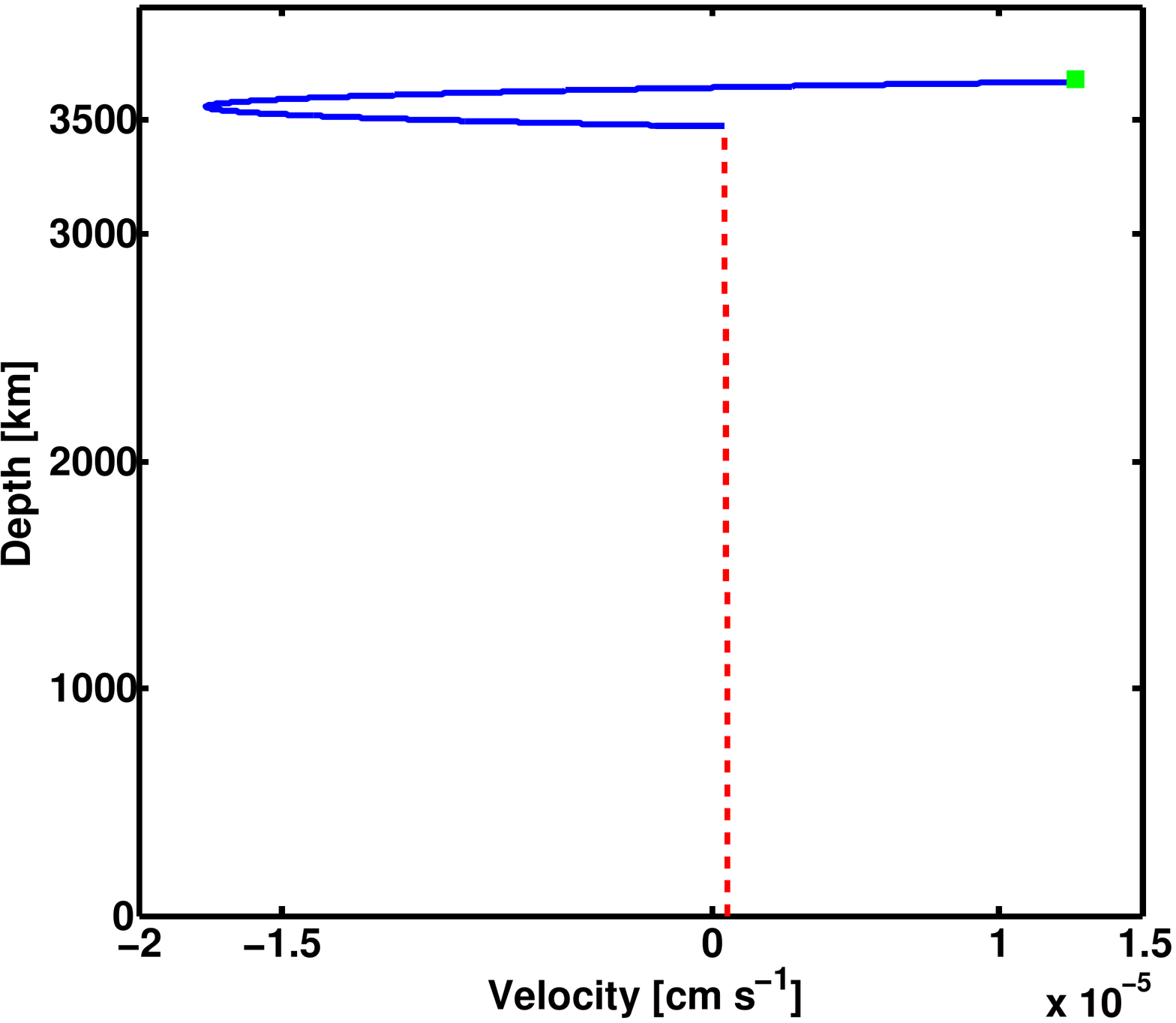}}}
\caption{\footnotesize{Depth profile of the horizontal velocity in a typical type I convection cell (left panel) and type II convection cell (right panel), for a fast moving plate and a viscosity contrast of $10^{-5}$. The green square represents the lithospheric plate, the solid blue curve represents the asthenosphere and the dashed red curve represents the lower mantle. A depth of zero corresponds to the silicate-ice mantle boundary.}}
\label{fig:f15}
\end{figure}

With these insights into the behavior of the different forces acting on the lithospheric plate we can explain the final entry in tables \ref{tab:PlateTectonic25} and \ref{tab:PlateTectonic50}, the characterization of the two types of convection cells. When the asthenospheric viscosity is low and the action of the slab pull is dominant ($\tau_R<0$) a strong asthenospheric flow opposite to the plate direction of motion is required to keep the plate in a uniform motion. The lower the asthenospheric viscosity, the stronger the backward asthenospheric flow must be. In addition to this behavior the model of \cite{Crowley2012} also requires the conservation of mass through the vertical cross section of the convection cell. The requirement of mass conservation coupled with the strong backward flow in the asthenosphere may force the flow in the lower mantle to change direction resulting in two depth levels where the horizontal velocity vanishes. This we have denoted as a type II cell, for which a typical horizontal velocity profile with depth is given in the right panel of Fig.\,\ref{fig:f15}.
 
A type I convection cell is more "regular" in its behavior having a single depth level where the horizontal velocity vanishes (see for example Fig.\,\ref{fig:f15}). Only when the asthenospheric viscosity is very low would a type II cell become unavoidable for $\tau_R<0$. For example, even a viscosity contrast of $10^{-5}$ for $\tau_R<0$ may result in either a type I or type II cell. These conditions were actually adopted in producing Fig.\,\ref{fig:f15}. 

\begin{figure}[ht]
\centering
\includegraphics[scale=0.5]{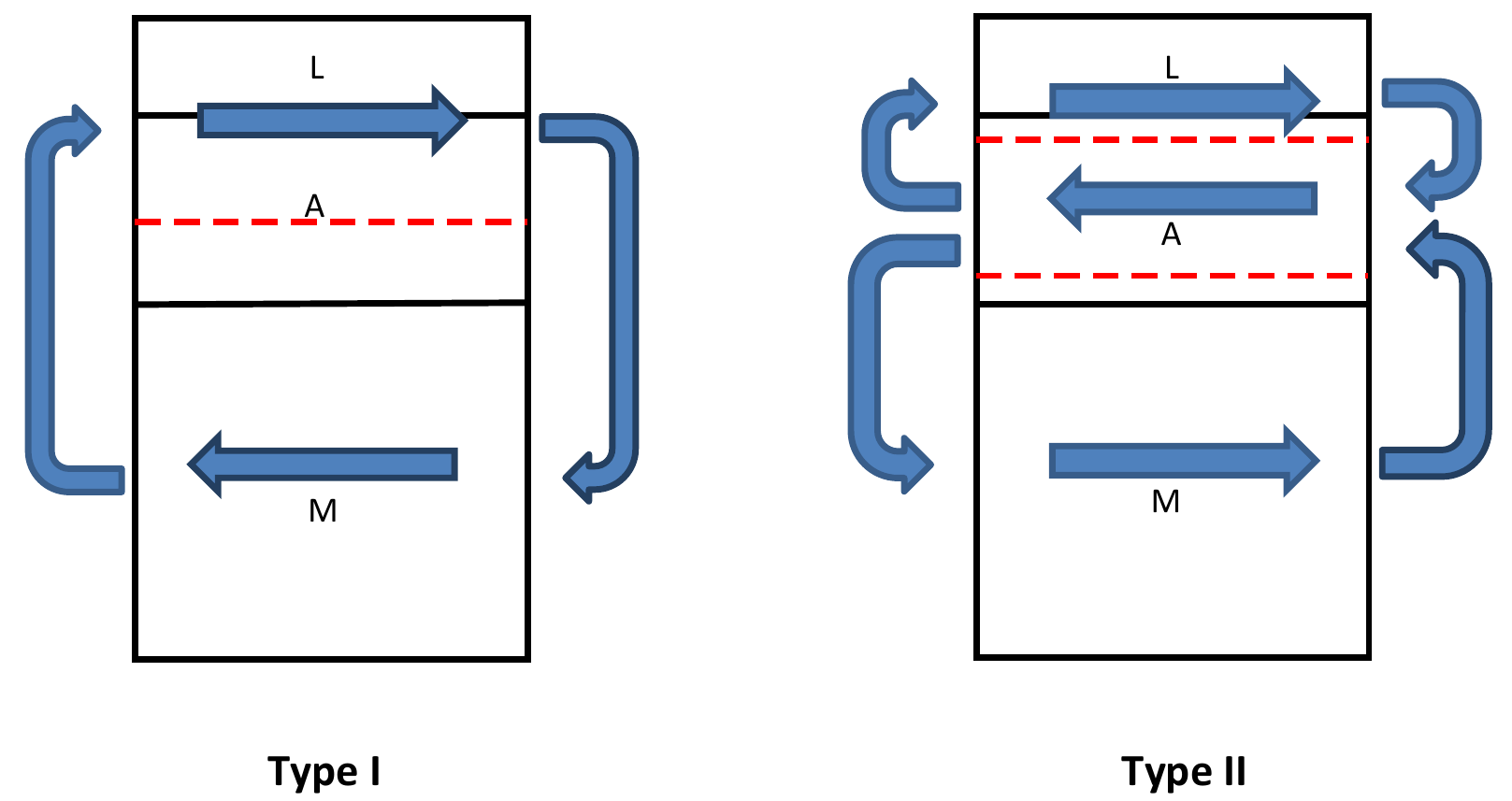}
\caption{\footnotesize{Schematic diagram of flow in a type I and type II cell. The dashed (red) curve represents the depth level where the horizontal velocity vanishes.}}
\label{fig:f17}
\end{figure}

A schematic diagram describing the consequences on the flow of a type I or type II behavior is shown in Fig.\,\ref{fig:f17}. From the diagram it seems a type II flow may promote convection cell partitioning and have partial asthenospheric downwelling under a lithospheric ridge, where the plates are spreading.  Although the parameter space we have solved for is not the parameter space occupied by planet Earth it still is interesting to point out that asthenospheric downwelling beneath a lithospheric ridge occurs anomalously on Earth in what is known as the Australia-Antarctic Discordance \citep{Stern2007}, though probably for different reasons.  

The idea that a low viscosity layer may contribute to convection cell partitioning was numerically tested by several authors \citep[e.g.][]{Cserepes1997}. Although a phase change boundary, by itself, will promote cell partitioning, the boundary layer formed will cause a strong increase in temperature.  This will result in a sharp decrease in local viscosity. \cite{Cserepes1997} suggest that in this low viscosity zone strong horizontal flow may develop which in turn will spread thermally unstable mass parcels, and hinder the formation of an avalanche of material through the phase change boundary. This will tend to increase the lifetime of the partitioning.

In the situation we have solved for, the low viscosity zone does not lie under a distinctive phase change boundary and therefore direct conclusions from existing numerical investigations should not be drawn.  Rather, a stability analysis for our particular case is in order. Such an analysis is beyond the scope of this work, and may actually require modification to the internal workings of the tectonic model we have adopted here. Nonetheless a simple argument suggests that a type II cell is less stable than the type I configuration: Partitioning of the cell would create a conductive thermal boundary layer at the partition, raising the temperatures below. This will reduce the lower mantle viscosity and thus the viscosity contrast. Lowering the viscosity contrast tends to establish a type I convection flow. \textit{A mechanism may therefore exist, that can both limit viscosity contrasts and convection cell partitioning}.    

As a final remark we wish to emphasize that the purpose of this exploratory study is to characterize the possible tectonic behavior for our studied planets. We do not suggest that the mechanism of plate tectonics is eternal.  Rather it has been shown that the geodynamic mechanism responsible for heat loss is subject to evolution \citep{Sleep2000}. A quantitative description of this evolution for the case of water planets is beyond the scope of this work.

\section{CH$_4$ OUTGASSING FLUX}

What impact do the tectonic modes explored above have on the flux of volatiles into the atmosphere?  In this section we give a simple illustration, using a binary ice composition, of the outgassing process for a particular scenario.  From table \ref{tab:MeltFraction} we see that for the fast moving plates the ascending ice experiences relatively little melting. For the $50\%$ ice mass fraction the partial melting is only a few percent. We can idealize this scenario by assuming that partial melting is negligible and so methane predominantly remains enclathrated in the ice as it ascends to form the new lithosphere.  
 
At the ridge the newly formed upper part of the crust is exposed to the conditions at the planetary surface. In other words, masses of ice that were once pressurized deeper in the mantle are depressurized when they are transformed to crustal material. The depressurization may trigger methane hydrate dissociation in the upper crust releasing its methane and transforming to hexagonal ice. This dissociation front will sweep the newly exposed upper crust down to a depth where the pressure provided by the overlying hexagonal ice column combined with the atmospheric pressure become sufficient for stabilizing methane hydrates. Since the newly formed crust originates from material once highly pressurized it is not likely to be substantially porous. 

Several experimental studies have tried to quantify the flux of methane from a destabilized column of methane hydrate
\citep[][]{Takeya2001,Sun2006,Gainey2012}, though for conditions not entirely consistent with those for water planets. Therefore, we develop physical arguments to help us extrapolate the experimental data.  The experiments of \cite{Takeya2001} suggest the kinetics of methane hydrate dissociation proceeds in two steps. In the first step the bulk of the hydrate mass is composed of hydrate ice grains which, when destabilized, rapidly release methane through their grain boundaries while their outer water shell transforms from empty hydrate into ice Ih. In the second stage any further dissociation of the hydrate grain becomes dependent on the ability of methane to diffuse through the grain's exterior ice Ih shell. This second step makes the process of hydrate dissociation diffusion limited and raises the issue of kinetic hindrance.

\cite{Takeya2001} found, for grain diameters of $70$~$\mu$m, a diffusion coefficient, $\tilde{D}$, of methane through ice Ih of $2.2\times 10^{-7}$~cm$^2$\,s$^{-1}$ at $189$\,K and $9.6\times 10^{-8}$~cm$^2$\,s$^{-1}$ at $168$\,K. These are very high diffusion coefficients in comparison to measurements of the diffusion of $N_2$ and $O_2$ through ice Ih for which the diffusion coefficient is found to be in order of $10^{-15}$~cm$^2$\,s$^{-1}$ \citep{Satoh1996}. The authors explain this discrepancy by the hypothesis that as the clathrate hydrate water lattice transforms to ice Ih, voids are introduced into the resulting lattice because the clathrate hydrate water lattice is less dense than that of ice Ih. This mechanism may be regarded as a self pore formation mechanism aiding the diffusion of methane.  We fit these laboratory results to an Arrhenius type equation which is a natural form for diffusion by thermal fluctuations, giving:
\begin{equation}
\tilde{D}=1.67\times 10^{-4}e^{-1254/T}\quad {\rm cm^2\,s^{-1}}
\end{equation} 
Here $T$ is in kelvins.

As explained in subsection $2.4$ the destabilized crustal methane hydrate column may reach a maximum depth in order of $\ell \sim 100$~m. Using the last relation for the diffusion coefficient we may estimate the time it takes methane to diffuse through the layer and into the atmosphere as $\ell^2/\tilde{D}$ which gives $10$~Ma and $3$~Ma for temperatures of $200$~K and $250$~K respectively.  Interestingly, even though the diffusion coefficients adopted here are very large and take into account the self pore formation mechanism, \textit{the resulting time scales are on the order of the resurfacing time scale for the fast moving plates}. It is therefore inaccurate to assume the entire column of destabilized methane hydrate would simply lose all its methane to the atmosphere. Rather, the problem of hydrate dissociation and methane diffusion under a concentration gradient must be more delicately quantified. The purpose of the following calculation is to estimate how much methane is lost to the atmosphere from a column of destabilized methane hydrate. 

At first all the methane molecules in the ice column are stored in methane hydrate grains which start to dissociate when destabilized. The concentration of methane in the column, $c$, is higher than its concentration in the overlying atmosphere and the density gradient drives the diffusion according to Fick's first law:
\begin{equation}\label{FickLaw1}
j=-\tilde{D}\frac{\partial c}{\partial z}
\end{equation}
where $j$ is the flux of methane and $z$ is the vertical position along the destabilized column.  In subsection $2.2$ we discussed possible grain diameters assuming grains have a geological time to ripen. Considering a grain size of $300$~$\mu$m, after a time of order of $(300~\mu{\rm m})^2/\tilde{D}\sim 1$~hr the diffusion limited stage of clathrate hydrate grain dissociation is over. Our assumed initial condition is therefore, that all methane molecules are already diffusing through an ice Ih column and are no longer stored in clathrate hydrate grains.  This means that at $t=0$ there are no sources, and mass conservation assumes the following form:
\begin{equation}\label{massconserv}
\frac{\partial c}{\partial t}=-\frac{\partial j}{\partial z}
\end{equation} 
Combining Eqs.\,($\ref{FickLaw1}$) and ($\ref{massconserv}$) we obtain the diffusion equation governing the flux of methane in the column:
\begin{equation}
\frac{\partial c}{\partial t}=\tilde{D}\frac{\partial^2c}{\partial z^2}
\end{equation}
where we have assumed a constant diffusion coefficient. This last assumption is valid since the thermal conductivity of ice Ih is much higher than that for clathrates and won't allow large temperature gradients over a $100$~m length scale.
At the head of the column ($z=h$), i.e. the planetary surface, the density is kept constant at the value of the atmospheric density, $c_{atm}$. The column base ($z=0$) marks the transition to the clathrate hydrate stability zone so there is no flux of methane through it. The initial and surface conditions are therefore:
$$
c(z,t=0)=c_0
$$
$$
c(z=h,t)=c_{atm}
$$
\begin{equation}
\left(\frac{\partial c}{\partial z}\right)_{z=0}=0
\end{equation}    
For the initial condition we assume a uniform density throughout the column, approximated by the density of methane in the initial clathrate hydrates.  Here there are $8$ methane molecules in a cubic unit cell whose length is $12\times 10^{-8}$~cm giving $c_0=4.6\times 10^{21}$~molec~cm$^{-3}$.  We use a classical separation for $c$ of the form:
\begin{equation}
c(z,t)=c_1(z)+c_2(z,t)
\end{equation}  
Using this separation we may divide our problem into the following two simpler problems:

%\begin{table}[ht]
\begin{center}
\begin{tabular}{ll}

$\tilde{D}\frac{d^2c_1}{dz^2}=0$ & $\frac{\partial c_2}{\partial t}=\tilde{D}\frac{\partial^2c_2}{\partial z^2}$ \\

$c_1(z=h)=c_{atm}$ & $c_2(z=h,t)=0$  \\

$\left(\frac{dc_1}{dz}\right)_{z=0}=0$ & $\left(\frac{\partial c_2}{\partial z}\right)_{z=0}=0$ \\
 
 & $c_2(z,t=0)=c_0-c_1(z)$  \\

\end{tabular} 
\end{center}
%\caption{\footnotesize{}}
%\label{tab:}
%\end{table}
The solution for the time independent set of equations is:
\begin{equation}
c_1(z)=c_{atm}
\end{equation} 
The solution for the set of equations governing $c_2$ is given in \cite{Carslaw1959} as:
$$c_2(z,t)= \frac{4}{\pi}(c_0-c_{atm})\sum_{n=0}^{\infty}\frac{(-1)^n}{2n+1}e^\frac{-\tilde{D}(2n+1)^2\pi^2t}  {4h^2}\cos\left(\frac{(2n+1)\pi z}{2h}\right)$$
\begin{equation}
= c_0-c_{atm} - (c_0-c_{atm})\sum_{n=0}^{\infty}(-1)^n\left\lbrace erfc\frac{(2n+1)h-z}{2\sqrt{\tilde{D}t}} + erfc\frac{(2n+1)h+z}{2\sqrt{\tilde{D}t}}\right\rbrace
\end{equation}
where
\begin{equation}
erfc(x)=1-erf(x)=\frac{2}{\sqrt{\pi}}\int_x^\infty e^{-\xi^2}d\xi
\end{equation} 
The reason there are two different series representing the solution for $c_2$ is that the trigonometric series is converging fast only for large values for the time, whereas the error function series is adequately convergent for time periods immediately after initiation of diffusion \citep{Crank1956}. It was stated by \cite{Carslaw1959} that for the time criterion $t<10^{-2}h^2/\tilde{D}$ the error function series is more adequate. When the time criterion is violated the trigonometric series ought be used. This understanding is carried throughout this subsection and for convenience we define $t_c\equiv 10^{-2}h^2/\tilde{D}$.  
The solution for the concentration of methane molecules in the column ($c$) may therefore be written as:
$$c(z,t) = c_{atm}+\frac{4}{\pi}(c_0-c_{atm})\sum_{n=0}^{\infty}\frac{(-1)^n}{2n+1}e^\frac{-\tilde{D}(2n+1)^2\pi^2t}  {4h^2}\cos\left(\frac{(2n+1)\pi z}{2h}\right)$$
\begin{equation}
 = c_0-(c_0-c_{atm})\sum_{n=0}^{\infty}(-1)^n\left\lbrace erfc\frac{(2n+1)h-z}{2\sqrt{\tilde{D}t}} 
+ erfc\frac{(2n+1)h+z}{2\sqrt{\tilde{D}t}}\right\rbrace
\end{equation}
The flux of methane molecules from the head of the column (i.e. planetary surface) may now be derived:
$$j_{surface}=-\tilde{D}\left(\frac{\partial c}{\partial z}\right)_{z=h}= \frac{2\tilde{D}}{h}(c_0-c_{atm})\sum_{n=0}^{\infty}e^{-\frac{\tilde{D}(2n+1)^2\pi^2t}{4h^2}}$$
\begin{equation}
=\left(c_0-c_{atm}\right)\sqrt{\frac{\tilde{D}}{\pi t}}\sum_{n=0}^\infty(-1)^n\left[e^{-\frac{n^2h^2}{\tilde{D}t}}-e^{-\frac{(n+1)^2h^2}{\tilde{D}t}}\right]
\end{equation}
where the upper series is for $t>t_c$ and the lower series is for $t\leq t_c$.

\begin{figure}[ht]
\centering
\includegraphics[scale=0.5]{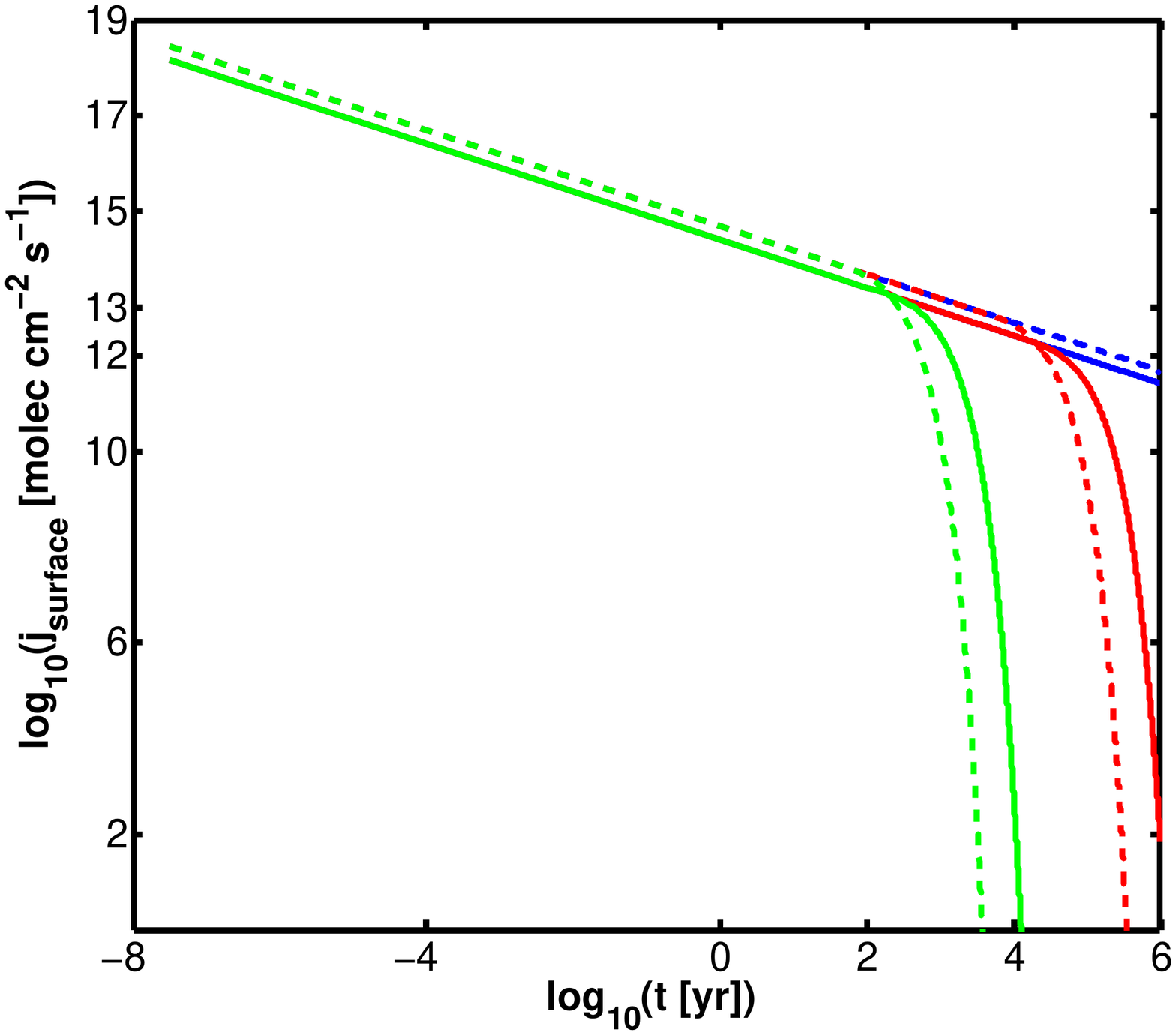}
\caption{\footnotesize{Flux of methane molecules from the head of a destabilized methane hydrate column, i.e. the planetary surface. Blue, red and green curves are for columns whose depths are: $100$~m, $10$~m and $1$~m, respectively. The solid and dashed curves are for an assumed temperature of $200$~K and $250$~K respectively.}}
\label{fig:f18}
\end{figure}

In Fig.\,\ref{fig:f18} we solve for three column depths ($h=100$\,m, $10$\,m and $1$\,m). For each choice of $h$ we solve for depth averaged temperatures of $200$\,K (solid curves) and $250$~K (dashed curves). Generally, higher temperatures give larger diffusion coefficients and larger initial methane fluxes. This also means a faster exhaustion of the methane stored in the water ice. In the first $100$\,yr since destabilization of the three columns, i.e. emergence of the three columns from the ridge, their planetary surface flux of methane is similar.  Even the difference between the two temperature cases is not great.

Since the methane hydrate dissociation begins at $t=0$, a specific column of destabilized hydrate, with unit horizontal cross section and depth $h$, would have released $\tilde{q}(t_d)$ methane molecules into the atmosphere after time $t_d$, where: 
$$\tilde{q}(t_d)=\int_0^{t_d}j_{surface}dt$$
\begin{equation}
 = \left(c_0-c_{atm}\right)\frac{h}{\sqrt{\pi}}\sum_{n=0}^\infty(-1)^n\left[n\Gamma\left(-\frac{1}{2},\frac{n^2h^2}{\tilde{D}t_d}\right)-(n+1)\Gamma\left(-\frac{1}{2},\frac{(n+1)^2h^2}{\tilde{D}t_d}\right) \right]
\end{equation}
for $t_d<t_c$, and
\begin{equation}
\begin{split}
= & \left(c_0-c_{atm}\right)\frac{h}{\sqrt{\pi}}\sum_{n=0}^\infty(-1)^n\left[n\Gamma\left(-\frac{1}{2},\frac{n^2h^2}{\tilde{D}t_c}\right)-(n+1)\Gamma\left(-\frac{1}{2},\frac{(n+1)^2h^2}{\tilde{D}t_c}\right) \right]\\
 & -\frac{8h}{\pi^2}\left(c_0-c_{atm}\right)\sum_{n=0}^\infty\frac{1}{(2n+1)^2}\left[e^{-\frac{\tilde{D}(2n+1)^2\pi^2t}{4h^2}}-e^{-\frac{\tilde{D}(2n+1)^2\pi^2t_c}{4h^2}}\right]
\end{split}
\end{equation}
 for $t_d>t_c$.  Here $\Gamma$ is the upper incomplete gamma function:
$$
\Gamma(a,\bar{x})=\int_{\bar{x}}^\infty t^{a-1}e^{-t}dt
$$

\begin{figure}[ht]
\centering
\includegraphics[scale=0.6]{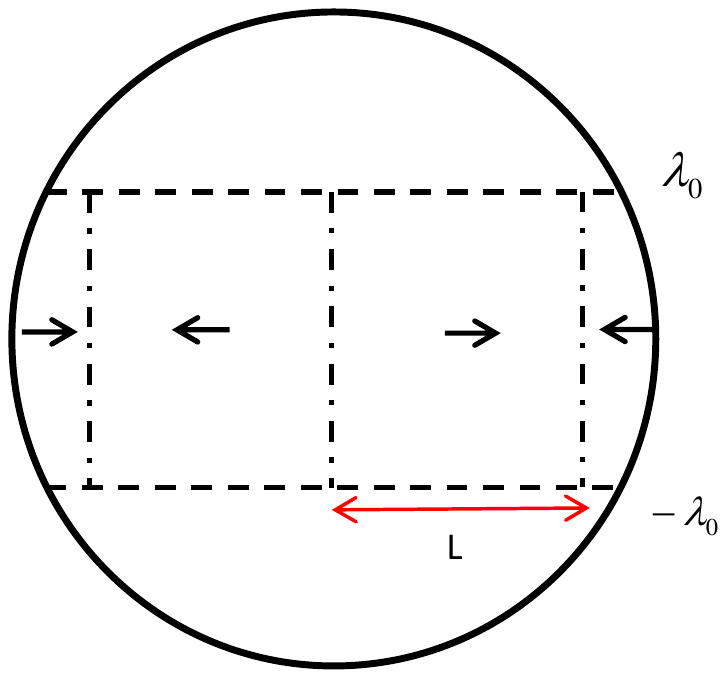}
\caption{\footnotesize{An approximated plate distribution used to estimate the global rate of methane release into the atmosphere. The latitude margin $[-\lambda_0\quad \lambda_0]$ is the area where methane clathrate hydrate is unstable on the planetary surface. $L$ is the width of the plate. Arrows represent plate direction of motion.}}
\label{fig:f19}
\end{figure}

We can now estimate the total number of methane molecules, $Q$, that enter the atmosphere per unit time.  An accurate answer requires an understanding of the feedback mechanisms between surface and atmosphere such as the greenhouse, and their influence on the planetary surface temperature. These have to be solved for each planet individually.   In addition, $Q$ depends on the spatial distribution, the size distribution, and the velocity distribution of the planetary tectonic plates, so that our estimate should be viewed as a first approximation.  To this end we adopt a plate distribution as shown in Fig.\,\ref{fig:f19}. 

We note that the time variable in the equations above may be translated to distance from the local ridge, $x$, via $x=U_pt$, where $U_p$ is the plate speed. In other words, instead of following a particular ice column as it moves with the plate and the temporal variation of the methane flux from the head of the ice column, we assume the plate is stationary with a spatial flux variation measured from the ridge. Let us assume a plate ridge is located along longitude $0$, then if $\lambda$ is the latitude and $R_p$ is the planet's radius we have $x=R_p\cos(\lambda)\phi$, where $\phi$ is the longitude. The variable of time may therefore be replaced by longitudes and latitudes:
\begin{equation}
t=\frac{R_p\cos(\lambda)\phi}{U_p}
\end{equation}
The time criterion, $t_c$, mentioned above may be translated to a longitude criterion as follows:
\begin{equation}
\phi_c=10^{-2}\frac{h^2(\lambda)}{\tilde{D}}\frac{U_p}{R_p\cos(\lambda)}
\end{equation}
where we have explicitly expressed the dependence of $h$ on the latitude via the variation in the surface temperature.  The flux of methane from the head of the ice column may now be written as a function of latitude and longitude. For $\phi>\phi_c$ we have:
\begin{equation}
j_{surface}^{>\phi_c}(\lambda,\phi) = \frac{2\tilde{D}}{h(\lambda)}(c_0-c_{atm})\sum_{n=0}^{\infty}e^{-\frac{\tilde{D}(2n+1)^2\pi^2R_p\cos(\lambda)\phi}{4h^2(\lambda)U_p}}
\end{equation}
while for $\phi\leq \phi_c$ we write:
$$j_{surface}^{\leq\phi_c}(\lambda,\phi) = $$
\begin{equation}\label{surfacefluxmajor}
\left(c_0-c_{atm}\right)\sqrt{\frac{\tilde{D}U_p}{\pi R_p\cos(\lambda)\phi}}\sum_{n=0}^\infty(-1)^n\left[e^{-\frac{n^2h^2(\lambda)U_p}{\tilde{D}R_p\cos(\lambda)\phi}}-e^{-\frac{(n+1)^2h^2(\lambda)U_p}{\tilde{D}R_p\cos(\lambda)\phi}}\right]
\end{equation}
We note that the density of atmospheric methane, $c_{atm}$, is here assumed constant. This is part of the idealization of  our problem. In a more realistic scenario $c_{atm}$ may evolve with time influencing the surface temperature. Such calculations though are beyond the scope of this work.
   
Assuming our plate is bounded by longitudes $0$ and $\phi_{t}$, representing the ridge and trench respectively, and that methane hydrate is unstable on the planetary surface between latitudes $[-\lambda_0\quad \lambda_0]$ around the equator, one may write:
$$
Q \sim N_{plate}\int_0^{\phi_t}\int_{-\lambda_0}^{\lambda_0}j_{surface}(\lambda,\phi)R^2_p\cos(\lambda)d\lambda d\phi 
$$
$$
=N_{plate}\int_0^{\phi_c}\int_{-\lambda_0}^{\lambda_0}j_{surface}^{\leq\phi_c}(\lambda,\phi)R^2_p\cos(\lambda)d\lambda d\phi
$$
\begin{equation}
+ N_{plate}\int_{\phi_c}^{\phi_t}\int_{-\lambda_0}^{\lambda_0}j_{surface}^{>\phi_c}(\lambda,\phi)R^2_p\cos(\lambda)d\lambda d\phi 
\end{equation} 
Since the integral accounts for the rate of methane loss to the atmosphere from a single plate (see approximated plate diagram in Fig.\,\ref{fig:f19}) we have multiplied it by $N_{plate}$, the number of such plates around the equatorial circumference, which we estimate as follows:
\begin{equation}
N_{plate}\sim\frac{2\pi R_p}{L}=\frac{2\pi R_p}{\hat{\xi}D_M}
\end{equation}
where $L$ is the plate length, $\hat{\xi}$ is the aspect ratio of the convection cell ($\approx 2$) and $D_M$ is the depth of the water mantle. For the $2M_E$ planet assuming $25\%$ ice mass fraction we have $N_{plate}=13$.

In reality, $h$ will depend on the latitude.  Since $h$ is the local depth where methane hydrates become stable and that the overlying layer is made of ordinary ice, one may write for the temperature, $T_h$, at the base of the column:
\begin{equation}\label{ColumnBaseTemp}
T_h-T_s=\frac{F_s}{\kappa_{Ih}}h
\end{equation}
where $T_s$ is the surface temperature, $F_s$ is the surface heat flux and $\kappa_{Ih}=3.3\times 10^5$~erg\,s$^{-1}$\,cm$^{-1}$\,K$^{-1}$ is the thermal conductivity of ice Ih.  
The pressure at the base of the column is:
\begin{equation}\label{PressureBottomColumn}
P^{CH_4}_{dis}(T_h)=P_s+\rho_{Ih}g_sh
\end{equation}
which is also equal to the methane hydrate dissociation pressure at the temperature $T_h$.  The surface gravity is $g_s$ and $\rho_{Ih}=0.917$~g\,cm$^{-3}$ is ice Ih bulk mass density.

\begin{figure}[ht]
\centering
\includegraphics[scale=0.5]{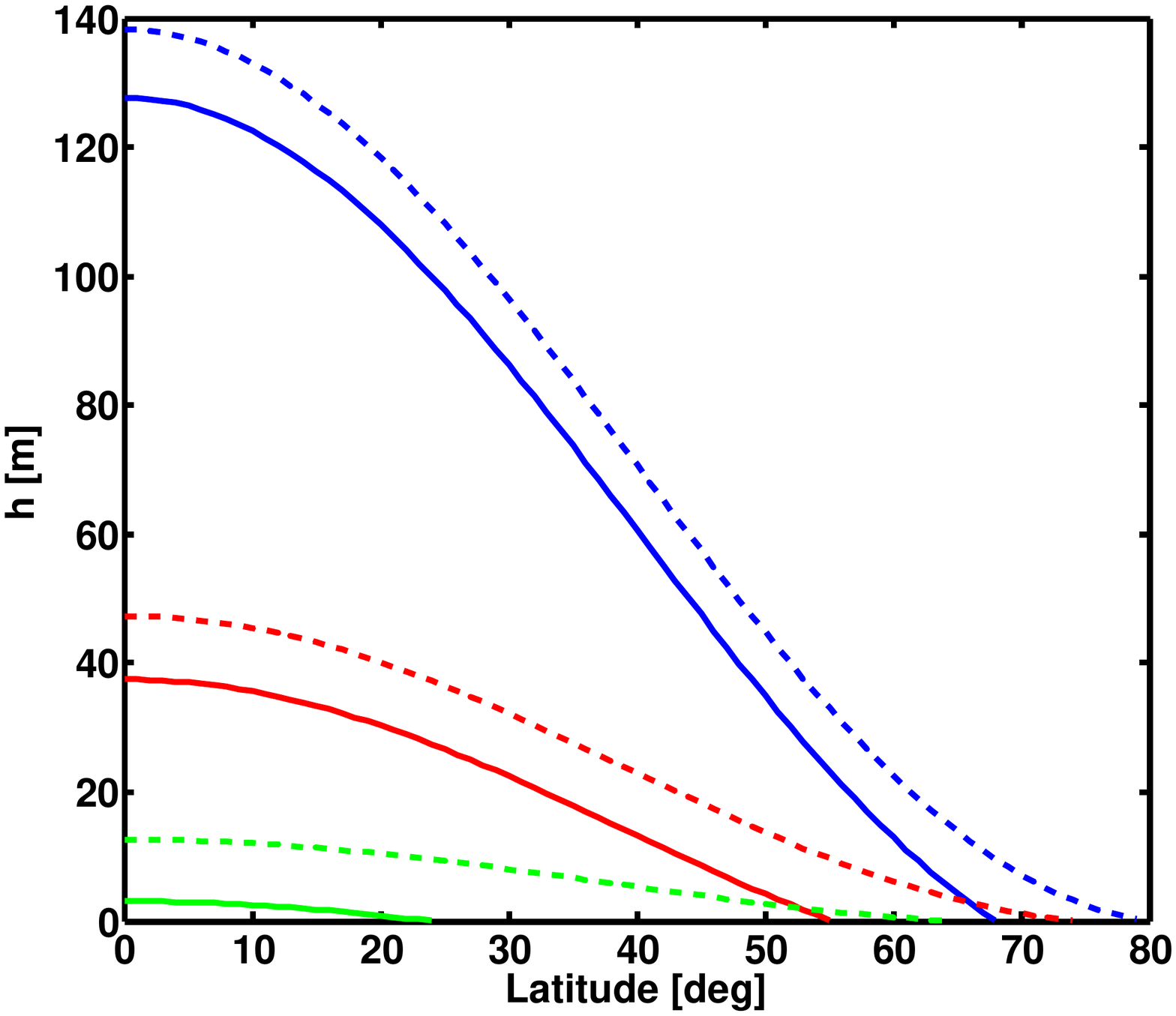}
\caption{\footnotesize{Depth of ice Ih mass column required to stabilize methane hydrates versus latitude, for the $2M_E$ planet assuming $25\%$ ice mass fraction. For a constant surface pressure the blue, red and green curves assume a $250$~K, $225$~K and $200$~K equatorial temperatures respectively. The solid (dashed) curves are for a $1$~bar ($0.1$~bar) planetary surface pressure. When $h$ diminishes to zero methane hydrates become stable on the planetary surface.}}
\label{fig:f20}
\end{figure}

The variation of surface temperature with latitude and longitude depends on the specifics of the planetary energy budget and should really be solved on a case by case basis. In what follows we assume the following approximate model:
\begin{equation}\label{SurfaceTemp}
T_s(\lambda)=T_{equator}\cos^{\frac{1}{4}}\lambda
\end{equation} 
where $T_{equator}$ is the average temperature around the equator of the planet, left here as a free parameter. In Fig.\,\ref{fig:f20} we solve Eqs.\,($\ref{ColumnBaseTemp}$) through ($\ref{SurfaceTemp}$) to yield the dependence of $h$ on the latitude. For a given surface pressure the higher surface temperature requires a higher pressure to stabilize hydrates and hence a deeper ice Ih column. For a given surface temperature decreasing the surface pressure needs to be compensated by further increasing the depth of the ice Ih column (see the phase diagram of methane hydrate in Fig.\,\ref{fig:f4} for reference).  

For our parameter space, when $t>t_c$, i.e. $\phi>\phi_c$, the gas flux from the head of the ice column diminishes substantially, as seen in Fig.\,\ref{fig:f18}, and its contribution to $Q$ may be neglected. The global rate of methane release into the atmosphere may therefore be approximated as: 
$$
Q \sim N_{plate}\int_0^{\phi_c}\int_{-\lambda_0}^{\lambda_0}j_{surface}^{\leq\phi_c}(\lambda,\phi)R^2_p\cos(\lambda)d\lambda d\phi\approx
$$
\begin{equation}\label{OutgassingRate}
\approx\frac{2}{5\sqrt{\pi}}N_{plate}\left(c_0-c_{atm}\right)U_pR_p\left\langle h\right\rangle_\lambda\lambda_0
\end{equation} 
where, in the last approximation, we represent the series of Eq.\,($\ref{surfacefluxmajor}$) using only its leading term for which $n=0$.  For this case the first exponential in the square brackets is unity and the second exponential is negligible. This assumption is justified by the fast convergence of this series. The average depth where methane clathrate hydrates become stable with respect to latitude, $\left\langle h\right\rangle_\lambda$, is obtained by averaging over curves as seen in Fig.\,\ref{fig:f20}. 

\begin{figure}[ht]
\centering
\includegraphics[scale=0.5]{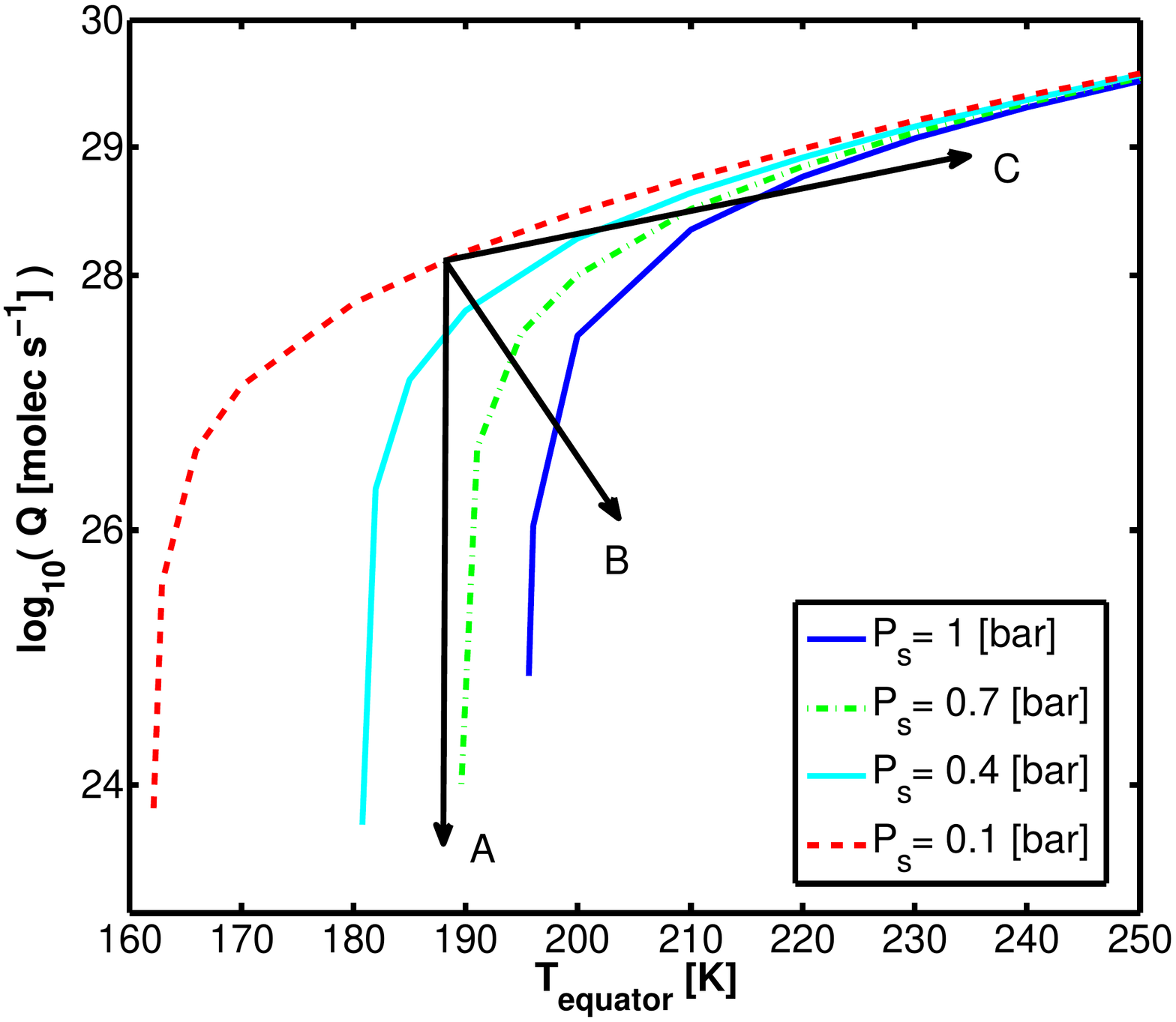}
\caption{\footnotesize{Global rate of methane release into the atmosphere as a function of the equatorial temperature. Curves represent constant surface pressures from $0.1$\,bar (red dashed curve) to $1$\,bar (solid blue curve). For each curve (isobar) there is a surface temperature where clathrate hydrates become stable throughout the planet's surface and $Q$ will drop to zero. Stabilizing clathrate hydrates on the surface means the surface pressure is equal to or greater than the local clathrate hydrate dissociation pressure. A higher surface pressure may stabilize clathrate hydrates at a higher surface temperature and therefore $Q$ will go to zero at a higher equatorial temperature. Arrows $A$, $B$ and $C$ are hypothesized evolutionary paths, see text for more detail.}}
\label{fig:f21}
\end{figure}

We assume $1$~m\,yr$^{-1}$ for the plate speed and solve for $Q$ as a function of the equatorial temperature (i.e. heliocentric distance, etc.), for four different surface pressures. The results are shown in Fig.\,\ref{fig:f21} where each curve represents a different constant surface pressure. For each surface pressure there is a surface temperature where clathrates become stable throughout the planetary surface ($\left\langle h\right\rangle_\lambda=0,~\lambda_0=0$ ) and $Q$ will drop to zero. Clathrate hydrates become stable on the planetary surface when the surface pressure is equal to or greater than the local dissociation pressure. A lower surface pressure will stabilize clathrate hydrates throughout the body's surface at a lower surface temperature so that a lower temperature will be required for $Q$ to go to zero.
 
In Fig.\,\ref{fig:f21} we also show three hypothetical evolutionary paths, arrows $A$, $B$ and $C$. These are quite general paths since the general behavior for $Q$ will be similar for clathrate hydrates hosting a mixture of guest molecules. In path $A$ gas is released into the atmosphere raising the surface pressure without raising the surface temperature. The consequent pressure build-up will continue until clathrate hydrates become stable throughout the planetary surface, forcing $Q$ to zero. Further clathrate hydrate decomposition may continue as a means of compensating for atmospheric escape or other sinks that deplete the atmosphere (e.g. polymerization to form aerosols). In path $B$ the gas released from clathrate hydrate dissociation also increases the surface temperature but the increase in pressure is still enough to continue to decrease $Q$ to negligible values. In path $C$ the release of gas increases the surface temperature drastically (e.g. a potent green house gas) so that even though the surface pressure increases so does the rate of global gas release, creating a runaway effect and probable surface melting. The distinction between these three paths again requires a case by case solution for each planet and is left for future work.   

We may gain some feeling for the numerical value found for $Q$ and consequently for the atmospheric stability by considering the ratio of the number of molecules in the atmosphere to the global rate of gas release. This can be seen as a measure of the time it takes to re-establish the atmosphere, with this outgassing mechanism. This atmospheric dynamic time scale is given by:
\begin{equation}
t_{atm}=\frac{4\pi R_p^2Hc_{atm}}{Q}
\end{equation} 
where $H$ is the atmospheric scale height. For our $2M_E$ planet with a $25\%$ ice mass fraction and the parameters adopted above, one finds $H\sim 10$~km.  Assuming an atmospheric density similar to that of the Earth, $c_{atm}=10^{19}$~molec\,cm$^{-3}$, we find $t_{atm}$ is $3$~Ga and $30$~Ma for $Q=10^{27}$~molec\,s$^{-1}$ and $10^{29}$~molec\,s$^{-1}$, respectively.  

Finally, we wish to solve explicitly for our proposed outgassing mechanism evolution with time. In this work we will not consider the greenhouse effect and so we restrict ourselves to path $A$ type behaviours, as described in fig.\ref{fig:f21}. The atmospheric surface pressure is related to the number of methane molecules in the atmosphere according to:
\begin{equation}
P_s(t)=\frac{m_{CH_4}g_s}{4\pi R^2_p}N^{atm}_{CH_4}(t)
\end{equation} 
where $m_{CH_4}$ is the molecular mass of methane and $N^{atm}_{CH_4}$ is the number of methane molecules in the atmosphere. 

We solve for the SI methane clathrate hydrate dissociation pressure using the theory of \cite{waalplat} and available experimental data. We find that the following formulation is appropriate for its three phase hydrate-ice Ih-vapour curve:
\begin{equation}\label{MethaneClathrateDisPress}
P^{CH_4}_{dis}(T)=\exp\left(A_{cc}-\frac{B_{cc}}{T}\right)\quad 
\end{equation}    
where $A_{cc}=25.086$ and  $B_{cc}=2199.7$\,K are the Clausius-Clapeyron coefficients giving the dissociation pressure in dyn\,cm$^{-2}$.
Combining Eqs.(\ref{MethaneClathrateDisPress} and \ref{ColumnBaseTemp}-\ref{SurfaceTemp}) yields after some algebraic steps:
\begin{equation}\label{LatExtentRelation}
\cos\lambda =\frac{1}{T^4_{equator}}\left[\frac{B_{cc}}{A_{cc}-\ln\left(P_s+\rho_{Ih}g_sh\right)}-\frac{F_s}{\kappa_{Ih}}h\right]^4
\end{equation}
From the relation $\lambda_0=\lambda(h=0)$ and the last equation one may obtain:
\begin{equation}
\cos\lambda_0 =\frac{1}{T^4_{equator}}\left[\frac{B_{cc}}{A_{cc}-\ln P_s}\right]^4
\end{equation}
From this relation we obtain $\lambda_0=\lambda_0(P_s,T_{equator})$, which together with Eq.($\ref{LatExtentRelation}$) yield the relation $\left\langle h\right\rangle_\lambda=\left\langle h\right\rangle_\lambda(P_s,T_{equator})$. These relations combined with Eq.($\ref{OutgassingRate}$) give $Q=Q(P_s,T_{equator})$. The evolution of the surface pressure with time may therefore be formulated inversely as:
\begin{equation}
t=\frac{4\pi R^2_p}{m_{CH_4}g_s}\int^{P_s(t)}_{P_s(t=0)}\frac{d\tilde{P}_s}{Q(\tilde{P}_s,T_{equator})}
\end{equation}

\begin{figure}[ht]
\centering
\mbox{\subfigure{\includegraphics[width=7cm]{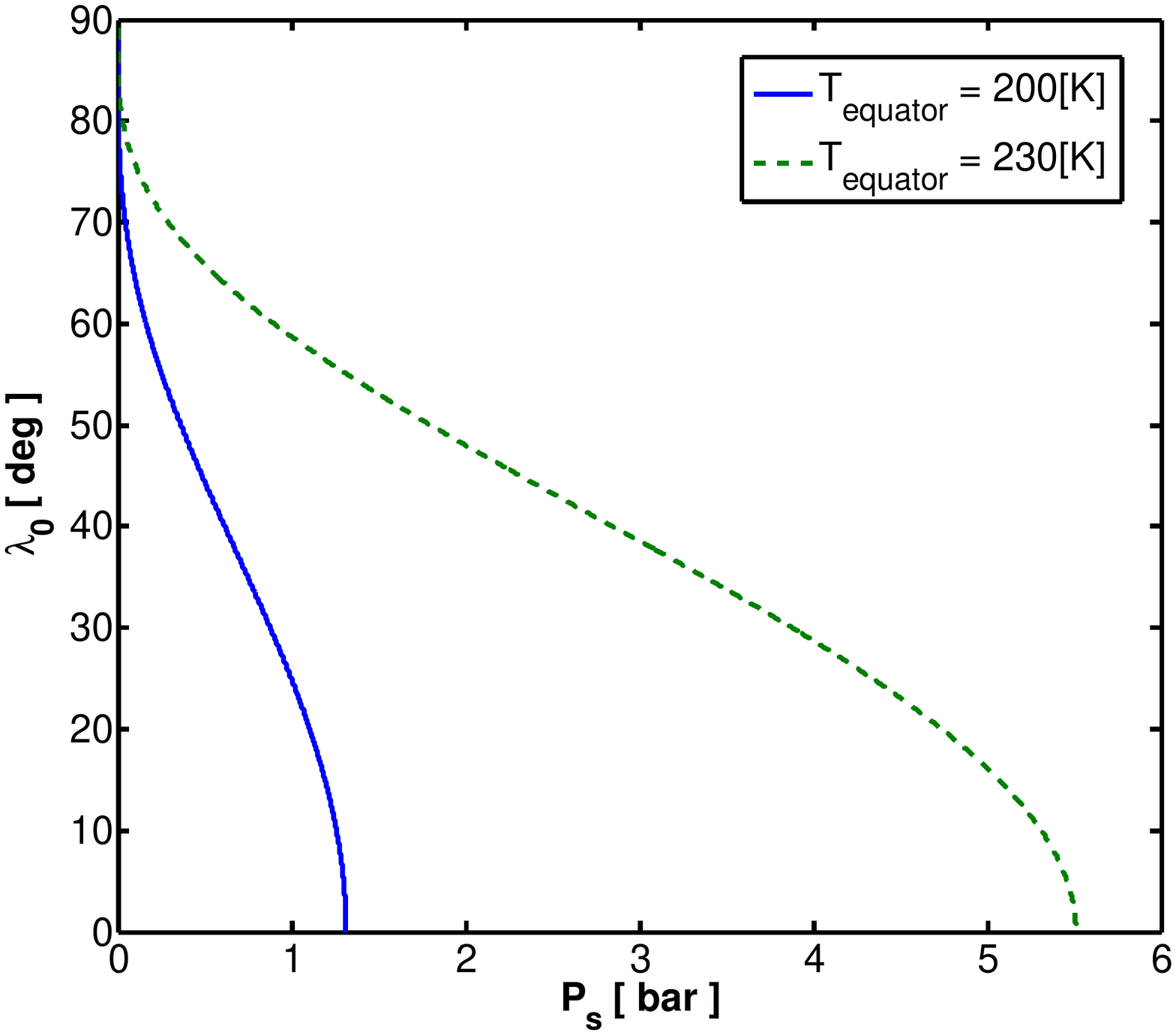}}\quad \subfigure{\includegraphics[width=7cm]{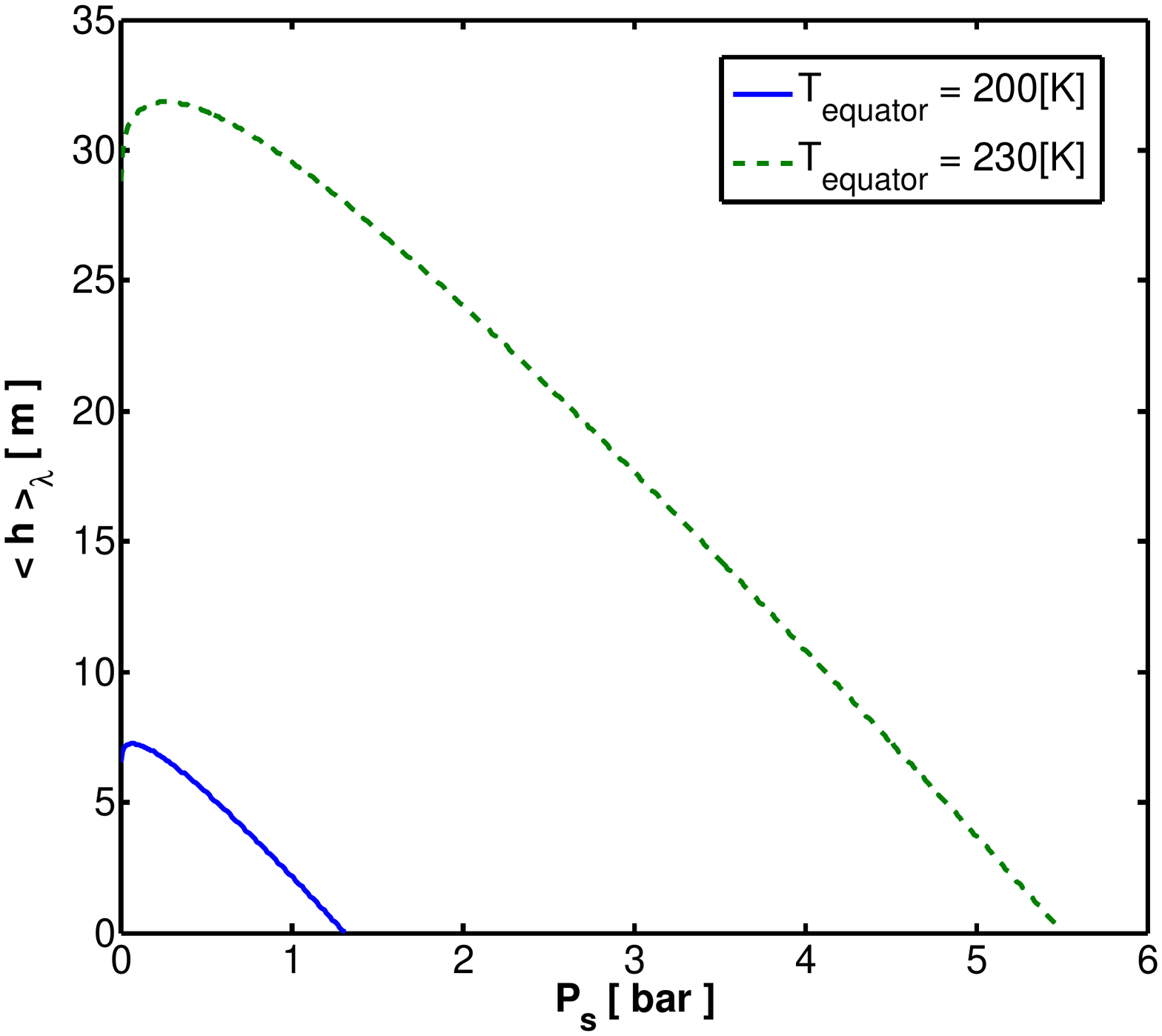}}}
\caption{\footnotesize{Latitudinal extent of SI methane clathrate hydrate instability on the planetary surface (left panel) and the depth to the SI methane clathrate hydrate stability zone from the planetary surface averaged over latitude (right panel), versus atmospheric methane surface pressure. We solve for two equatorial averaged temperatures: $200$\,K (solid blue curve) and $230$\,K (dashed green curve).}}
\label{fig:f25}
\end{figure}

In fig.\ref{fig:f25} we plot the dependences of $\lambda_0=\lambda_0(P_s,T_{equator})$ and $\left\langle h\right\rangle_\lambda=\left\langle h\right\rangle_\lambda(P_s,T_{equator})$ for two equatorial temperatures: $200$\,K and $230$\,K. When $\lambda_0=0$ and $\left\langle h\right\rangle_\lambda=0$ the methane clathrate becomes stable throughout the planetary surface. This happens for methane surface pressures of at least: $1.3$\,bar and $5.5$\,bar for $T_{equator}=200$\,K and $230$\,K respectively. As these methane surface atmospheric pressures are attained the outgassing will cease ($Q\rightarrow 0$). We note that our choice for the variation of the surface temperature with latitude (see eq.$\ref{SurfaceTemp}$) introduces an error into the model. Clearly the surface temperature near the planetary poles is not well represented. This is the reason the surface atmospheric pressure drops to zero for $\lambda_0=90^\circ$ (i.e. the case where methane clathrate is nowhere stable on the planetary surface).

\begin{figure}[ht]
\centering
\includegraphics[scale=0.5]{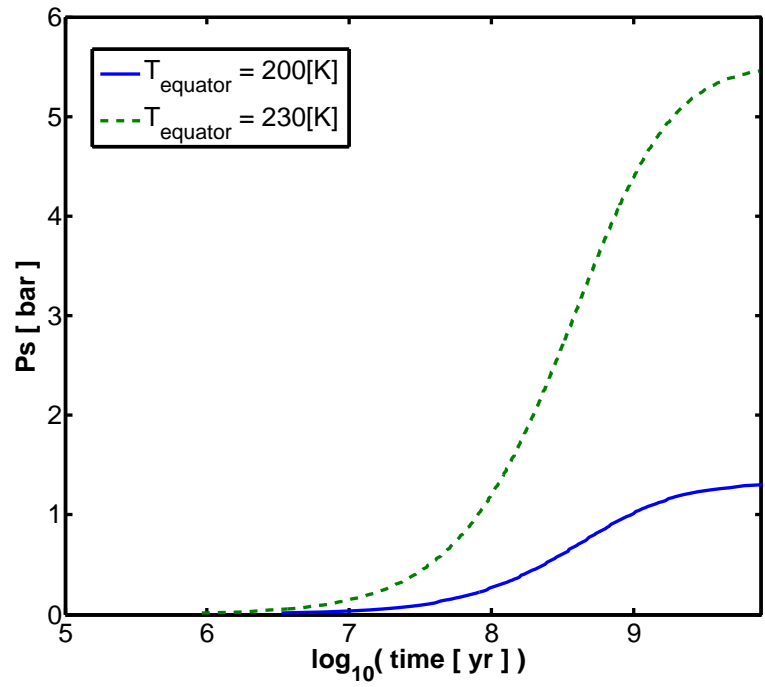}
\caption{\footnotesize{Surface methane atmospheric pressure increase with time due to outgassing, solved for two equatorial averaged temperatures: $200$\,K (solid blue curve) and $230$\,K (dashed green curve). }}
\label{fig:f27}
\end{figure}

In fig.\ref{fig:f27} we give the increase in the methane atmospheric surface pressure with time, again for two scenarios for the averaged equatorial surface temperature, $T_{equator}=200$\,K and $230$\,K. As time progresses and the surface pressure of methane builds up, methane clathrate hydrate becomes stable on an increasing fraction of the planetary surface. As a consequence of that the outgassing flux of methane diminishes asymptotically to zero. At the same time the atmospheric surface pressure of methane approaches asymptotically the SI methane clathrate hydrate dissociation pressure for the highest surface temperature, here assumed at the equator. For an equatorial temperature of $200$\,K ($230$\,K) we find that after $1$\,Ga the atmospheric surface pressure has reached a value of some $77\%$ ($80\%$) of the methane clathrate dissociation pressure at $200$\,K ($230$\,K). This asymptotic behaviour is a consequence of our choice of a type $A$ evolutionary path. 

Such an outgassing behaviour is not able to severely deplete methane reservoirs in the planetary ice mantle. Let us assume the abundance of methane in the ice mantle is in the range of $1\%-10\%$ by number (see a more in depth discussion on methane abundance below in the discussion section). Then, for a $2$M$_E$ water planet with a $25\%$ ice mass fraction the number of methane molecules outgassed to the atmosphere is approximately in the ranges of $0.05\%-0.005\%$ and $0.2\%-0.02\%$ of the  methane present in the ice mantle, for $T_{equator}=200$\,K and $230$\,K respectively.          

\section{DISCUSSION}
 
Viscosity is a key parameter when trying to model the dynamics inside a planet. We made use of two techniques to estimate viscosities for our binary methane-water solid solution, to compensate for the little experimental data available. One technique uses the homologous temperature, another implements molecular dynamic simulations to help constrain the viscosity activation enthalpy.  
The similar topology of iso-viscous plots and the melt curve indicates the ice rheology depends on the composition of the occluded hydrophobic molecules, both in cage clathrates and in filled ice. The way in which the rheology will change is not straightforward to predict. Massive hydrocarbons, if present, will lower the dissociation temperature indicating a decrease in the viscosity. However their larger size will reduce their ability to diffuse through the ice matrix, therefore increasing the viscosity related to interstitial diffusion (Newtonian viscosity).

In scaling the planetary surface heat flux we used surface heat flux measurements for the Earth and scaled according to the silicate and iron abundance in our planets, this introduces several caveats. One such caveat is the disregard of the primordial heat of accretion. This heating mechanism increases in importance as the mass of the planet considered is increased. The high abundance of water ice assumed for water planets with its lower latent heat of melting will probably help rid the planet faster of its accretional heating. Quantifying this mechanism requires an evolutionary perspective which we hope to address in the future. Nonetheless, this caveat implies our choice for the scaling of the surface heat flux is a lower bound value.   
Another problem that may arise stems from the above analysis being restricted to the ice mantle. A more global approach which takes into consideration both the internal silicate mantle and the metallic core is in order. The surface heat flux from the Earth is relatively large due to the existence of plate tectonics. It is not clear whether a highly pressurized internal silicate mantle (tens of GPa on the silicate-water boundary) will have an external thermal boundary layer and heat flux resembling those for the Earth. It is possible that the actual thermal boundary layer, between the silicate and ice mantles in water planets, conducts less heat than in our scaling assumption using Earth's surface. This effect would imply our choice for the planetary surface heat flux is an upper bound value. 

We have argued in subsection $2.4$ that crossing the melting curve for pure ice Ih into liquid water prior to forming clathrate hydrates results in a submerged clathrate hydrate layer which is gravitationally unstable. This is because the bulk density of methane clathrate hydrate is lower than the mass density of liquid water. The bulk density of a clathrate hydrate depends on the composition of the guest molecules and the degree of cage occupancy. In a more complex scenario, where guest molecules heavier than methane are also present and enclathrated a liquid water layer may be gravitationally stable on top of a hydrate layer \citep{McKay2003}. This may prove important for hot water-rich planets. 

We show that clathrate hydrate formation tends to restrain the thermal profile from penetrating the liquid water thermal stability field. This proposed inhibition mechanism acting against the formation of subterranean global oceans must be understood in the context of a real three dimensional planet. In a 3D planetary system, clathrate forming molecules will not be distributed evenly. Regions with a low concentration of hydrophobic molecules will not be able to stabilize clathrates throughout, resulting in the formation of local subterranean lakes. These subterranean lakes may have dissolved gas concentrations controlled by the presence of clathrates surrounding the liquid water reservoir, with probable astrobiological implications. 

It is important to test whether clathrates can indeed prevent the formation of subterranean oceans. One option for such a test is to ask whether a subterranean liquid layer will result in surface geological features differing from the features above a clathrate layer. The problem in trying to answer this question is that in the water-rich planets we studied, we always found a sub-layer confined to the clathrate melting curve (the DBL layer). Since the physical characteristics of such a layer are poorly known a comparison is difficult. Moreover, one may argue that a liquid sub-layer and a sub-layer whose conditions are close to melting should have some similarity making distinction a delicate task. Resolving the matter requires further research into the behaviour of on-melt layer dynamics.

A binary mixture of methane and water does not take into consideration the effects of other cosmochemically important constituents. Two such important molecules are ammonia and methanol. Observations of comets yield at most $6:100$ and $1:100$ for the methanol to water and ammonia to water abundance ratios \citep{Despois2005}. The reason we mention these particular molecules here is due to their antifreeze quality, i.e. they effectively lower the chemical potential of liquid water.  
It is also known that ammonia aqueous solutions and methanol aqueous solutions form stoichiometric compounds upon freezing rather than non-stoichiometric clathrate hydrates. For these reasons both ammonia and particularly methanol are canonically considered to be clathrate hydrate inhibitors \citep{Sloan2003}, though industrially not very efficient ones. 

\cite{lunsteve85} estimated the effect of adding ammonia to a water-methane mixture. They have argued that as a result of the addition of ammonia the thermodynamic stability field of liquid water will expand into the pure ice Ih stability field and the dissociation curve for clathrates will also shift to lower temperatures. Their analysis predicted no change in the hydrate-liquid-vapour three phase gradient ($dP/dT$) due to the addition of ammonia. The subterranean ocean inhibition effect we find is solely dependent on this gradient and therefore ought prevail the addition of ammonia to the solution. 

Besides expanding the thermodynamic stability field for liquid water, ammonia and methanol are usually thought to actively promote clathrate dissociation. This assumption comes from the idea that all guest molecules in the clathrate cages ought be hydrophobic. It is commonly believed that since ammonia and methanol can create hydrogen bonds with water they will force the formation of stoichiometric compounds. 
Recent experiments and molecular dynamic simulations however are starting to reveal the true nature of the effect that ammonia and methanol have on a water system \citep{Shin2012,Shin2013}. It appears that the role ammonia and methanol play changes dramatically when clathrate hydrate forming molecules are present.
It has been shown that ammonia and methanol act as catalysts, by increasing the reactivity of water ice surfaces, therefore accelerating the rate of clathrate hydrate formation. The hydrophobic molecules present in the mixture will then limit the ability of ammonia and methanol to hydrogen bond to the water network resulting in clathrate hydrates. These will capture in their cages both the ammonia, the methanol and the hydrophobic constituents. Ammonia and methanol therefore widen the stability field for liquid water but by no means inherently promote clathrate hydrate dissociation.

It is interesting to note that molecular dynamics simulations show it is energetically favourable to replace the methane molecules trapped in the large cages of a SI clathrate hydrate with methanol molecules \citep{Shin2013}. Only in the case the ice experienced melting along its history and the methanol is in an aqueous solution will the reaction above become unfavourable. This means spatially confined melting episodes resulting in local, possibly subterranean, lakes will tend to concentrate methanol within them. This will stabilize local liquid reservoirs and deprive methanol from their surrounding regions and therefore enhance their resistance to melting. The astrobiological consequences of such an effect could be interesting. 

\begin{figure}[ht]
\centering
\includegraphics[scale=0.5]{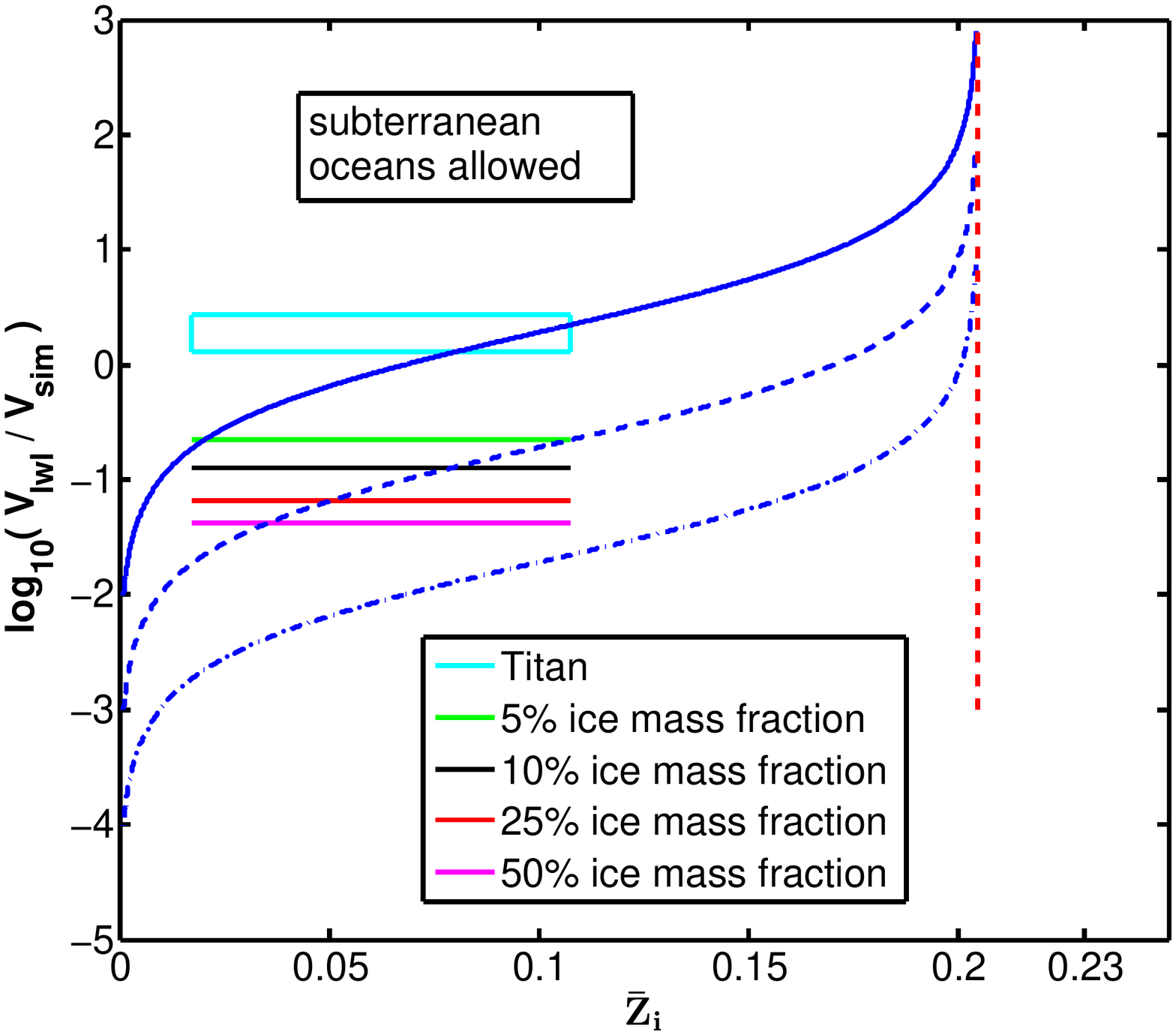}
\caption{\footnotesize{Ratio of the hypothesized subterranean ocean volume over the volume of the underlying solid ice mantle versus accreted methane to water ratio by number. Blue (solid , dashed and dashed-dotted) curves represent $\Re$ of $1$, $0.1$ and $0.01$ respectively. $\Re$ is the fraction out of the methane molecules initially in the internal solid ice mantle that can migrate outward and enrich the hypothesized subterranean ocean layer. Falling in the domain above (below) one of the blue curves means that for that given $\Re$ (i.e. restriction on methane transport) methane enrichment in the outer ice layer is not sufficient (is sufficient) for full enclathration of the global subterranean ocean. Cyan rectangle is the parameter range for Titan \citep[see][]{Fortes2012}. The parameter ranges for the $2$M$_E$ water planet assuming: $5\%$, $10\%$, $25\%$ and $50\%$ ice mass fractions are the: green, black, red and magenta line segments respectively. Vertical dashed (red) curve is the asymptote for $\bar{Z}_i=Z_{cs}$.}}
\label{fig:f24}
\end{figure}

The inhibition of a global subterranean ocean requires that enough clathrate forming molecules be available at the depth range where the planetary geotherm falls within the liquid water phase regime. The availability of sufficient clathrate forming molecules to this region is not easily determined. The first uncertainty is the abundance of clathrate forming molecules in the planetary water mantle. Ice compositions within the icy bodies of our solar system are not very well constrained. This problem becomes even less constrained when it comes to planets formed outside of our solar system. The second uncertainty concerns the ability of the clathrate forming molecules, that perhaps are initially evenly distributed within the ice mantle, to convectively redistribute. For example, CO$_2$ will encourage the formation of clathrates and therefore the inhibition of a subterranean ocean. CO$_2$ in comets is even more abundant than CH$_4$ (about $6:1$, \cite{Despois2005}). However, the straight forward conclusion that the added presence of CO$_2$ turns subterranean oceans even less likely than when considering methane alone must first address the ability of CO$_2$ to migrate from the interior to the region where liquid water phase is stable. In this work we have addressed this issue for the case of methane, though, this problem is not yet solved neither for CO$_2$ nor for other clathrate forming molecules. With these caveats in mind we wish to make an approximate first attempt to quantify the parameter range that either enables or hinders the formation of a subterranean ocean. In this attempt we narrow the plethora of clathrate forming molecules and consider only the case of methane. We will further restrict the methane to water abundance ratio to values commonly adopted for our solar system, recalling that in other solar systems this ratio may attain different values. 
Let us consider a water planet that has accreted a methane to water ratio, by number, of $\bar{Z}_i$. We also assume this ratio is initially uniform throughout the ice mantle. In order to enclathrate the entire region where the geotherm falls within the liquid water phase a methane to water ratio of $Z_{cs}$ is required there. Beneath this liquid water layer (denoted henceforth as $lwl$) lies the internal solid ice mantle (denoted $sim$). Then according to our assumptions the initial methane to water ratios in both these layers are:
\begin{equation}\label{MethaneWaterRatio1}
\bar{Z}^{lwl}_i=\bar{Z}_i=\frac{N^{lwl}_{CH_4,i}}{N^{lwl}_{H_2O}}\quad,\quad \bar{Z}^{sim}_i=\bar{Z}_i=\frac{N^{sim}_{CH_4,i}}{N^{sim}_{H_2O}}
\end{equation}  
Here $N^{p}_{CH_4,i}$ and $N^{p}_{H_2O}$ are the initial number of methane molecules in the $p$ layer and water molecules in the $p$ layer, respectively.
Now let us assume that there is a maximum in the number of methane molecules that can be redistributed from the $sim$ to the $lwl$ layer, therefore enriching it. In other words, there may be restrictions on how many methane molecules can the internal ice mantle lose in favour of enriching the outer ice mantle, where water may be liquid. Let us say this maximum number is the $\Re$ fraction of the total number of methane molecules initially in the $sim$ layer. Therefore, following this redistribution, and in case the $lwl$ was enriched in methane just enough to enclathrate it, then the methane to water ratio in the $lwl$ becomes:
\begin{equation}\label{MethaneWaterRatio2}
\bar{Z}^{lwl}_{redistributed}=\frac{N^{lwl}_{CH_4,i}+\Re N^{sim}_{CH_4,i}}{N^{lwl}_{H_2O}}=Z_{cs}
\end{equation}
Eqs.($\ref{MethaneWaterRatio1}$) and ($\ref{MethaneWaterRatio2}$) yield the relation: 
\begin{equation}\label{WaterFraction}
\frac{N^{sim}_{H_2O}}{N^{lwl}_{H_2O}}=\frac{1}{\Re}\left(\frac{Z_{cs}}{\bar{Z}_i}-1\right)
\end{equation} 
The ratio of the number of water molecules composing the two layers may be approximated by the layers' bulk mass densities and volumes, as:
\begin{equation}
\frac{N^{sim}_{H_2O}}{N^{lwl}_{H_2O}}\approx \frac{\rho_{sim}V_{sim}}{\rho_{lwl}V_{lwl}}
\end{equation} 
Which combined with eq.($\ref{WaterFraction}$) gives:
\begin{equation}\label{ParameterSpaceOcean}
\frac{V_{lwl}}{V_{sim}}\approx \Re \frac{\rho_{sim}}{\rho_{lwl}}\frac{\bar{Z}_i}{Z_{cs}-\bar{Z}_i}
\end{equation}    
A fully occupied SI methane clathrate hydrate requires $Z_{cs}=1/5.75$. A SH clathrate hydrate can be stabilized by methane molecules occupying its smaller cages and a bigger guest molecule occupying its large cage. Since we consider only methane then SH clathrate stability requires multi-occupancy ($2-3$ methane molecules) of its large cage \citep{Susilo2008}. We therefore adopt an average value for $Z_{cs}$ of $0.2046$ to account for both the SI and SH clathrate formation on the expense of the subterranean ocean. For the bulk mass density ratio we adopt a value of $2$. In fig.\ref{fig:f24} we solve Eq.($\ref{ParameterSpaceOcean}$) for three cases of $\Re$: $1$, $0.1$ and $0.01$. We also show in the figure the volume ratios for several bodies, where, following \cite{iro03} we vary the possible $\bar{Z}_i$ values for each body between $0.017$ and $0.107$. This range of values for $\bar{Z}_i$ corresponds to CO over CH$_4$ ratios in the solar nebula between $40$ and $5$, respectively. The volume ratio range for Titan is derived from the internal structures suggested for both the light and dense ocean cases in \cite{Fortes2012} (see cyan rectangle in fig.\ref{fig:f24}). From the figure it is clear that, considering only methane, a full enclathration of the subterranean ocean layer in Titan will require both an accretion of ice highly enriched in methane (large $\bar{Z}_i$) and almost a complete migration of methane initially locked in the internal ice V and VI mantle ($\Re=1$) into the outer water layer. The case of water planets is quite different. The full transport of all internal methane to the outer water layer where the subterranean ocean can form will overwhelm it with clathrate forming molecules resulting in complete enclathration. If only $10\%$ of the internal mantle methane reservoir can migrate outward and enrich the outer water layer then the $2$M$_E$ water planet with $5\%$ ice mass fraction may end up with a subterranean ocean. For the larger ice mass fractions a non uniformity of just a few percent in the distribution of methane within the ice mantle, in favour of enriching its outer layer, will suffice to completely enclathrate the hypothesized subterranean ocean layer.           

We have estimated the thermal profile extending the entire ice mantle for various planetary masses and ice mass fractions. By following the adiabat to high pressures we can determine where the filled ice becomes unstable and where super-ionic and reticulating phases are introduced. The determination of these phase changes is important for understanding the transport of carbon within an ice mantle and how it changes with the planetary mass and structure. This requires an accurate thermal expansivity for filled ice which is yet unknown. When more accurate thermodynamic parameters for filled ice become available it will be possible to determine for each planetary mass the threshold ice mass fraction where the ice mantle transitions from a molecular solid mantle to one which is partly super-ionic or reticulating. The thermodynamic parameters we have adopted show that a $2$M$_E$ water planet will have a molecular solid ice mantle throughout, whereas a $5$M$_E$ water planet may posses a super-ionic lower ice mantle. Changes made to our assigned filled ice thermal expansivity will not change the qualitative picture described. 

When modelling the internal thermal profile one has to properly account for the different phases. Phase changes may cause convection partitioning and the different phases may have different adiabatic slopes and rheologies. Generally a phase diagram is a three dimensional pressure, temperature and composition entity. These 3D diagrams are scarcely known and this is also the case for the carbon-water system. A question may arise whether methane filled ice is stable at high pressure for different water to methane ratios. It is known experimentally that filled ice forms a crystal structure with a $2:1$ water-methane ratio \citep{loveday01}. 
But what would happen in regions that are not so rich in methane? A recent experiment up to $11$GPa shows that if water is much more abundant than methane, the available methane would still prefer to be incorporated in the water matrix as filled ice, while the extra water would simply form ice VII, rather then separating into pure solid methane and ice VII \citep{Ohtani2010}.  

We have explored the dynamic and thermal characteristics of several tectonic modes possibly active in water planets.
The plates considered were treated as isolated systems, although plates on Earth are known to interact. The analysis should therefore be considered as representative of an average plate behaviour.

We find that clathrates and filled ice support the existence of an asthenosphere, whose affect on the lithospheric thermal and dynamic behaviour is large. The theory, as explained in section $5$, demands dividing the ice mantle into three sections: a lower mantle, an asthenosphere and a lithosphere. Each section is assumed to have a constant viscosity. A more realistic viscosity profile with depth (see fig.\,\ref{fig:f11}) would demand a finer division of the mantle. Such a finer division though logical will become appropriate only when better viscosity estimations become available.

Another caveat is that the theory of \cite{Crowley2012} is derived for Newtonian viscosities whereas the viscosities may be non-Newtonian in the filled ice lower mantle. This means the lower mantle viscosity could be stress-dependent. We have made sure our solution is physically consistent by deriving stress profiles in the convection cell. These were averaged and compared with the initial choice for the deviatoric stress tensor. 

In deriving the partial melting under spreading centres in section $5$ we considered methane as the only hydrophobic constituent. In case other clathrate hydrate forming molecules are present the partial melting percentages will likely change. This is because different volatile compositions will change the position of the melt curve along the adiabat of ascending ice and will also change the heat of fusion. However, different volatile compositions are not expected to change the general understanding that fast moving plates require only partial melting to account for the full radioactive budget, whereas partial melting will cause sluggish plates to overheat.

In section $6$ we have estimated the outgassing flux of methane to the atmosphere, in case that is controlled by clathrate hydrate dissociation and methane diffusion. The diffusion coefficient would be highly dependent on the size of the diffusing molecule and its interaction with the water lattice. This implies the outgassing flux is not necessarily representative of the composition of volatiles in the ice matrix. Different compositions of clathrate hydrate forming molecules would also change the depth and latitudinal extent of clathrate hydrate stability.
Therefore the quantitative results of section $6$ may change somewhat when varying the volatile composition, but we do not expect a qualitative change to our conclusions. 
Although our method for estimating the outgassing and global volatile release is approximate it may still teach us about global trends of the system, such as: the outgassing efficiency of the mechanism proposed, its ability to replenish the atmosphere and possible trends in surface-atmosphere interactions. In this respect it is a useful tool to probe the outgassing nature of an "average" cold water planet. Taking the analysis another step forward will require a substantial increase in complexity. One will have to account for tectonic plate dynamics and spatial distribution in addition to exact surface-atmosphere interactions. The first of these is at the moment unobtainable both theoretically and observationally, the second may be resolved using atmospheric observations that will become available in the future and theoretical atmospheric models.

\section{SUMMARY}      
 
Water super Earths are ubiquitous and are now accessible to remote sensing. Interpreting the spectroscopic observations of their atmospheres requires an understanding of the processes of transport of gases in the water rich mantle. As a first step in modelling these processes we have calculated the transport of methane including a new high pressure form for the binary solid solution. 

The main conclusion we reach here is that methane from the interior will reach the atmosphere and provide significant fluxes for a range of conditions. We find that the incorporation of methane changes substantially the structural properties of the water mantle.
The bulk of the mantle stabilizes a high pressure form of ice (filled ice) with different thermal and mechanical properties than those of pure ice VII and other polymorphs. We find that a methane clathrate hydrate subsurface layer can inhibit the formation of a global subterranean ocean, resulting in increased stresses on the crust causing modes of ice plate tectonics. 

An important implication of this work is that the ice mantle dynamics is critical to modelling the composition of the atmosphere. 
Our model predicts a global outgassing rate of: $10^{27}-10^{29}$\,molec\,s$^{-1}$ for a $2$M$_{E}$ planet (see section $6$). We expect other gases capable of forming clathrate hydrates to behave in a similar fashion. Once such multicomponent mixtures of water and volatiles are computed consistently we should be able to interpret future spectroscopic observations of water super Earths.

This work begins by examining the characteristics of a planetary crust composed of methane clathrate hydrate (see section $2$). In addition to the thermal properties of methane clathrate hydrate SI we study its rheology, both Newtonian and non-Newtonian. By drawing a viscosity map for the planetary near surface layer we find that diffusion creep is the adequate solid state creep for describing the crust. For this solid state creep mechanism we test for both stagnant lid and small viscosity contrast solutions.

We find there are five types of crustal regimes, differing in their structure and thermodynamic behaviours (see subsection $2.4$). These regimes depend on the surface temperature and heat flux. One difference between the crustal types is whether methane clathrate hydrate is stable on the planetary surface. If it is not then it underlies a ice Ih layer whose thickness is at most an order of $100$\,m, for our studied planets. Another difference between crustal types is whether the crust conductive thermal profile crosses the clathrate hydrate dissociation curve. If the heat flux is sufficiently low the crust will become unstable to convection within the clathrate hydrate stability field so that a convective cell will form below the conductive crust, in the clathrate hydrate layer. But if the heat flux is high enough, the clathrate hydrate dissociation (i.e. melt) temperature is reached before initiation of convection and the temperature profile begins to follow the hydrate dissociation curve. In this latter scenario the upper part of the crust is conductive while its lower part is set by its on-melt confined behaviour. We show that this effect hinders the formation of a subterranean ocean beginning at shallow depths. 

In section $3$ we describe the assumed internal structure for our studied water planets. The super Earths we study consist of an iron core, a perovskite (MgSiO$_3$) mantle, and an icy outer shell. For the icy mantle we adopt a fine structure of: methane filled ice Ih for the higher end pressures and methane clathrate hydrate (both SI and SH). We consider masses of $2$, $5$, and $10$\,M$_E$ and water fractions of $25\%$ and $50\%$. For the $2$\,M$_E$ planet we allow the water mass fraction to range from $3\%$ to $60\%$.

In section $4$ we derive the thermal profiles spanning the icy mantles of our studied water planets. We find that for all of our studied cases an on-melt layer, underlying the upper conductive part of the crust, is present. This on-melt layer (named DBL in the text) is approximately $1$\,km wide and may be avoided only for very low heat fluxes or due to the absence of clathrate forming volatiles. The conductive part of the crust is found to be no more than about $1$\,km thick (see table $\ref{tab:CrustandDBL}$ for more detail).

Following a adiabatic thermal profile in the methane clathrate hydrate part of the mantle convection cell we find another, deeper, on-melt layer. This layer is confined to the methane clathrate hydrate SH dissociation curve, beginning at a pressure level of approximately $1-1.5$\,GPa. Therefore, under our assumptions, and in accordance with the lower pressure segment of the thermal profile, the presence of methane hinders the formation of a subterranean ocean. This is a general consequence of the topography of the dissociation slope of clathrates. This slope has a "normal" behaviour as opposed to the anomalous melt curve for ice Ih which decreases in temperature for increasing pressure. Also, the dissociation curve gradient, is not as steep as the adiabatic gradient in a liquid water convection cell when higher pressure phases such as SH clathrate can form. We further show that the hindrance of a subterranean ocean due to clathrate formation is far more likely in water planets then in the icy satellites of our solar system. 

We derive the Clausius-Clapeyron curve (see appendix A) for the clathrate to filled ice phase transition and show that no phase change induced partitioning of the ice mantle convection cell is expected. We therefore continue to follow an adiabat from the clathrate and into the filled ice part of the mantle, down to the thermal boundary layer separating the ice and silicate mantles. The characteristics of this bottom boundary layer are studied using convective instability analysis. For this aim we have derived the viscosity and thermal conductivity for a highly pressurized ($\sim 100$\,GPa) solid solution of water and methane at high temperature ($\sim 1000$\,K).
   
We find the higher thermal expansivity of the filled ice, compared with that for water ice VII, results in shallower adiabtic profiles and hotter interiors, compared to those derived for a pure water mantle (see fig.\ref{fig:f9}). This effect may promote conditions in the lower ice mantle that are within the stability field of the super-ionic and reticulating phases (see discussion in section $4$). This is for water planets much less massive ($5-10$\,M$_E$) than Uranus or Neptune and that lack a substantial H/He atmosphere. The consequences of these phases on the lower mantle volatile composition and their transport to the atmosphere are left for future work. For the case of the $2$\,M$_E$ water planet we find it likely that its entire icy mantle is confined to the molecular solid filled ice phase.  

Our results suggest a relatively large stress on the lithosphere can be supported, thus activating modes of ice plate tectonics. In section $5$ we map the dynamic characteristics of the possible tectonic modes for a $2$\,M$_E$ water planet and for two ice mass fractions of $25\%$ and $50\%$. The multiple solution nature of tectonics (see discussion in section $5$) implies that the planetary evolution must be traced in order to determine the active tectonic mode. We suggest that the different tectonic modes may result in different methane outgassing regimes and may therefore be observationally distinguishable.

Our estimates for the viscosity throughout the ice mantle indicate that a low viscosity layer exists between the planetary lithosphere and lower mantle. We refer to this layer as the asthenosphere and argue it should probably be a common feature in water planets due to the affect volatiles have on the thermal profile and solid ice phase diagram. The asthenospheres of our studied planets will have viscosities at least three orders of magnitude lower than that of the lower mantle. The impact of this low viscosity mid-layer on the tectonic modes that may develop is prominent.
Based on the rheological profiles we derive for the crystals composing our studied planets we set the boundary between the asthenosphere and the lower mantle at the clathrate to filled ice phase transition (pressure level of approximately $2$\,GPa).
We further find that a very low asthenospheric viscosity may promote partitioning of the ice mantle convection cell though we suggest a mechanism that can counteract this effect.

For this three-layered ice mantle model: having a lithosphere, an asthenosphere and lower mantle, we calculate the dynamical characteristics of plate tectonics. This is done for different assumed ratios of the asthenospheric viscosity to lower mantle viscosity (ratio ranging $10^{-6}-10^{-3}$). By analysing the forces acting on an icy tectonic plate we find that the energy dissipation due to plate bending at subduction is not enough to counteract the effect of slab pull, which promotes plate motion. This is due to the lower lithospheric viscosity of the icy plate compared to the viscosity of a silicate plate. 
We find that the tectonic plate speeds for the case of a frozen water planet are higher than for the plates on Earth. Sluggish plates have speeds of order $1$\,cm\,yr$^{-1}$ and fast plates have speeds in the range of $1-10$\,m\,yr$^{-1}$, where higher ice mass fractions have faster plates (see tables $\ref{tab:PlateTectonic25}$ and $\ref{tab:PlateTectonic50}$ ).

The low viscosity asthenosphere reduces the stress applied on the base of the lithosphere by the underlying convection cell. So much so that breaking of the lithosphere into the thicker sluggish plates may encounter some difficulty in case the lithosphere experiences a large influx of silicate dust. No such difficulty exists for the fast, and thin, plates. 
The asthenosphere, representing a deviation from isoviscosity, yields large differences between the mantle overturn time scale and the planetary resurfacing time scale. Relaxing the assumption of isoviscosity for the icy mantle may result in an overturn time scale in the order of $100$\,Ma (see table $\ref{tab:TimeScale}$). This is an order of magnitude larger than the overturn time scale derived assuming an isoviscous model. The derived overturn times depend on the ice mass fraction and asthenospheric to lower mantle viscosity ratio (see table $\ref{tab:TimeScale}$). The resurfacing time scale, for the fast tectonic plate solution, is in order of $1$\,Ma for viscosity contrasts down to $10^{-5}$ and as low as $0.1$\,Ma for a viscosity contrast of $10^{-6}$. A resurfacing time of $100$\,Ma is found for the sluggish tectonic plate solution.  
 
The conductive cooling across the lithosphere, for the $2$\,M$_E$ planet and both the $25\%$ and $50\%$ ice mass fractions, is not sufficient to remove the planetary radiogenic heating. For the purpose of estimating the conductive cooling through the plates we account for plate thickening with age and discuss the possibility of hot spots as a mechanism responsible for keeping the plate thin with age.
We quantify the percent of melt required, beneath spreading centres, in order to account for the difference between the radiogenic budget and the ability of the lithospheric plates to cool conductively. We find that a few percent melt is enough for the fast tectonic plate solutions, and so these solutions may very well over-cool the planet resulting in a thickening of the lithosphere. In turn, this thickening will force a tectonic mode change into the sluggish plate solution resulting in over-heating and lithospheric thinning and vice verse (see table $\ref{tab:MeltFraction}$ for more detail). The larger the planetary ice mass fraction the less partial melting is needed.

In section $6$ we explore the relation between the dynamics of the active tectonic mode and the outgassing flux of methane to the atmosphere. We propose a mechanism where during plate resurfacing fresh methane clathrate hydrate is exposed to the planetary surface conditions, followed by its dissociation. We show that even when one considers self porosity in the upper most layer of the plate, still, diffusional limitations will not allow all the methane molecules liberated from the dissociated clathrate cages to actually reach the atmosphere. We calculate the concentration of methane as a function of time and depth in the upper part of the planetary crust and from that we obtain the surface flux of methane to the atmosphere. We find, assuming a surface temperature in the range of $200$\,K-$250$\,K, the initial methane flux at the ridge is about $10^{18}$\,molec\,cm$^{-2}$\,s$^{-1}$ and decreasing with ageing of the crust (see fig.\ref{fig:f18}). A destabilized methane clathrate hydrate column, newly exposed to the surface conditions, will exhaust its methane supply to the atmosphere in a time depending on the depth of clathrate hydrate destabilization. For example, a $10$\,m column will require $1$\,Ma to become exhausted, in order of the resurfacing time scale. We further formulate the outgassing flux of methane as a function of latitude and longitude across the planetary surface. From that we derive the global rate of methane release into the atmosphere. 
We map the global rate of methane release as a function of atmospheric surface pressure and equatorial (assumed maximal) temperature (see fig.\ref{fig:f21}). For example, we show that for a maximal surface temperature of $250$\,K (temperature decreasing towards poles) and atmospheric surface pressure up to $1$\,bar the global rate may reach $10^{29}$\,molec\,s$^{-1}$. We also demonstrate, qualitatively, how the outgassing mechanism may vary between stable and runaway-like due to its influence on the surface temperature, i.e. the greenhouse effect. 
Assuming the outgassing does not alter substantially the surface temperature we integrate the outgassing rate over time, once assuming an equatorial temperature of $200$\,K and once $230$\,K. We find that after $1$\,Ga a methane atmosphere of $1.0$\,bar and $4.4$\,bar is outgassed for the lower and higher equatorial temperatures, respectively. Methane is therefore expected to be a significant component in water planets' atmospheres.

\section{ACKNOWLEDGEMENTS}

We convey our thanks to Prof. Richard J. O'Connell for helpful conversations. We also thank the anonymous referee for his constructive suggestions. This work was supported by the Origins of Life Initiative at Harvard University.

\section{APPENDIX}

\subsection{NOMENCLATURE}

\begin{deluxetable}{ll}
\tablecolumns{2}
\tablewidth{0pc}
\tablecaption{Nomenclature}
\tablehead{
\colhead{Symbol} & \colhead{Physical quantity} }
\startdata
$A$                 &      pre-exponential factor in the viscosity \\
$a$                 &      crystal grain diameter \\
$a_{cl}$            &      methane clathrate crystal grain diameter \\
$a_{cavity}$        &      radius of the water made hydration cell (i.e. cavity) \\
$a_{cage}$          &      radius of the water made cage in the solid phase that traps methane \\
$\textbf{b}$        &      Burgers vector \\
$B$                 &      bulk modulus \\
$B_{FI}$            &      bulk modulus of methane filled ice Ih \\
$B_{cl}$            &      bulk modulus of methane clathrate \\
$B^w_{liq}$         &      bulk modulus of liquid water \\
$B^{FI}_{eff}$      &      effective bulk modulus of a lattice vacancy in methane filled ice Ih \\
$\tilde{B}$         &      pressure derivative of the bulk modulus \\
$\tilde{B}_{cl}$    &      pressure derivative of the bulk modulus of methane clathrate \\
$\tilde{B}^w_{liq}$ &      pressure derivative of the bulk modulus of liquid water \\
$\tilde{B}_{FI}$    &      pressure derivative of the bulk modulus of methane filled ice Ih \\
$B_U$               &      fitted parameter for the Umklapp phonon scattering mechanism \\
$C^{cl}_p$          &      isobaric heat capacity of methane clathrate \\
$C_p$               &      isobaric heat capacity \\
$C_{p,bbl}$         &      isobaric heat capacity at the boundary layer between the water and \\ 
                    &      silicate mantles \\
$C_{p,l}$           &      isobaric heat capacity of the lithosphere \\
$C_{p,total}$       &      averaged isobaric heat capacity representing the entire convective cell \\
                    &      in the ice mantle \\
$C$                 &      factor often used in formulating the viscosity (defined in eq.$\ref{strainrate}$) \\
$C_{FI}$            &      value of $C$ for methane filled ice Ih \\
$C_{cl}$            &      value of $C$ for methane clathrate \\
$C^{FI}_{Ki}$       &      Langmuir constant of a $K$ type guest molecule in an $i$ type cage in \\
                    &      methane filled ice Ih \\
$C^{Clath}_{Ki}$    &      Langmuir constant of a $K$ type guest molecule in an $i$ type cage in \\
                    &      clathrate hydrate \\
$C_s$               &      sound speed \\
$c$                 &      concentration of methane in the upper crust \\
$c_{atm}$           &      atmospheric concentration of methane near the planetary surface \\
$c_0$               &      concentration of methane in the upper crust when newly formed \\
                    &      at ridge \\
$d$                 &      convective length scale \\
$d_{bbl}$           &      radial dimension of the boundary layer between the water and \\
                    &      silicate mantles \\
$d_l$               &      maximal thermal thickness of the lithosphere \\
$d_{total}$         &      depth scale of the entire convective cell in the ice mantle \\
$D$                 &      lattice diffusion coefficient \\
$D_{CH_4}$          &      diffusion coefficient of methane in its clathrate \\
$D_{H_2O}$          &      water self diffusion coefficient in a clathrate lattice \\
$D^{(SI)}_{ave}$    &      weighted lattice diffusion coefficient for methane clathrate \\
$\tilde{D}$         &      diffusion coefficient of methane in a ice Ih matrix formed from \\
                    &      dissociated clathrate \\
$D_M$               &      depth of the planetary water layer \\
$E^*$               &      activation energy in viscosity \\
$E^*_{cl}$          &      activation energy in viscosity for methane clathrate \\
$E1$                &      energy barrier for creating an opening in the hydrogen bonded \\
                    &      network of liquid water \\
$E2$                &      energy barrier for methane thermal jumping into the liquid water \\
                    &      bulk through an opening in the latter hydrogen bonded network \\
$E3$                &      energy of a hydrophobic solute molecule encapsulation in a liquid \\
                    &      water hydration cell \\                   
$\dot{e}_{ij}$      &      strain rate tensor \\
$F$                 &      heat flux entering a convection cell \\
$F_s$               &      planetary surface heat flux \\
$F_{s,planet}$      &      planetary surface heat flux due to radioactive decay, scaled using \\
                    &      data from Earth\\
$F_{s,Earth}$       &      surface heat flux for Earth \\
$F_{bbl}$           &      heat flux entering the boundary layer between the water \\
                    &      and silicate mantles \\
$F_{plate}$         &      surface heat flux as allowed by plate tectonic conduction \\
$F_{in}$            &      flux of methane molecules impinging on the liquid water \\
                    &      phase and dissolving within it \\
$F_{out}$           &      flux of methane molecules dissolved in water that are \\
                    &      liberated from the solution \\
$f_K$               &      fugacity of substance composed of type $K$ molecules \\
$g$                 &      acceleration of gravity \\
$g_{s}$             &      planetary surface acceleration of gravity \\
$g_{s,E}$           &      Earth's surface acceleration of gravity \\
$g_{bbl}$           &      acceleration of gravity at the boundary layer between the \\
                    &      water and silicate mantles \\
$H^*$               &      activation enthalpy \\
$H^*_{cl}$          &      activation enthalpy for methane clathrate \\
$H^{\beta,FI}_{H_2O}$   &  enthalpy per water molecule in the empty methane filled ice \\
                    &      Ih crystal \\
$H^{\beta,Clath}_{H_2O}$&  enthalpy per water molecule in the empty methane clathrate \\
                    &      crystal \\
$H$                 &      atmospheric scale height \\ 
$h_{slab}$          &      length of slab attached to the lithospheric plate \\
$h$                 &      depth from planetary surface where methane clathrate \\
                    &      becomes stable \\
$j$                 &      flux of methane in the upper crust toward the atmosphere \\
$j_{surface}$       &      flux of methane molecules from the planetary surface \\
$k$                 &      Boltzmann's constant \\ 
$L_{ex,sI}$         &      linear thermal expansivity for a SI clathrate \\
$L_{ex}$            &      linear thermal expansivity \\
$L$                 &      length of the lithospheric plate \\
$L_{GR}$            &      global ridge length \\
$m$                 &      Burgers vector over grain size exponent (defined in eq.$\ref{strainrate}$) \\
$m_{H_2O}$          &      mass of a water molecule \\
$m_{CH_4}$          &      mass of a methane molecule \\
$m_{guest}$         &      mass of a guest molecule entrapped in the water ice lattice \\
$M_{p}$             &      planetary mass \\
$M_{E}$             &      mass of planet Earth \\
$\dot{M}$           &      total rate of clathrate mass crossing the solidus under a ridge \\
$n$                 &      stress exponent in viscosity \\
$n_{cl}$            &      stress exponent in viscosity for methane clathrate \\
$n_{FI}$            &      stress exponent in viscosity for methane filled ice Ih \\
$\hat{n}$           &      power of temperature (eq.$\ref{ThermalConductivityClathrate}$) \\
$n_g$               &      number density of methane molecules in the gas phase \\
$n_{cavity}$        &      number density of hydration cells (i.e. cavities) that can \\
                    &      encapsulate a methane molecule in the aqueous solution \\
$n_{H_2O}$          &      number density of water molecules in the liquid phase \\
$\tilde{n}$         &      number density of resonating guest molecules which scatter \\
                    &      phonons in the water ice crystal \\
$N_{cavity}$        &      number of hydration cells (i.e. cavities) capable of \\
                    &      encapsulating a methane molecule  \\
$N_{H_2O}$          &      number of water molecules \\
$N_{CH_4}$          &      number of methane molecules dissolved in aqueous solution \\
$N^{atm}_{CH_4}$    &      number of methane molecules in the planetary atmosphere \\                 
$N_{plate}$         &      number of plates around the equatorial circumference \\
                    &      outgassing methane \\                   
$P$                 &      pressure \\
$P0$                &      reference pressure \\
$P_{b,crust}$       &      pressure at the bottom of the planetary crust \\
$P_s$               &      pressure at the planetary surface \\ 
$P_{Si-H_2O}$       &      silicate and water mantle boundary pressure \\
$P_{center}$        &      pressure at the planetary center \\
$P_{Fe-Si}$         &      iron core and silicate mantle boundary pressure \\
$P_{b,DBL}$         &      pressure at the base of the dissociation boundary layer \\
$P^{up}_{bbl}$      &      pressure at the outer boundary of the bottom \\
                    &      boundary layer (water to silicate transition) \\
$P^{CH_4}_{dis}$    &      pressure at the boundary between the near surface \\
                    &      ice Ih layer and deeper SI methane clathrate hydrates\\                
$Pr$                &      probability a methane molecule impinging on liquid \\
                    &      water will become dissolved \\                     
$\left\langle Q_{adv}\right\rangle_l$ & advective rate of heat transfer through a horizontal cross \\
                    &      section of the lithosphere averaged along the lithospheric depth\\
$\left\langle Q_{adv}\right\rangle_{total}$ & advective rate of heat transfer through a horizontal cross \\
                    &      section of the ice mantle averaged along the ice mantle depth\\
$Q_{deficiency}$    &      global energy rate difference between that released due to \\
                    &      radioactive decay and that conducted away from the \\
                    &      lithospheric plates\\                     
$Q$                 &      total number of methane molecules that enter the \\
                    &      atmosphere per unit time \\                    
$\tilde{q}$         &      number of methane molecules released to the atmosphere \\
                    &      in a given time from \\ 
                    &      a unit planetary surface as a function of $h$\\                     
$Ra$                &      Rayleigh number \\
$Ra_{crit}$         &      critical Rayleigh number \\
$Ra_{bbl}$          &      Rayleigh number of the boundary layer between the water \\
                    &      and silicate mantles \\
$R$                 &      Kihara core dimension for a methane molecule \\
$R_{E}$             &      Earth's radius \\
$R_{p}$             &      planetary radius \\
$R_{core}$          &      iron core radius \\
$R_{Si-H_2O}$       &      distance from the planetary center to the water mantle \\
$R_{curv}$          &      radius of curvature of the bent lithosphere at subduction \\
$\Re$               &      Fraction of internal ice mantle methane that can redistribute \\
                    &      and enrich the upper ice mantle \\ 
$S$                 &      convection cell surface area through which the heat enters \\
$t$                 &      time \\
$t_{overturn}$      &      time to convect material from the bottom to \\
                    &      the surface of the ice mantle \\
$t_{resurface}$     &      time for the renewal of a tectonic plate \\
$t_{isovis}$        &      time scale of convection assuming isoviscosity \\
$t_c$               &      convergence time criterion \\
$t_{atm}$           &      time scale for establishing the planetary atmosphere \\
$T$                 &      temperature \\
$T_m$               &      melting temperature \\ 
$T_{m,cl}$          &      melting temperature for methane clathrate, \\
                    &      the hydrate-aqueous solution-vapour three phase \\
$T_{b,crust}$       &      temperature at the base of the planetary crust \\
$T_s$               &      temperature at the planetary surface \\
$T_{ad}$            &      adiabatic temperature characterizing the convecting \\
                    &      sub-layer \\ 
$\bar{T}$           &      mid-layer temperature used in the small viscosity \\
                    &      contrast approximation \\
$T_{Si-H_2O}$       &      temperature at the boundary between the water and \\
                    &      silicate mantles \\ 
$T^{FI}_{ad}$       &      adiabatic temperature profile in the methane filled ice \\
                    &      convection cell \\
$\bar{T}_{bbl}$     &      average temperature at the boundary layer between the \\
                    &      water and silicate mantles \\
$T_h$               &      temperature at the boundary between the near surface \\
                    &      ice Ih layer and deeper \\ 
                    &      methane clathrate hydrates \\                  
$T_{equator}$       &      average temperature around the equator of the planet \\
$\tilde{T}$         &      reference temperature (eq.$\ref{ThermalConductivityClathrate}$) \\
$u$                 &      solid state creep velocity \\
$U_p$               &      horizontal speed of the lithospheric plate \\
$V^*$               &      activation volume in viscosity \\
$V^*_{cl}$          &      activation volume in viscosity for methane clathrate \\
$V^*_{0,FI}$        &      activation volume in viscosity for methane filled \\
                    &      ice Ih \\
$V_M$               &      maximal vertical flow velocity in the water mantle \\
                    &      convection cell \\
$V_{sim}$           &      volume of the internal solid ice mantle \\
$V_{lwl}$           &      volume of the subterranean ocean that would have \\
                    &      existed in case it was not completely enclathrated \\
$v$                 &      macroscopic volume \\
$v_m$               &      molecular volume of bulk solvent, derived from bulk \\
                    &      parameters \\
$v_{cavity}$        &      hydration cell (i.e. cavity) average velocity \\
$\bar{v}$           &      average molecular thermal velocity in the gaseous phase \\  
$\tilde{v}$         &      scale of atomic or molecular volume \\
$\tilde{v}_{CH_4}$  &      volume per methane molecule \\
$\tilde{v}^{fluid}_{CH_4}$  &      volume per methane molecule in a fluid phase \\
$\tilde{v}^*_{CH_4}$  &      hypothetical volume per methane molecule, \\
                    &      often estimated at infinite dilution \\
$\tilde{v}^{\beta,FI}_{H_2O}$    & volume per water molecule in the empty methane \\
                    &      filled ice Ih crystal \\
$\tilde{v}^{\beta,Clath}_{H_2O}$ & volume per water molecule in the empty methane \\
                    &      clathrate crystal \\
$v_{isovis}$        &      convective velocity in a cell assuming isoviscosity \\
$V_{ascent}$        &      speed of mass ascent across the solidus \\
$W$                 &      spreading center width at the depth of the solidus \\
$\left\langle W^{FI}_{CH_4}\right\rangle$ & spatially averaged potential energy of a methane \\
                    &      molecule in the filled ice Ih water lattice \\
$\left\langle W^{Clath}_{i,CH_4}\right\rangle$ & spatially averaged potential energy of a methane \\
                    &      molecule in a $i$ type cage of \\                   
                    &      a methane clathrate \\ 
$W_{cc}$            &      energy required to create a hydration cell (i.e. cavity) in \\
                    &      liquid water \\ 
$W_{gain}$          &      potential of interaction between a solute molecule \\
                    &      and its surrounding solvent \\                   
$X_{CH_4}$          &      fraction of unoccupied cages in methane clathrate \\
$\tilde{X}_{CH_4}$  &      mole fraction of methane in solution with water \\    
$X^{Si+Fe}_p$       &      planetary mass fraction of silicates and metals \\
$X_{melt}$          &      melt fraction of mass crossing the solidus \\
$x$                 &      horizontal distance from ridge \\
$y^{FI}_{Ki}$       &      probability a $K$ type molecule occupies a type $i$ \\
                    &      cage in filled ice Ih \\                                    
$y^{Clath}_{Ki}$    &      probability a $K$ type molecule occupies a type $i$ \\
                    &      cage in clathrate \\                                    
$z$                 &      depth coordinate in the crust \\
$Z$                 &      number of water molecules forming the hydration \\
                    &      cell around a methane molecule \\
$Z_{cage}$          &      number of water molecules in the solid phase that \\
                    &      form the cage that traps methane \\
$\bar{Z}_i$         &      methane to water ratio by number, as accreted by \\
                    &      the water planet \\
$Z_{cs}$            &      Averaged methane to water ratio by number \\
                    &      required to stabilize SI and SH methane hydrate \\ 
$\alpha$            &      thermal diffusivity \\
$\alpha_{cl}$       &      methane clathrate thermal diffusivity \\
$\alpha_{bbl}$      &      thermal diffusivity at the boundary layer between \\
                    &      the water and silicate mantles \\
$\alpha_l$          &      thermal diffusivity of the lithosphere \\
$\alpha_{total}$    &      averaged thermal diffusivity representing the entire \\
                    &      convective cell in the ice mantle \\
$\hat{\alpha}$      &      fitted parameter for the Umklapp phonon scattering \\
                    &      mechanism \\
$\Gamma_{cc}$       &      gradient of the Clausius-Clapeyron curve \\
$\gamma_{CH_4}$     &      activity coefficient of methane in solution with \\
                    &      water \\ 
$\Delta T$          &      temperature difference driving the convection \\
$\Delta T_{reo}$    &      rheological temperature difference \\
$\Delta T_{bbl}$    &      temperature difference across the boundary layer \\
                    &      between the water and silicate mantles\\
$\Delta T_l$        &      temperature difference across the lithosphere \\
$\Delta H$          &      energy required to dissociate a unit mass of \\
                    &      methane clathrate \\
$\delta$            &      thermal boundary layer \\
$\delta_{crust}$    &      radial dimension of the planetary crust \\
$\delta_{DBL}$      &      radial dimension of the layer confined to the \\
                    &      methane clathrate dissociation curve\\
$\epsilon$          &      Kihara energy parameter between methane and \\
                    &      water \\
$\epsilon_{LJ}$     &      Lennard-Jones energy parameter between \\
                    &      methane and water \\
$\eta$              &      chemical potential \\
$\eta^{FI}_{H_2O}$  &      chemical potential of water in methane filled ice Ih \\
$\eta^{Clath}_{H_2O}$        &   chemical potential of water in methane clathrate \\
$\eta^{FI}_{CH_4}$  &      chemical potential of methane in methane filled \\
                    &      ice Ih \\
$\eta^{Clath}_{CH_4}$        &   chemical potential of methane in methane clathrate \\
$\eta^{pure}_{CH_4}$         &   chemical potential of homogeneous methane bulk \\
$\eta^{\beta,FI}_{H_2O}$     &   chemical potential of water in an empty methane \\
                    &      filled ice Ih crystal \\
$\eta^{\beta,Clath}_{H_2O}$  &   chemical potential of water in an empty methane \\
                    &      clathrate crystal \\
$\eta^{fluid}_{CH_4}$        &   chemical potential of methane in its fluid phase \\
$\eta^{solution}_{CH_4}$     &   chemical potential of methane in solution with water \\
$\eta^*_{CH_4}$     &      reference chemical potential for methane \\   
$\theta$            &      logarithm of the ratio of viscosities across \\
                    &      the cold boundary layer \\
$\Theta_D$          &      Debye temperature \\ 
$\kappa_{cl}$       &      thermal conductivity of methane clathrate \\
$\kappa$            &      thermal conductivity \\
$\kappa_{bbl}$      &      thermal conductivity at the boundary layer between \\
                    &      the water and silicate mantles \\
$\kappa_l$          &      thermal conductivity of the lithosphere \\
$\kappa_{Ih}$       &      thermal conductivity of ice Ih \\                   
$\tilde{\kappa}$    &      reference thermal conductivity (eq.$\ref{ThermalConductivityClathrate}$) \\
$\Lambda$           &      Poisson ratio \\
$\lambda$           &      latitude \\
$\lambda_0$         &      latitudinal domain where methane clathrate is \\
                    &      unstable on the planetary surface \\                          
$\hat{\mu}$         &      shear modulus \\
$\mu$               &      dynamic viscosity \\
$\mu_A$             &      dynamic viscosity of the asthenosphere in the ice \\
                    &      mantle \\
$\mu_M$             &      dynamic viscosity of the lower ice mantle; \\
                    &      total ice layer$-$lithosphere$-$asthenosphere \\
$\mu_L$             &      dynamic viscosity of the lithosphere \\
$\nu$               &      kinematic viscosity \\
$\nu_{cl}$          &      methane clathrate kinematic viscosity \\
$\nu_{FI}$          &      methane filled ice kinematic viscosity \\
$\nu_{bbl}$         &      kinematic viscosity at the boundary layer between \\
                    &      the water and silicate mantles \\
$\nu^{FI}_i$        &      ratio of type $i$ cages to water molecules in filled \\
                    &      ice Ih \\
$\nu^{Clath}_i$     &      ratio of type $i$ cages to water molecules in SI \\
                    &      methane clathrate \\  
$\xi$               &      pressure exponent in $\chi$ (see eq.$\ref{ExpansivityGeneral}$) \\
$\hat{\xi}$         &      aspect ratio of the convection cell in the water \\
                    &      mantle \\
$\tilde{\xi}$       &      solvent volume packing efficiency \\
$\rho_{cl}$         &      bulk mass density of methane clathrate \\
$\rho_{FI}$         &      bulk mass density of methane filled ice Ih \\
$\rho^w_{liq}$      &      bulk mass density of liquid water \\
$\rho$              &      bulk mass density \\
$\rho_{bbl}$        &      bulk mass density at the boundary layer \\
                    &      between the water and silicate mantles \\
$\rho_{Ih}$         &      bulk mass density of ice Ih \\
$\rho_m$            &      Kihara length parameter between methane \\
                    &      and water \\
$\rho_{sim}$        &      averaged bulk mass density of the internal \\
                    &      solid ice mantle \\ 
$\rho_{lwl}$        &      averaged bulk mass density of the liquid \\
                    &      water layer that would have existed in case \\
                    &      it was not completely enclathrated \\ 
$\sigma$            &      horizontal stress applied on a vertical cross \\
                    &      section of the lithosphere \\
$\sigma_{tens}$     &      tensile strength of the lithosphere \\
$\sigma^{hs}_{CH_4}$&      hard sphere diameter of a methane molecule \\
$\sigma^{hs}_{H_2O}$&      hard sphere diameter of a water molecule \\
$\sigma_{LJ}$       &      Lennard-Jones length parameter between \\
                    &      methane and water \\
$\tau_{ij}$         &      deviatoric stress tensor \\
$\tau$              &      second invariant of the deviatoric stress tensor \\
$\tau_p$            &      shear stress operating on the base of the \\
                    &      lithospheric plate \\
$\tau_R$            &      net resistive stress on the lithospheric plate \\
$\tau_{bend}$       &      effective stress associated with plate bending \\
                    &      at subduction \\
$\tau_F$            &      fault stress from interaction with overlying \\
                    &      plate at subduction \\
$\tau_{sp}$         &      stress on a lithospheric plate associated with \\
                    &      slab pull \\
$\tau_{total}$      &      averaged total relaxation time for phonon scattering \\ 
$\tau_U$            &      Umklapp phonon scattering relaxation time \\
$\tau_{Res}$        &      resonance relaxation time for phonon scattering \\
                    &      due to methane vibrations in the water ice crystal \\
$\phi$              &      longitude \\
$\Phi_M$            &      kinetic energy dissipation in the lower mantle \\
                    &      convective cell \\  
$\chi$              &      volume thermal expansivity \\
$\chi_{cl}$         &      methane clathrate volume thermal expansivity \\
$\chi^w_{liq}$      &      liquid water volume thermal expansivity \\
$\chi_{bbl}$        &      volume thermal expansivity at the boundary layer \\
                    &      between the water and silicate mantles\\
$\chi_l$            &      volume thermal expansivity of the lithosphere \\
$\chi_{total}$      &      averaged volume thermal expansivity representing \\
                    &      the entire convective cell in the ice mantle \\
$\psi$              &      coefficient describing water-methane interaction \\
                    &      for the phonon resonant scattering model \\ 
$\Omega$            &      dimensionless constant relating $H^*$ to $T_m$, \\
                    &      defined in eq.$\ref{OmegaHomologous}$ \\  
$\Omega_{cl}$       &      value of $\Omega$ for methane clathrate \\
$\omega$            &      frequency \\ 
$\omega_D$          &      Debye frequency \\
$\omega_0$          &      guest molecule translational vibration frequency \\
$\hbar$             &      Planck constant over $2\pi$ \\
                       
\enddata
\tablecomments{\footnotesize{}}
\label{tab:Nomenclature}
\end{deluxetable}

\subsection{APPENDIX A: PHASE CHANGE INSTABILITY ANALYSIS} 

The stability of a convecting region with a phase transition is discussed in \cite{schubert1971}.  An important factor is the sign of the gradient of the Clausius-Clapeyron curve, $\Gamma_{cc}$.  A descending slab reaching a phase transition curve with $\Gamma_{cc}>0$ will reach the phase transition before the surrounding mantle material, since it started with an anomalously lower temperature. The slab will experience a phase change induced density increase relative to its surrounding, invigorating its descent. This phenomenon is known as the phase boundary distortion. Since the process will be exothermic the slab will be heated and thus will expand, thereby lowering its density, which will act to inhibit further descent.  The opposite will happen when a descending slab reaches a phase transition with $\Gamma_{cc}<0$. 

It was shown \citep[see e.g.][]{christensen1985,olson1982} that the phase boundary distortion is the dominant effect. Therefore, a positive $\Gamma_{cc}$ phase transition will encourage further instability (i.e. prevent cell partitioning), whereas a negative $\Gamma_{cc}$ phase transition, beyond a critical value, may result in a leaking partitioned cell.  There is not enough experimental data to deduce $\Gamma_{cc}$ for the classical clathrate and filled-ice Ih phase transition. We shall therefore try to estimate its value on theoretical grounds.

We view the phase transition as one between two phases of a water and methane solution, where water is the solvent and methane is the solute. At the phase boundary, equilibrium demands an equality of chemical potentials, $\eta$, between the water in both phases and methane in its phases. Since more methane (per water molecules) may be dissolved in filled ice-Ih than in classical cage clathrates we need to consider the chemical potential for pure methane as well. The following equations should then be obeyed along the phase curve:  
$$
\eta_{H_2O}^{FI}=\eta_{H_2O}^{Clath}
$$
\begin{equation}
\eta_{CH_4}^{FI}=\eta_{CH_4}^{Clath}=\eta_{CH_4}^{pure}
\end{equation}
where the upper indices $FI$ and $Clath$ refer the value to either filled ice-Ih or a classical cage clathrate respectively.
Using the theory of solutions and the statistical mechanical model for clathrates, derived by \cite{waalplat}, the last set of equations may be formulated as:
$$
\eta_{H_2O}^{\beta,FI}+kT\sum_i\nu_i^{FI}\ln\left(1-\sum_Ky_{Ki}^{FI}\right)=\eta_{H_2O}^{\beta,Clath}+kT\sum_i\nu_i^{Clath}\ln\left(1-\sum_Ky_{Ki}^{Clath}\right)
$$
$$
y_{Ki}^{FI}=\frac{f_KC_{Ki}^{FI}}{1+\sum_Jf_JC_{Ji}^{FI}}
$$
\begin{equation}
y_{Ki}^{Clath}=\frac{f_KC_{Ki}^{Clath}}{1+\sum_Jf_JC_{Ji}^{Clath}}
\end{equation}
where the index $\beta$ refers the parameter to the empty solvent crystal. Here $k$ is Boltzmann's constant, $\nu_i$ is the hydration number, which is the ratio between number of type $i$ cages to number of water molecules in a crystal unit, $y_{Ki}$ is the probability of finding a $K$ type guest molecule in an $i$ type cage, $f_K$ is the fugacity of the $K$ type molecules pure material and $C_{Ki}$ is the Langmuir constant of a type $K$ guest molecule in an $i$ type cage.

If we assume a single type of guest molecule, taken to be methane, and a single type of opening in the filled ice-Ih structure, the last set of equations may be written as:
$$
\frac{\eta_{H_2O}^{\beta,FI}}{kT} +\nu^{FI}\ln\left(1-\frac{f_{CH_4}C_{CH_4}^{FI}}{1+f_{CH_4}C_{CH_4}^{FI}}\right) =              
$$
\begin{equation}
 \frac{\eta_{H_2O}^{\beta,Clath}}{kT}+\sum_i\nu_i^{Clath}\ln\left(1-\frac{f_{CH_4}C_{i,CH_4}^{Clath}}{1+f_{CH_4}C_{i,CH_4}^{Clath}}\right)
\end{equation}
To obtain the Clausius-Clapeyron curve we need to differentiate the last equation, keeping in mind that both the fugacity of methane and its Langmuir constants are functions of both the pressure and the temperature. For the algebraic procedure and the appropriate substitutions required we refer the reader to paper I where a similar procedure was used to calculate the phase boundary between filled ice-Ih and water ice VII. We get:
$$
\Gamma_{cc}\equiv\frac{dP}{dT}=
$$
\begin{equation}\label{ClausiusClapeyron}
\frac{-\frac{\Delta H^*}{T}-\frac{1}{T}\left(\nu^{FI}\left\langle W_{CH_4}^{FI}\right\rangle-\sum_i\nu_i^{Clath}\left\langle W_{i,CH_4}^{Clath}\right\rangle\right)-k\left(\sum_i\nu_i^{Clath}-\nu^{FI}\right)}{-\Delta \tilde{v}+\tilde{v}_{CH_4}\left(\nu^{FI}-\sum_i\nu_i^{Clath}\right)-\nu^{FI}\left\langle\left(\frac{\partial W_{CH_4}^{FI}}{\partial P}\right)_T\right\rangle+\sum_i\nu_i^{Clath}\left\langle\left(\frac{\partial W_{i,CH_4}^{Clath}}{\partial P}\right)_T\right\rangle}
\end{equation}
where we have defined:
$$
\Delta H^*\equiv H_{H_2O}^{\beta,FI}- H_{H_2O}^{\beta,Clath}
$$
\begin{equation}
\Delta \tilde{v}\equiv \tilde{v}_{H_2O}^{\beta,FI}- \tilde{v}_{H_2O}^{\beta,Clath}
\end{equation}
In the last set of equations, $H_{H_2O}^{\beta,FI}$ and $H_{H_2O}^{\beta,Clath}$ are the $\beta$ phase enthalpies, per water molecule, of the filled ice-Ih and cage clathrate empty structures respectively. $\tilde{v}_{H_2O}^{\beta,FI}$ and $\tilde{v}_{H_2O}^{\beta,Clath}$ are the volumes, per water molecule, in the two empty solvent structures. $\tilde{v}_{CH_4}$ is the volume per methane molecule, $\left\langle W_{CH_4}^{FI} \right\rangle$ and  $\left\langle W_{i,CH_4}^{Clath}\right\rangle$ are the spatially averaged potential energies of methane within its water surrounding in the filled ice and type $i$ clathrate cage respectively. $\left\langle\left({\partial W_{CH_4}^{FI}}/{\partial P}\right)_T\right\rangle$ and $\left\langle\left({\partial W_{i,CH_4}^{Clath}}/{\partial P}\right)_T\right\rangle$ are the partial derivatives with respect to pressure of the potential energies just mentioned.

For Eq.\,(\ref{ClausiusClapeyron}) we note that filled ice-Ih is a distortion on the water ice Ih crystal \citep{loveday01}. We can thus approximate the enthalpy of the former using values for the latter. The difference in enthalpy between ice Ih and methane clathrate hydrate is known from experiment to give $-2.3062\times 10^{-14}$~erg\,molec$^{-1}$ \citep{haghighi09}.  For a description of how the potential energy of a methane molecule in a cage of a classic clathrate behaves we refer the reader to \cite{mckoysinan}. From their work we deduce potential well depths of  $-4.502\times 10^{-13}$ and $-3.832\times 10^{-13}$~erg\,molec$^{-1}$ for the small and large cage respectively. From paper I we derive a depth of $-4.184\times 10^{-13}$~erg\,molec$^{-1}$ for the potential well of methane in the filled ice-Ih structure. For the clathrate hydration numbers we take $\nu_1^{Clath}=2/46$ and $\nu_2^{Clath}=6/46$, whereas for filled ice-Ih we assume $\nu^{FI}=1/2$. For the volume difference we find $\Delta\tilde{v}=-1.9\times 10^{-24}$~cm$^3$\,molec$^{-1}$, where we have estimated the volume of filled ice-Ih with an approximate equation of state, at $2$~GPa and room temperature, and the volume for the clathrate hydrate at the same pressure and temperature conditions. For the volume of a methane molecule, in a solid, we use the data of \cite{hazen80} which suggests a volume of $4.032\times 10^{-23}$~cm$^3$ molec$^{-1}$ for a pressure of $2$~GPa. 

Using these values while neglecting the derivatives with respect to pressure of the potential energy we find a value of $3.90$~MPa\,K$^{-1}$ for $\Gamma_{cc}$, at $300$~K. From paper I we estimate $\left\langle\left({\partial W_{CH_4}^{FI}}/{\partial P}\right)_T\right\rangle\approx -3.425\times 10^{-24}$~cm$^3$\,molec$^{-1}$. We further calculate that increasing the pressure on a clathrate cage from $1$~GPa to $2$~GPa will decrease the cage radius from $4.15\times 10^{-8}$~cm to    
$3.99\times 10^{-8}$~cm which yields an estimate of $\left\langle\left({\partial W_{CH_4}^{Clath}}/{\partial P}\right)_T\right\rangle\approx -6.38\times 10^{-24}$~cm$^3$\,molec$^{-1}$. These small effects will change the value of the Clausius-Clapeyron curve to $\Gamma_{cc}=3.75$~MPa\,K$^{-1}$.  Since the value of $\Gamma_{cc}$ is estimated to be positive we argue that the filled ice-Ih and cage clathrate phase transformation will not cause convective cell partitioning.

\subsection{APPENDIX B: METHANE SOLUBILITY IN LIQUID WATER}
 
The degree of solubility of methane in liquid water changes the latter's chemical potential and therefore the methane-water phase diagram topology.  Here we estimate the solubility of methane in liquid water as a function of pressure. The experimental data for methane solubility in liquid water extends up to $2\times 10^8$~Pa \citep[see][and references within]{Duan2006} so that we need to extrapolate over an order of magnitude in the pressure.  In equilibrium the chemical potential of methane in the fluid phase will equal its chemical potential in solution:
\begin{equation}
\eta_{CH_4}^{fluid}=\eta_{CH_4}^{solution}=\eta_{CH_4}^\ast+kT\ln\left(\gamma_{CH_4}\tilde{X}_{CH_4}\right)
\end{equation}
Here we use the convention of representing the chemical potential of methane in solution by a superposition of a hypothetical reference methane chemical potential at the T-P conditions of the solution (marked here with an asterisk) and the logarithm of the mole fraction, $\tilde{X}_{CH_4}$. Deviations from an ideal solution are introduced via the activity coefficient of methane in water, $\gamma_{CH_4}$ \citep{Denbigh}.

Since the partial derivative of the chemical potential per particle with respect to pressure is the volume per particle, we get:
\begin{equation}\label{convth}
\gamma_{CH_4}\tilde{X}_{CH4}=\gamma^0_{CH_4}\tilde{X}^0_{CH4}\exp\left(\frac{1}{kT}\int^P_{P_0}\left[\tilde{v}^{fluid}_{CH_4}-\tilde{v}^\ast_{CH_4}\right]dP\right)
\end{equation}  
where the upper index $0$ refers the quantity to a reference pressure, $P_0$. The first volume appearing in the integrand is the volume per methane molecule in the fluid, which has to be evaluated from the equation of state when experimental data is scarce, which is often the case at high pressures. The second volume (denoted by an asterisk) is a hypothetical one. Some authors assign it a polynomial form whose coefficients are chosen to agree with experimental solubility data \citep[e.g.][]{Duan2006}. Such a fitting procedure will certainly prove risky when trying to extrapolate far beyond the experimental data.

Another common practice is to choose an ideal solution as the reference state, for which the activity coefficient is unity and the energy of mixing is zero. Since ideality is a good approximation in the limit of a very dilute solution the hypothetical volume becomes a volume at infinite dilution \citep{Denbigh}.  This latter volume is also unknown theoretically and introduces additional uncertainty. For example, \cite{Sawamura2007} found experimentally that the volume of infinite dilution may have a negative isothermal compressibility at low pressures and a positive value at high pressures for the same substance, indicating an intricate behavior. Even relatively small errors in the determination of the above-mentioned volumes will accumulate in the integral and their weight will increase exponentially when evaluating the mole fraction of methane in solution. In addition, commonly used activity coefficients, such as the Wilson formalism \citep{Wilson1964} and the universal quasi-chemical equation \citep{Abrams1975}, inherently lack pressure dependencies which are thought to be small at low pressures \citep{Abrams1975} but whose influence may become important for the case of high pressure extrapolations.

The problems we have just mentioned have led authors in the past to deduce, based on a formalism similar to Eq.\,(\ref{convth}), that at high pressure ($\sim 1$\,GPa) the solubility of hydrophobic substances in liquid water may increase dramatically. This means that hydrophobic substances are easily carried in water currents under the application of high pressure.  However we know that this is not the case for the solubility of diatomic nitrogen in liquid water.  There the solubility exhibits a maximum at around $2.7\times 10^8$\,Pa \citep[see][and references therein]{Prausnitz}. A maximum in solubility in liquid water is also found for several aromatic hydrocarbons, around a pressure of $10^8$~Pa \citep{Sawamura2007}. It is indeed argued in \cite{Prausnitz} that the volume of infinite dilution is virtually constant with pressure in contrast to the fluid volume. Therefore, as pressure increases, the fluid volume which is initially larger than the volume of infinite dilution, becomes smaller than the latter, resulting in a maximum in solubility.  Practically, an insufficient accuracy of the fluid equation of state results in having to fictitiously manipulate the volume of infinite dilution to match the data points, weakening the ability to extrapolate beyond the experimental data to high pressures. 

Given the above difficulties we argue that a different approach is in order; a simple kinetic model, which we believe may be more reliable when extrapolating to high pressures where data is scarce or uncertain.  In equilibrium the flux of methane gas molecules arriving at the liquid water surface and dissolving within the water, $F_{in}$, must equal the flux of dissolved methane molecules that reach the liquid-gas boundary layer and are liberated back to the gas phase, $F_{out}$.  For the incoming flux we write:
\begin{equation}\label{Fin}
F_{in}=\frac{1}{4}n_g\bar{v}Pr
\end{equation}
where $n_g$ is the number density of methane molecules in the gas phase and $\bar{v}$ is the average molecular velocity in the gas. We therefore assume for the inward flux the usual flux of gas molecules impinging on a surface, only that we weight it by $Pr$, the probability a methane molecule from the gas phase impinging on the liquid water surface will dissolve in the water rather then scatter or equilibrate and thermally jump back to the gas phase.

In the classical theory of solutions a hydrophobic solute molecule becomes encapsulated in a hydration cell (i.e. cavity) formed of water molecules.  The hydration cell around the hydrophobic solute requires both restructuring and some loss of hydrogen bonding on the water molecules' part \citep[see][and references therein]{Ruckenstein2003}. As water molecules are mobile, particularly in the liquid phase, forming a hydration cell demands that a solute molecule enter immediately or the cell will collapse. In this respect solubility is different from adsorption where a probability of cavity occupation is considered.  In the case of a solution we expect every cavity to be occupied by exactly one methane molecule and the number density of dissolved methane molecules will equal the number density of appropriate cavities. The outward flux may thus be written as:
\begin{equation}\label{Fout}
F_{out} = n_{cavity}v_{cavity}(1-Pr)
\end{equation}  
where $n_{cavity}$ is the number density of cavities or dissolved methane molecules and $v_{cavity}$ is an average cavity velocity. The flux term is multiplied by the probability for cavity collapse and release of the methane back to the gaseous phase. 

For the number density of cavities we write:
\begin{equation}\label{cavitynumberdensity}
n_{cavity}=\frac{N_{cavity}}{N_{H_2O}}n_{H_2O}=\frac{N_{cavity}}{N_{H_2O}}\frac{\rho^w_{liq}}{m_{H_2O}}
\end{equation} 
where $N_{cavity}$ and $N_{H_2O}$ are the total numbers of cavities and water molecules in the liquid system respectively, and $\rho^w_{liq}$ and $m_{H_2O}$ are the bulk water mass density and molecular mass respectively.  The fraction of methane in solution is:
\begin{equation}\label{molfraction1}
\tilde{X}_{CH_4}=\frac{N_{CH_4}}{N_{CH_4}+N_{H_2O}}
\end{equation}
where $N_{CH_4}$ is the total number of dissolved methane molecules. Assuming every cavity is occupied by exactly one methane molecule ($N_{cavity}=N_{CH_4}$) we have from combining Eqs.\,(\ref{cavitynumberdensity}) and (\ref{molfraction1}):
\begin{equation}\label{cavitynumberdensity2}
n_{cavity}=\frac{\rho^w_{liq}}{m_{H_2O}}\frac{\tilde{X}_{CH_4}}{1-\tilde{X}_{CH_4}}
\end{equation} 
Equating the incoming flux (Eq.\,\ref{Fin}) with the outward flux (Eq.\,\ref{Fout}) and substituting for the cavity number density of Eq.\,(\ref{cavitynumberdensity2}) we find after some algebra:
\begin{equation}\label{Methanefraction}
\tilde{X}_{CH_4}=\frac{1}{1+4\left(\frac{kT\rho^w_{liq}}{\hat{f}_{CH_4}m_{H_2O}}\right)\left(\frac{v_{cavity}}{\bar{v}}\right)\left(\frac{1}{Pr}-1\right)}
\end{equation} 
where we have exchanged the methane gas number density with the ratio of its partial pressure to thermal energy. To account for intermolecular interactions which grow in importance when an ideal gas is pressurized we use the fugacity of methane instead of its partial pressure.

We hypothesize a three step mechanism: for a methane molecule impinging on the liquid water surface.  First an opening in the hydrogen bonds must be created through which the methane molecule can pass. Then the methane molecule enters.  Finally, when the hydrophobic solute molecule is in the water, a hydration shell forms around it to complete the process of dissolution.   The probability, $Pr$, that a methane molecule impinging the surface is dissolved, is then given by the product of the Boltzmann probabilities for each of these three steps:
\begin{equation}
Pr=\frac{e^{-\frac{E1}{kT}}}{1+e^{-\frac{E1}{kT}}}\frac{e^{-\frac{E2}{kT}}}{1+e^{-\frac{E2}{kT}}}\frac{e^{-\frac{E3}{kT}}}{1+e^{-\frac{E3}{kT}}}
\end{equation}       
The activation energy, $E1$, for creating an opening in the hydrogen bonded network of liquid water and the activation energy for methane diffusion through this opening, $E2$, are not known. Hence we estimate for them the values derived for the clathrate hydrate network (see Eqs.\,\ref{methanediffusion} and \ref{waterdiffusion}) of $6.9708\times 10^{-13}$ and $6.042\times 10^{-13}$~erg, respectively.  

The evaluation of the energy associated with the solute molecule encapsulation in a liquid water cavity, $E3$, is somewhat more complicated. It is composed of the work required to create the cavity in liquid water and the energy gain from the solute-solvent interactions.  \cite{Graziano1998} has shown, using a solid sphere molecular model, that the work for creating a cavity in liquid water has the following form:
\begin{equation}
W_{cc}=kT\left[\ln\frac{1}{1-\tilde{\xi}}+\frac{3\tilde{\xi}}{1-\tilde{\xi}}\frac{\sigma^{hs}_{CH_4}}{\sigma^{hs}_{H_2O}}+\frac{3\tilde{\xi}(2+\tilde{\xi})}{2(1-\tilde{\xi})^2}\left(\frac{\sigma^{hs}_{CH_4}}{\sigma^{hs}_{H_2O}}\right)^2+\frac{\tilde{\xi} Pv_m}{kT}\left(\frac{\sigma^{hs}_{CH_4}}{\sigma^{hs}_{H_2O}}\right)^3\right]
\end{equation}
where $\sigma^{hs}_{CH_4}$ and $\sigma^{hs}_{H_2O}$ are the hard sphere diameters of methane and water respectively. For the hard sphere diameter of a water molecule we simply assume a value of $2.75\times 10^{-8}$\,cm from hydrogen bond length. The hard sphere diameter for the methane molecule is taken to be a free parameter, for which a value of $5.08\times 10^{-8}$\,cm is found to adequately represent the experimental solubility data.  
$v_m$ is the molecular volume in the bulk solvent, derived from the bulk density and molar mass, for which we estimate a value of $3.0\times 10^{-23}$~cm$^3$\,molec$^{-1}$. The solvent volume packing efficiency, $\tilde{\xi}$, is the ratio of the hard sphere volume of water to $v_m$.  It is important to note that the work required for cavity creation increases linearly with the pressure, $P$.

The hard sphere diameters themselves are functions of the temperature and pressure. We take that into consideration by adjusting the hard sphere diameter to pressure by the multiplication factor:
\begin{equation}\label{multipfactor}
\left(1+\frac{\tilde{B}}{B}P\right)^{-1/3\tilde{B}}
\end{equation} 
where for the bulk modulus, $B$, and its pressure derivative $\tilde{B}$ we assume typical clathrate cage values of $8$\,GPa and $7.61$, respectively. We will not reach temperatures high enough to necessitate thermal corrections to the hard sphere value.

For the energy gain, $W_{gain}$, from solvent-solute interactions we use the formalism of guest-host interactions derived for clathrate hydrates. This, we assume, is a good estimation as the water hydration cavity, formed around a solute molecule in liquid water, is found to have a clathrate-like cage geometry \citep{Glew1962}. The guest-host potential of interaction for a spherical guest molecule was derived by \cite{mckoysinan} by averaging a Kihara pair potential on all space angles summing over all pairs:
$$
W_{gain}(r)=\frac{Z\epsilon}{2}\left[\frac{\rho_m^{12}}{a_{cavity}^{11}r}\left(\delta^{10}+\frac{R}{a_{cavity}}\delta^{11}\right) - \frac{2\rho_m^{6}}{a_{cavity}^{5}r}\left(\delta^{4}+\frac{R}{a_{cavity}}\delta^{5}\right)\right]
$$
\begin{equation}\label{cagepotential}
\delta^N=\frac{1}{N}\left[\left(1-\frac{r}{a_{cavity}}-\frac{R}{a_{cavity}}\right)^{-N} - \left(1+\frac{r}{a_{cavity}}-\frac{R}{a_{cavity}}\right)^{-N}\right]
\end{equation}  
Here $Z$ is the number of water molecules building the cavity.  We take $Z=20$ \citep{Jorgensen1985}. For the spherical cavity radius we assume $a_{cavity}=3.74\times 10^{-8}$~cm based on the carbon-oxygen distance found via molecular cluster simulations \citep{Ruckenstein2003}. The cavity radius is also adjusted to pressure using the multiplication factor of Eq.\,($\ref{multipfactor}$). $R$ is the Kihara core dimension for a methane molecule and $\epsilon$ and $\rho_m$ are the Kihara energy and distance parameters respectively, between methane and water. We apply combining rules on the pure substance parameters taken from \cite{hirschcurbird} to obtain the Kihara mixed potential parameters.

The energy of a hydrophobic solute molecule encapsulation in a liquid water hydration cell is:
\begin{equation}
E3=W_{cc}+W_{gain}
\end{equation} 
For the thermal velocity, $\bar{v}$, of a methane molecule we assume the following form:
\begin{equation}
\bar{v}=\sqrt{\frac{8kT}{\pi m_{CH_4}}}
\end{equation}
where $m_{CH_4}$ is the mass of a methane molecule.  We keep the cavity velocity, $v_{cavity}$, as a free parameter, which we expect to be a function of temperature with some activation energy representing the energy required in the water reconstruction during the process of cavity migration.

\begin{figure}[ht]
\centering
\mbox{\includegraphics[scale=0.5]{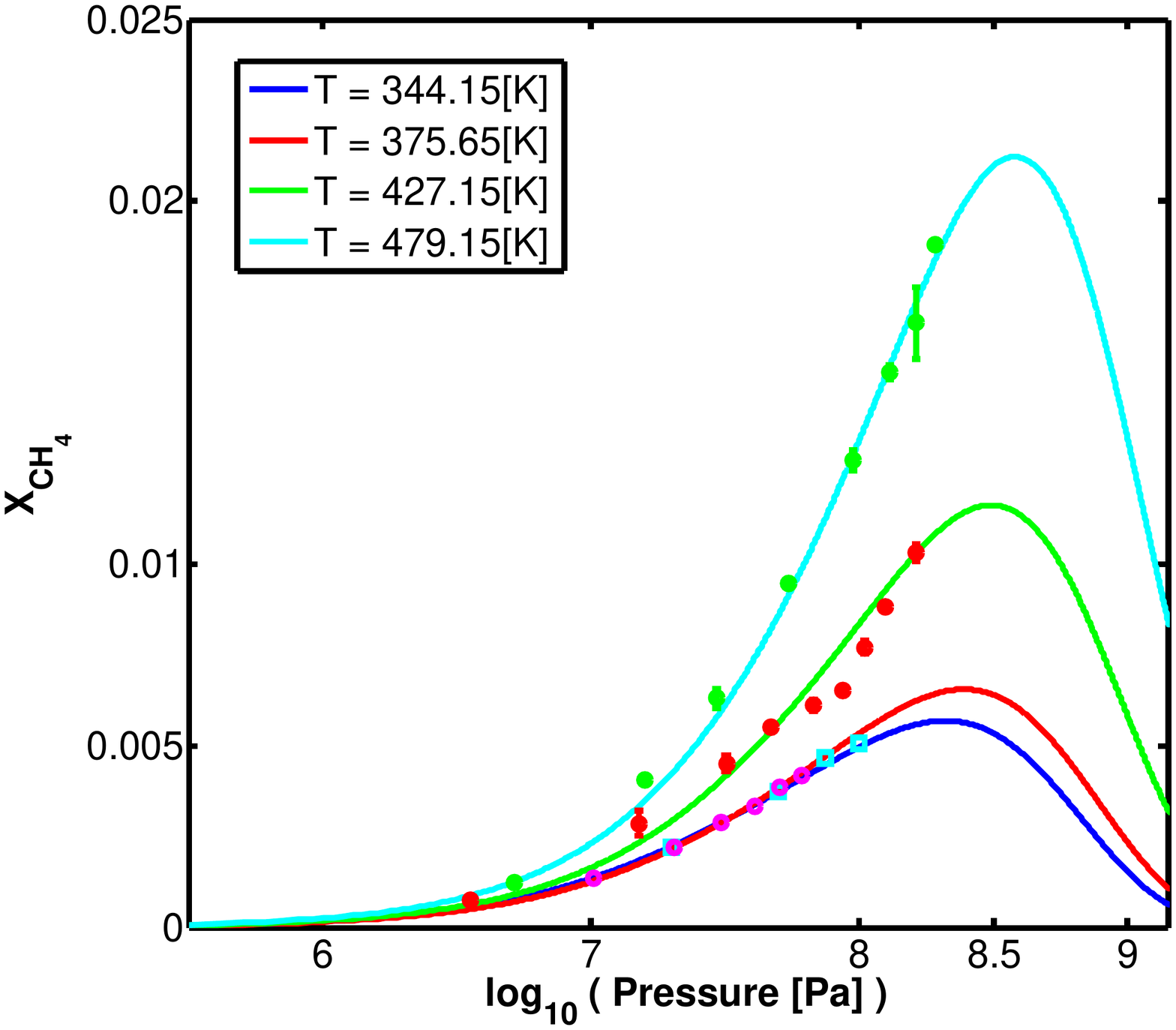}}
\caption{\footnotesize{Estimated methane solubility, in abundance, in a methane-water solution as a function of pressure for four isothermal scenarios: $T=479.15, 427.15, 375.65$ and $344.15$\,K. Solubility data points for $T=479.15$\,K are from \cite{Price1979} (green dots plus error bars in the on-line version). Solubility data points for $T=427.15$\,K are from \cite{Price1979} (red dots plus error bars in the on-line version). Solubility data points for $T=375.65$\,K are from \cite{Osullivan1970} (hollowed magenta circles). Solubility data points for $T=344.15$\,K are from \cite{Dhima1998} (hollowed cyan squares). Solid curves are the theoretical model predictions for the four isotherms tested for.}}
\label{fig:f22}
\end{figure}  

In Fig.\,\ref{fig:f22} we plot our modeled methane solubility in a solution with water as a function of system pressure and compare it with experimental solubility data points for four different temperatures. For each temperature data set we search for the optimal value for the cavity velocity, $v_{cavity}$, by minimizing the average absolute deviation of our model from the data points. Our model fits the data points with an absolute mean deviation of $2.2\%$, $2.0\%$, $10.0\%$ and $4.2\%$ for the cases of the $344.15$\,K, $375.65$\,K, $427.15$\,K and $479.15$\,K isotherms respectively. At first as the pressure increases so does the solubility of methane. Since the work required to create a cavity also increases with pressure and appears in the exponent of the probability, the probability for a hydration cell forming around a hydrophobic solute decreases with pressure, forcing a maximum in the solubility. 

As expected we find that $v_{cavity}$ increases with increasing temperature. Assuming the cavity velocity is a diffusional mechanism with some activation energy, we interpolate on the four values of the cavity velocity for the four data set temperatures, and find for the activation energy a value of $7.6\times 10^{-13}$~erg. This value is larger (though of the same order) than the value we have adopted for $E_1$. This is reasonable as a cavity movement requires the movements of several molecules.

\subsection{APPENDIX C: THERMAL CONDUCTIVITY IN THE BOTTOM BOUNDARY LAYER} 

Estimating the length scale and the temperature difference across the boundary layer connecting the filled water ice mantle and a hypothesized silicate interior (BBL) requires knowledge of the local thermal conductivity. The BBL in our model comprises the deepest part of a methane filled-ice Ih mantle. Since no experimental data exists for the thermal conductivity of this crystal structure a theoretical estimation is in order. Although this order of magnitude estimate will limit our ability to give a precise description of the BBL, its effect on the global problem of methane transport is expected to be very mild.   

A common model for phonon transport and scattering in a crystal assumes a Debye model for the density of modes \citep{Callaway1959}:
\begin{equation}\label{thermconductivityint}
\kappa(T,P)=\frac{1}{2\pi^2C_s}\int_0^{\omega_D}\tau_{total}\frac{\hbar \omega^4}{kT^2}\frac{e^{\frac{\hbar\omega}{kT}}}{\left(e^{\frac{\hbar\omega}{kT}}-1\right)^2}d\omega
\end{equation}  
Here $C_s$ is an average sound speed and $\omega_D$ is the Debye frequency. $\tau_{total}$ is an averaged total relaxation time, representing an averaged time scale for all the phonon scattering mechanisms acting to re-establish an equilibrium Bose-Einstein phonon distribution, for which the phonon transport vanishes. The most important phonon scattering mechanism is the one that can re-establish an equilibrium phonon distribution the fastest, i.e. the one having the shortest relaxation time. Therefore, assuming the different scattering mechanisms are independent to first order, the averaging may be formulated as \citep{Callaway1959}:
\begin{equation}\label{timesum}
\frac{1}{\tau_{total}}=\sum_i\frac{1}{\tau_i}
\end{equation}
where $\tau_i$ is the relaxation time of the $i$th phonon scattering mechanism.
  
\cite{Tse1988} have shown, for a clathrate hydrate of tetrahydrofuran (THF), that the phonon scattering mechanisms originating from the hosting water lattice and those originating from the entrapped guest molecules may be decoupled. We adopt this assumption for the case of methane filled-ice as well. 

First, we assume for the host water lattice of the filled ice Ih the crystal structure of water ice Ih, based on the similarity between both structures \citep{loveday01}. As the temperatures in the deep water mantle are high ($>>10$~K) we neglect boundary scattering of phonons and choose Umklapp scattering as the sole mechanism operating in the water host lattice \citep{ashcroft}.   
The dependencies of the Umklapp relaxation time on temperature and frequency are:
\begin{equation}\label{umklapp}
\frac{1}{\tau_{U}}=B_U\omega^2Te^{\frac{-\Theta_D}{\hat{\alpha} T}}
\end{equation}
where $\omega$ is the mode frequency and $\Theta_D$ is the Debye temperature. The parameters $B_U$ and $\hat{\alpha}$ are determined by substituting $\tau_U$ for $\tau_{total}$ in Eq.\,($\ref{thermconductivityint}$) and fitting the theoretical curve to the experimental data for the thermal conductivity of ice Ih \citep{Tse1988}. We assume $\Theta_D=226$~K and $C_s=2637$~m\,s$^{-1}$ for ice Ih and use the experimental data for the thermal conductivity of ice Ih from \cite{slack1980}. We obtain $B_U=1.4\times 10^{10}$~s\,K$^{-1}$ and estimate $\hat{\alpha}\approx 6.55$. We adopt these values for the host lattice of the filled-ice Ih as well.

The guest molecules, for the case of THF clathrate hydrate, were shown to behave as a resonance phonon scattering mechanism. The translational vibrations of the entrapped guest molecules have frequencies comparable with those of the lattice acoustic phonons, resulting in resonance scattering. It is the translational rather then rotational degrees of freedom of the guest
molecules that yield the resonance effect since the glass-like behavior resulting from this resonance is also seen in monoatomic guest species \citep[see][for a more in depth discussion]{Tse1988}.  \cite{Krivchikov2005b} also suggests that the guest methane molecule behaves as a resonant phonon scatterer, but not to the same degree as for THF clathrate hydrates . 

The phenomenological expression for the relaxation time of a resonant scatterer is \citep{Tse1988}:
\begin{equation}
\frac{1}{\tau_{Res}}=\tilde{n}\psi\frac{\omega_0^2\omega^2}{\left(\omega_0^2-\omega^2\right)^2}
\end{equation}     
where $\tilde{n}$ is the number density of guest molecules in the water lattice, $\psi$ is a coefficient related to the water-methane interaction potential and $\omega_0$ is the translational vibration frequency of the guest responsible for the resonant phonon scattering. Actually, there could be several resonant frequencies at play and $\omega_0$ is therefore a weighted average of them all. It may be that the single resonant model is more adequate for THF clathrate hydrate then for methane hydrate, for which a multiple resonant model should be tested. Such complications are avoided here in view of the lack of experimental data. 

In order to estimate the guest molecule translational frequency, $\omega_0$, we use the cell theory of Lennard-Jones and Devonshire. \cite{mckoysinan} have shown, that treating the guest molecule as an oscillating point mass spring results in the following translational frequency:
\begin{equation}\label{transfreq}
\omega_0=\sqrt{\frac{k_{spring}}{m_{guest}}}=\sqrt{ \frac{2Z_{cage}\epsilon_{LJ}}{m_{guest}a_{cage}^2}}\left[88\left(\frac{\sigma_{LJ}}{a_{cage}}\right)^{12}-20\left(\frac{\sigma_{LJ}}{a_{cage}}\right)^6\right]^{\frac{1}{2}}
\end{equation} 
where $k_{spring}$ is the spring constant, $m_{guest}$ is the mass of the entrapped molecule, $Z_{cage}$ is the number of water molecules building the cage, $a_{cage}$ is the cage radius, and $\epsilon_{LJ}$ and $\sigma_{LJ}$ are the Lennard-Jones potential parameters between the guest molecule and a water molecule.

Using published data for the SI methane clathrate hydrate crystal structure \citep{davidson71} we take $a_{cage}=4.22\times 10^{-8}$~cm for the average cage radius, $Z_{cage}=23$, and $2.66\times 10^{-23}$~g for the mass of a methane molecule. We also take $\epsilon_{LJ}=2.29\times 10^{-14}$~erg and $\sigma_{LJ}=3.14\times 10^{-8}$~cm. These yield $\omega_0=4.4\times 10^{12}$~s$^{-1}$, for the translational frequency of a methane molecule in a SI clathrate hydrate cage. 

We combine $\tau_{Res}$ and $\tau_U$, according to Eq.\,($\ref{timesum}$), and derive the theoretical thermal conductivity for a SI methane clathrate hydrate with the help of Eq.\,($\ref{thermconductivityint}$). Fitting the theoretical curve to the experimental data for the SI methane clathrate hydrate thermal conductivity gives $\psi$. We assume that there are $8$ methane molecules per cubic unit cell in a SI clathrate hydrate, and that the side length is $12\times 10^{-8}$~cm. For the SI clathrate hydrate we take $\Theta_D=218$~K and $C_s=2100$~m s$^{-1}$ \citep{Krivchikov2005b}. This gives $\psi=10^{19}$~cm$^{-3}$~s$^{-1}$. Although the latter numerical value was derived using a fit to the SI clathrate hydrate data we shall adopt it for the methane filled-ice as well.
 
In methane filled-ice the concentration of methane molecules is larger than that in the SI clathrate hydrate, and we estimate its value to be $1.3\times 10^{22}$~cm$^{-3}$ based on the data in \citep{lovedaynat01}.  \cite{loveday01} further points out that the oxygen-carbon distances in methane filled-ice are some $0.5\times 10^{-8}$~cm shorter than the oxygen-carbon distances in SI methane clathrate hydrate. Subtracting this length change from $a_{cage}$ and keeping all other parameters in Eq.\,($\ref{transfreq}$) the same as for SI clathrate hydrate yields $\omega_0=1.1\times 10^{13}$~s$^{-1}$ for filled-ice.

Determining the thermal conductivity of methane filled-ice Ih, at the bottom boundary layer, still requires the sound speed and Debye temperature at a pressure of $\sim 100$~GPa and temperatures as high as $\sim 1000$~K.  From the equation of state for methane filled-ice (see paper I) we find a value of $2.3$~g\,cm$^{-3}$ for the bulk mass density at the relevant pressure, and a speed of sound of approximately:
\begin{equation}
C_s\approx\sqrt{\frac{B_{FI}+\tilde{B}_{FI}P}{\rho_{FI}}}=13.3~{\rm km\,s^{-1}}
\end{equation} 
Therefore, the Debye temperature at the BBL is given by:
\begin{equation}
\Theta_D=\frac{\hbar C_s}{k}\left(\frac{6\pi^2\rho_{FI}}{\bar{m}}\right)^{1/3}=1713~{\rm K}
\end{equation}

Here $\bar{m}$ is an average of the molecular masses of water and methane, weighted according to their abundance in filled ice.

\begin{figure}[ht]
\centering
\includegraphics[scale=0.5]{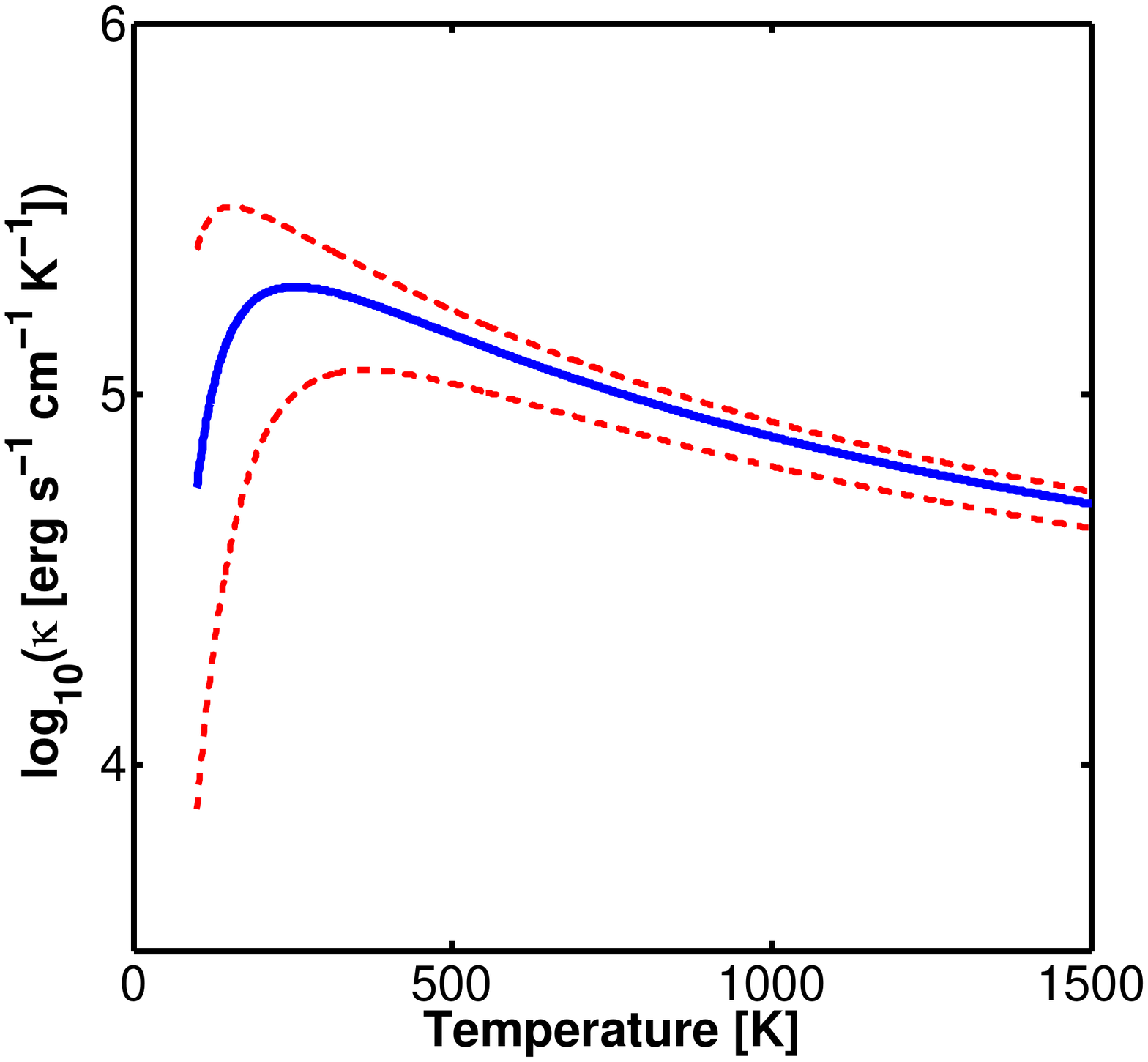}
\caption{\footnotesize{Estimated thermal conductivity for methane filled-ice Ih, at a pressure of $100$\,GPa, appropriate for the conditions at the water mantle and silicate interior boundary layer. Solid curve (blue on the on-line version) is the solution with the parameters given in the subsection text. The two dashed curves (red) are for $\psi$ an order of magnitude higher and lower than the value stated at the text.}}
\label{fig:f23}
\end{figure} 

In Fig.\,$\ref{fig:f23}$ we present our solution for the thermal conductivity at the BBL. Since $\psi$ was obtained using a fitting procedure to SI methane clathrate hydrate we check for its influence on the solution by varying it by an order of magnitude. It is seen that the exact value of $\psi$ is mostly important at temperatures lower than what we expect to find at the BBL. Therefore, errors due to the adoption of $\psi$ from SI clathrate hydrate to filled-ice are less significant for the thermal conductivity estimation at the BBL. We also see from Fig.\,$\ref{fig:f23}$ that while the high pressure acts to increase the thermal conductivity, as the temperature increases the relaxation time for the Umklapp process shortens and finally prevails.

\bibliographystyle{apj}
\bibliography{amitmemo} 

\end{document}